\documentclass[amsmath,amssymb,aps,remo,preprint,uperscriptaddress,eqsecnum]{revtex4-1}
\usepackage{hyperref}
\usepackage{graphicx,subfig}

\usepackage[utf8]{inputenc}
\pdfoutput=1
\begin{document}

\title{Statistical Mechanics of liquids and fluids in curved space}

\author{Gilles Tarjus$^1$}
\author{Fran\c{c}ois Sausset$^2$}
\author{Pascal Viot$^1$}
\bibliographystyle{apsrev4-1.bst}
\affiliation{$^1$ Laboratoire de Physique Th\'eorique de la Mati\`ere Condensée,
 Université Pierre et Marie Curie-Paris 6, UMR CNRS 7600, 4 place Jussieu, 75252 Paris Cedex 05, France\\}

\affiliation{$^2$ Department of Physics,Technion Haifa, 32000 Israel}

\maketitle
\tableofcontents
\section{Introduction} The statistical mechanics of liquids and fluids
is  by now a  mature field which has evolved  from the study of simple
liquids  to that  of  increasingly  complex  systems and phenomena (see e.g \cite{Barrat:2003}).
A research direction which has received only limited attention so far is
that of  liquids embedded in  curved spaces. Although  the topic might
appear  at  first sight  as a   purely theoretical  curiosity  or only
relevant to  general  relativity,  it  arises  in situations  in which
particles are  adsorbed  or  confined on   a substrate with  nonzero
curvature, be it the wall of a porous material, the surface of a large
solid particle or an interface in an oil-water emulsion. In the latter
cases,  the fluid  is  confined to   a curved  manifold   which is  
two-dimensional. Statistical-mechanical systems in three-dimensional  curved spaces and manifolds  are not
encountered  in our everyday life  experience,  but they are useful as
templates or models to study properties whose  direct investigation in  Euclidean three-dimensional space remains
difficult or   inconclusive:   curvature and  geometry    then provide
additional control parameters to envisage the behavior of a system.

In this    article, we review the  progress   made on  the statistical
mechanics of  liquids  and fluids embedded in  curved  space. Our main
focus will be   on  two-dimensional   manifolds of  constant    nonzero
curvature and on the influence of the latter  on the phase behavior,
thermodynamics and  structure  of simple  liquids.  Reference will also be
  made to existing work  on  three-dimensional curved space and
two-dimensional  manifolds with varying  curvature. On the other hand,
we exclude from  the scope of  the article substrates with fluctuating
geometry, such as membranes\cite{Bowick2001,Nelson:2004}. The geometry will always be
considered  as  frozen,  providing the background   whose metrical and
topological characteristics  affect   the behavior  of  the embedded
fluids, but with no feed-back influence from the latter.

The rest of the paper is organized as follows: in Sec.~\ref{sec2}, we review 
physical examples where the curvature of substrate modify the physical properties of systems
in comparison with those observed in in Euclidean space. In Sec.~\ref{sec3}, we discuss the specificities
of thermodynamics for liquids coming from the finiteness of space (constant positive curvature) or strong 
influence of the boundary effects (constant negative curvature). We introduce in Sec.~\ref{sec4} elements of liquid state theory  for describing 
the structure and the thermodynamic properties of fluids in curves \emph{spaces}. Section~\ref{sec5} is devoted
to the influence of the curvature on the structure of liquids as well as the modifications of the Coulomb interaction
We consider in Sec.~\ref{sec6} ill-ordered dense phases where the geometric frustration prevents the appearance of
an ordering transition and slows down drastically the relaxation dynamics of liquids driving the system to 
a glassy behavior. In Sec.~\ref{sec7}, the low-temperature regions of phase diagram is analyzed by means the elastic
theory of defects, which reveals the structure of topological defects generally in excess in curved spaces. Concluding
remarks are drawn in Sec.~\ref{sec8}.

\section{Theoretical           motivations       and          physical
realizations}\label{sec2} 
In  this section we expand  a little more on the reasons for studying 
fluids in curved spaces.  Before moving on to
more   solid  grounds, let us   first   acknowledge that non-Euclidean
geometries are  fun   and fascinating. Musing    about the differences
between spherical and  hyperbolic worlds, i.e.  behavior  in spaces of
constant  positive and negative curvature  respectively, is an exciting
intellectual experience! At a fundamental level, a nonzero curvature
introduces  (at   least)   one extrinsic   lengthscale   (a ``radius  of
curvature'')   in the  behavior of   systems embedded   in such curved
spaces. One then expects  that the ``long-distance'' properties of the
system, namely those involving lengths much larger than the radius (or
radii) of  curvature, are  modified  whereas ``local''  ones should be
rather insensitive to   curvature in general.  Curving space  may also
change  the  topology of  the  substrate, going   for instance from an
infinite flat plane to a sphere or a torus, which also affects some properties of
the  embedded  system  such as the    nature of its  ordered condensed
phases.

From a theoretical point of view, curving space provides an additional
control parameter (the Gaussian  curvature or the associated radius of
curvature for a homogeneous  space of constant curvature) for studying
the properties  of a  fluid or a  liquid, in  addition to  the  common
thermodynamic    parameters. This  may  prove  interesting  in several
situations. First, there  are  cases  for  which  in the    standard
Euclidean space, boundary conditions matter. This is true for instance
for  Coulombic systems in  which charged  particles interact through a
long-range  Coulomb  potential or in  the vicinity  of  a gas-liquid 
critical   point  where  correlations  extend  over  the  whole system
size.  The long-range  character    of the  interactions  or  of   the
correlations  entail the use of  boundary conditions, usually periodic
boundary conditions, in standard statistical-mechanical treatments. An
alternative is  provided  by the    so-called ``spherical    and
hyperspherical boundary conditions''\cite{Kratky1980,PhysRevLett.43.979,Caillol1991}, which  amount to considering the
system on the surface of a sphere (for  a two-dimensional space) or a
hypersphere (for a three-dimensional  space).  A (hyper)sphere is  a
homogeneous  and  isotropic  manifold of   constant  positive Gaussian
curvature. It is of finite extent so that  no boundary conditions need
to be specified. The Euclidean space is  then recovered by letting the
radius of the (hyper)sphere go to infinity.

Another domain in  which  curving space  proves to be of  major theoretical
importance   is what   goes   under  the   concept   of  ``geometric
frustration''\cite{Sadoc:1999}. The  latter  describes an incompatibility  between  the
preferred local order in a system and  the tiling of  the whole space. Geometric
frustration has emerged in the  theoretical description of glasses and
amorphous solids from the  consideration of local icosahedral order in
metallic  glasses.   Icosahedral,  or more generally  poly-tetrahedral
order, is favored for local arrangement of  atoms\cite{Frank:1952} but cannot extend to
form a periodic  tiling of three-dimensional  Euclidean space\cite{Sadoc:1999,Nelson:2002}. It  has first
been  realized by   Kl{\'e}man and Sadoc\cite{Kl'eman1979}, and  further developed by several
groups\cite{Sadoc:1999,Nelson:2002},  that perfect tetrahedral/icosahedral order  can  exist on the
surface of  a hypersphere  with  an appropriately  chosen radius. More
generally, curving space  can be seen as  a way to reduce  (as in the above
example) or   increase  the amount  of  geometric  frustration in  a
liquid.  Increasing frustration is helpful  for  liquids of spherical
particles  in  two dimensions. Indeed, the   ordinary  ``flat`` space then
leads to no  frustration.  The locally preferred hexagonal order can
tile space to produce a  triangular lattice. As a result, one-component
atomic  liquids on  a Euclidean plane  rapidly  order under cooling (or
compressing) and  do not form  glasses. To  study glass  formation  and
amorphous  packings in such systems,  one must introduce curvature. As
suggested by Nelson  and  coworkers\cite{Nelson1983,Nelson:2002,PhysRevB.28.6377}, one could then  mimic frustrated
icosahedral order in  ordinary three-dimensional  space by considering
frustrated  hexagonal  order  in  the  hyperbolic  plane, which   is a
homogeneous   and   isotropic  two-dimensional   manifold of  constant
negative Gaussian curvature.

Up to now,  we have   focused on  purely theoretical motivations   for
studying curved  spaces. As mentioned  in the introduction, there are
also  physical  realizations  of   fluids  on two-dimensional   curved
substrates.  Such   situations  occur  in    adsorption and    coating
phenomena.  Generally, the adsorbing solid   substrate is not flat (it
could  be locally cylindrical or spherical)   and an equilibrium fluid
monolayer  can form on  its surface  if  the fluid-solid attraction is
strong enough.  There are cases where the  curvature can even be quite
strong, with the associated length only moderately larger than that of
the adsorbate.  This happens for instance in the  adsorption of gas molecules inside
the  cavities of zeolite  molecular     sieves or that  of   colloidal
particles  on  larger   spherical  particles  in  the   phenomenon  of
heterofloculation\cite{Post1986}.

Fluid  and condensed-phase behavior is  also observed at the oil-water
interface  in  emulsions\cite{Binks2002,Aveyard2003}. The presence  of   colloidal particles at the
interfaces actually stabilize the emulsions, then often referred to as
Pickering emulsions\cite{pickering1907}. The colloidal particles are irreversibly adsorbed
but  still mobile at  the interface which is usually  the surface of a
spherical droplet. Up  to now, most of  the existing experiments focus
on   dense,  crystal-like  arrangements   of  colloidal  particles  on
spherical droplets\cite{Binks2002,Aveyard2003,Subra2005,Subramaniam2006,Bausch2003,Lipowsky:2005}, but    one  could imagine   studying less   dense,
liquid-like   behavior\cite{Tarimala2004}. One  could  also  consider  adsorbed colloidal
particles  on more exotic   interfaces such  as the  infinite periodic
minimal  surfaces formed for instance  by amphiphilic molecules in the
presence of water\cite{meunier87,Sadoc1988,Hajduk1994}. Such  surfaces have negative, but varying, curvature
and are periodic in  all three directions.  On the  other hand, we  do
not  discuss in this  review situations  in  which the geometry of the
substrate  changes   when coated    by colloidal
particles, as may occur for bubbles in foams for instance\cite{PhysRevLett.99.188301,subra2005b}. In addition,
we only briefly allude to crystalline-like order on curved substrates,
as  this important  topic has  been  recently extensively  reviewed by
Bowick and Giomi\cite{Bowick2009}.

\section{Thermodynamics and boundary effects}\label{sec3}
\subsection{Thermodynamic     limit   and     boundary     conditions}
Thermodynamics,    and  more generally   all   dynamic   and structural
properties of  a macroscopic system, are  retrieved with statistical mechanics
by taking  the  thermodynamic limit in which  one  let the number  of
particles and the  volume of the  system go to  infinity while keeping
the density finite. This also  ensures equivalence between the various
thermodynamic statistical  ensembles. Such a  procedure is, at least
conceptually, easy to     implement  in Euclidean  space   when   both
interactions   and  spatial    correlations    between particles   are
short-ranged. So  long as the system  is finite,  the properties under
study  depend on the  choice of the boundary  conditions, e.g. free or
periodic, but as one takes the thermodynamic limit the contribution of
the  boundary vanishes irrespective of  the chosen conditions (the rate
of  convergence   does however depend on   the  boundary conditions). The same behavior
 also characterizes manifolds  of  varying curvature that  are of
infinite  extension and can be embedded  in Euclidean space. Note however, that in addition to
the subtleties  occurring for long-range,  e.g. Coulombic, interactions
or long-range correlations near a critical point, some care must be exerted
for anisotropic and multi-connected manifolds as the spatial extent of
the system can be made infinite only in certain directions.

Quite  different is the  behavior  of homogeneous  spaces of  constant
nonzero  curvature. Spherical  geometry   associated  with a  positive
curvature indeed leads to spaces of finite extent whereas hyperbolic geometry
associated with   a negative  curvature  allows for   infinite spatial
extent but with a strong effect of the boundary conditions. In what
follows, we discuss  these two main geometries, specializing for simplicity to the
two-dimensional case.
\subsection{Spherical   geometry} In  two   dimensions, the  space  of
constant positive curvature  is the surface  $S_2$ of a sphere. If $R$
is the radius of the sphere, $1/R^2$ is its  Gaussian curvature and its
area $4\pi  R^2$ is obviously finite.   Taking the thermodynamic limit
$R\rightarrow \infty$ ``flattens'' the space and one then recovers the
Euclidean plane. The  same is true  in higher dimensions. If one
wishes   to consider   fluids on   spherical   substrates of  constant
curvature   $1/R^2$, one must resort  to  the thermodynamics of finite
systems.   This  topic     has  been  (ands    still  is)  extensively
discussed.    For small  systems,   one  route is   provided by Hill's
formulation\cite{hill1963} of the thermodynamics of  small systems: one starts with a
finite  canonical system  and consider a   special  ensemble formed by
replicas of this system. The intensive properties of the system depend
separately on the  number of particles $N$ on  a sphere and  the total
area  $A=\pi  R^2$ of the sphere  instead   of being functions  of the
particle density   $\rho=N/A$    only. This formalism   leads   to the
introduction  of an additional   ``pressure'' on  top of  the familiar
spreading  pressure $p$ of  the fluid  on  the curved substrate, which
invalidates  for     small      enough  systems    the     Gibbs-Duhem
equation\cite{hill1963,Post1986}.

The  dependence on the finite size  and finite number of particles can
be illustrated by  looking at  the  virial coefficients of  the density
expansion of the spreading pressure. This expansion reads\cite{Kratky1980}
\begin{equation}\label{eq:eospost}
 \frac{\beta P}{\rho}=1+\sum_{j=2}^\infty B_j(N,R)\rho^{j-1},
\end{equation} 
where $\beta=1/k_BT$ and the first virial coefficients $B_j(N,R)$ for a canonical system are given by \cite{Kratky1980}
\begin{align}\label{eq:B2post}
 B_2(N,R)&=\left(1-\frac{1}{N}\right)\left(B_{2,M}(R)-\frac{R}{2}B'_{2,M}(R)\right),\\
B_3(N,R)&=\left(1-\frac{1}{N}\right)
\left[\left(1-\frac{2}{N}\right) \left(B_{3,M}(R)-\frac{R}{4}B'_{3,M}(R)\right) \right.\nonumber\\
 &+\frac{2}{N}\left. B_{2,M}(R) \left(B_{2,M}(R)-\frac{R}{2}B'_{2,M}(R)\right) \right],
\end{align}
with $B'_{j,M}(R)=\partial B_{j,M}(R)/\partial R$. The $B_{j,M}(R)$'s are the Mayer cluster integrals which are defined as usual, for a spherically 
symmetric pair interaction potential 
$u(r)$ by
\begin{equation}
 B_{2,M}(R)=-\frac{1}{2A(R)}\int_{A(R)}\int_{A(R)}dS_1 dS_2 f_{12}
\end{equation}
and
\begin{equation}
 B_{3,M}(R)=-\frac{1}{3A(R)}\int_{A(R)}\int_{A(R)}\int_{A(R)}dS_1 dS_2 dS_3 f_{12} f_{13} f_{23},
\end{equation}  with $f_{ij}=\exp(-\beta u(r_{ij}))-1$, $r_{ij}$ being
the geodesic distance between points $i$ and $j$ on $S_2$, $A(R)=4 \pi R^2$, and
$dS$ is the differential area on  $S_2$. The first virial coefficients
have been computed for hard disks on $S_2$ (and hard spheres on $S_3$)
and the resulting approximate equation of state for the spreading pressure
compared  to   simulation results. This     will be discussed  later
on.  Note  that  we  have interpreted   above the interaction  between
particles as acting  in  curved space and  therefore depending  on the
geodesic distance on $S_2$. This corresponds to the ``curved line-of-force'' case\cite{Post1986}. 
When the curved  space can be embedded in a
higher-dimensional Euclidean  space   (here, the  sphere  $S_2$ in  the
$3$-dimensional Euclidean  space    $E_3$),  one  can    also  envisage
interactions  acting  through  the   embedding Euclidean space,  which
corresponds to  the ``Euclidean  line-of-force'' case\cite{Post1986}. Which
case   arises   depends    on   the physics    of  the   system  under
consideration. Most existing studies use  the ``curved line-of-force''
description and this is what we shall do throughout this article.

It is worth stressing again that spherical geometry does not allow one
to study   finite-size  effects at  both constant  curvature and  constant
particle density. This however may be  viewed as a  boon if one wishes
to study the  {\it flat} case   in the thermodynamic limit.  Spherical
boundary conditions with spheres of increasing  radius indeed offer an
alternative  to the  more  common periodic boundary conditions   implemented
directly in the Euclidean space.
\subsection{Hyperbolic geometry}  If finiteness is the characteristic
of spherical substrates, a quite different and singular behavior takes
place  in  hyperbolic world.  The hyperbolic plane   $H_2$, which  is a
homogeneous,  simply  connected two-dimensional manifold  of  constant
negative curvature,  is of  infinite extent\cite{Hilbert:1952,Coxeter:1969},  (see Appendix A).  The thermodynamic
limit,  at  constant curvature   and   constant particle density,  can
therefore be taken in such a space. However, a peculiar feature of the
hyperbolic geometry  is that because  of the exponential  character of
the metric at large distance, the boundary of a finite region of $H_2$
grows as  fast as the  total area of this region  when  the size of the
latter increases.  To be more specific,  consider the hyperbolic plane
of Gaussian curvature  $K=-\kappa^2$.  The metric in  polar  coordinates
($r,\phi$) is given by

\begin{equation}
 ds^2=dr^2+\left(\frac{\sinh (\kappa r)}{\kappa}\right)^2d\phi^2.
\end{equation}

This form makes  apparent the connection  with the spherical metric of
the  sphere   $S_2$ which  is obtained   by  replacing in   the  above
expression $\kappa$ by  $iR^{-1}$. See Appendix A. We  shall
refer  to  $\kappa^{-1}$  as   the ``radius of   curvature''  of  the
hyperbolic plane. Still in polar coordinates, the differential area is
expressed as
\begin{equation}
 dS=\frac{\sinh(\kappa r)}{\kappa}drd\phi,
\end{equation} 
which again is  the counterpart  of the expression  for
$S_2$ with the  replacement of $\kappa$ by $iR^{-1}$.  It  is a simple
exercise in hyperbolic geometry  to show that the area  of a disk of radius
$r$ is given  by $A(r)=2\pi\kappa^{-2}(\cosh(\kappa r)-1)$ whereas its
perimeter is equal  to  $P(r)=2\pi\kappa^{-1}\sinh(\kappa r)$ (see Appendix A).  For  a
large radius such that $\kappa r\gg 1$, one then finds as announced above
that both $A(r)$ and $P(r)$ grow exponentially with the distance $r$,
as $\exp(\kappa r)$.

As a   result of the above   property,  the  boundary  effects  are  never
negligible, even    in  the thermodynamic  limit!  An  illustration is
provided by the  ideal gas behavior\cite{modes:235701,modes:041125}.  The canonical
partition  $Q_1^{id}$ can be   obtained from the eigenenergies of  the
Schrödinger equation for a single  particle in a large domain $\Sigma$
of   $H_2$.    This   requires determining    the    spectrum  of  the
Laplace-Beltrami operator  with specified  boundary conditions. For an
open domain  $\Sigma$, the result,  obtained  as an expansion  in the
inverse of   the de Broglie  thermal  wavelength $\lambda_T=\sqrt{2\pi
\hbar^2/(mk_BT)}$ is as follows\cite{kean67,modes:235701,modes:041125}.
\begin{equation}
 Q_1^{id}(T)=\frac{A(\Sigma)}{\lambda^2_T}\pm \frac{P(\Sigma)}{4\lambda_T}+\frac{1}{12\pi}\left[\int_\Sigma dS\,K +\int_{\partial \Sigma} ds\, K_g \right]+O(\lambda_T),
\end{equation}
where $K=-\kappa^2$  is the Gaussian curvature,  $K_g$
the geodesic curvature of the boundary  $\partial \Sigma$, and the plus
or minus  signs of the second term  correspond to Neumann or Dirichlet
boundary conditions  for  the wavefunction  on $\partial  \Sigma$. The
perimeter $P(\Sigma)$ being of the same order  as the area $A(\Sigma)$
for a large domain, the above expression proves  the dependence of the
ideal  gas  partition  function  on the  boundary  conditions  in  the
thermodynamic limit.

If one    nonetheless  insists on    defining ``bulk''  thermodynamic
properties  in  $H_2$,  a  solution is to   consider periodic boundary
conditions. The  procedure however  is  quite complex  in $H_2$ as one
must change  the boundary condition  when increasing the  area of the
primitive cell  in which the  fluid is  embedded. This is  detailed in
Ref.\cite{Sausset:2007} and is  briefly   explained in  Appendix B.   Imposing  periodic
boundary conditions amounts to replicating a  primitive cell (chosen as
a regular  polygon)  to realize   a  tiling of the  whole   hyperbolic
plane. From  a topological  point of view,  this procedure  leads to a
compact ''quotient space`` which is obtained by identifying in a specific
manner the edges of the primitive cell by pairs. In the Euclidean case,
it is well known that one obtains in this way for square and hexagonal
periodic  boundary conditions  a  one-hole  torus. For  the hyperbolic
case, tori  of  genus $g$ (i.e.   with $g$ holes)   with  $g\ge 2$ are
generated\cite{Sausset:2007}. Under such  periodic boundary conditions, the boundary terms
associated with $\partial   \Sigma$ (and $P(\Sigma)$)   disappear as the
manifold is now compact (boundaryless) and one is left with
\begin{equation}
 Q_1^{id}(T)=A(\Sigma)\left[\frac{1}{\lambda^2_T}-\frac{\kappa^2}{12\pi}\right],
\end{equation} which is  positive and proportional  to the area of the
system  if the de Broglie wavelength  is small  enough compared to the
radius  of curvature $\kappa^{-1}$. Note that  due to the Gauss-Bonnet
theorem\cite{Sadoc:1999}, the  area $A$ of a  primitive cell associated  with a compact
quotient space of genus $g$ is fixed and equal to
\begin{equation}
A=2\pi \kappa^{-2}(g-1),
\end{equation} 
where, we recall, $g\ge 2$ on $H_2$.

This, rather long, detour via the ideal-gas limit illustrates that (i)
boundary effects always have  to be  considered in hyperbolic geometry,
even  in  the thermodynamic   limit,  and (ii)   ``bulk`` thermodynamic
quantities     can   be  defined      by   using   periodic   boundary
conditions.  Actually,  a poor    man's  way of  studying   such  bulk
properties is to  follow a procedure  commonly employed in  analytical
studies  of  statistical-mechanical systems  defined  on  a  so-called
''Bethe lattice``\cite{Domb1960,M'ezard2001}.  This amounts to  focusing on the ''deep interior``
of a very large domain, far enough from  the boundary and restricting in all
spatial integrals that appear in the calculation of the fluid properties
to  this deep   interior.   In  practice,  one should   exclude   from
computations  a boundary region,  which is taken  as a region of 
width (or width whose ratio to the linear size  of the system goes to
zero  when  the  latter goes   to infinity  (near  the  boundary of an
otherwise very large  domain), and subsequently take  the thermodynamic
limit. We shall come back to this point in the following sections.

\section{Liquid state theory in spaces of constant nonzero curvature}\label{sec4}
\subsection{Statistical  mechanics}\label{sec41} We consider a  liquid
at equilibrium in  canonical conditions (fixed temperature $T$, volume
$V$  and  number of  particles  $N$)   in a  $d-$dimensional  Riemannian
manifold. The latter is equipped with a metric that can be expressed in
a $d-$dimensional set of coordinates ${\bf x}=(x_1,..,x_d)$ as
\begin{equation}\label{eq:ds} 
 ds^2=\sum_{i,j=1}^{d}g_{ij}({\bf x})dx^idx^j,
\end{equation} where $g_{ij}({\bf x})$ is the metric tensor, from which one can also derive
by   standard  differential   geometry  the  geodesic  equation,   the
Levi-Civita  connection and    Riemann    (curvature)  tensor\cite{Terras:1985,goetz1970,Sadoc:1999}.    Thee
infinitesimal element of volume then reads
\begin{equation}\label{eq:dS}
 dS=\sqrt{|g({\bf x})|}\prod_{i=1}^d dx_i
\end{equation} where    $g({\bf x})$   is the  determinant  of   the  metric
tensor. The   geometry is assumed  to  be frozen,  i.e. $g_{ij}({\bf x})$ is
fixed in each point  of  the manifold  and is  not influenced by   the
behavior  of the liquid. For simplicity,  we restrict the presentation
to a one-component atomic liquid  with pairwise additive  interactions
$u(r)$ that only depend on the geodesic  distance between atoms computed with the metric in Eq.~(\ref{eq:ds})
(''curved line of force``, see  Sec.~\ref{sec3}). The canonical  partition function is
given by
\begin{equation}\label{eq:partitionini}
 Q_N(V,T)=Q^{id}_N(V,T)\frac{Z_N(V,T)}{V^N}
\end{equation}
where $Q^{id}_N(V,T)=\frac{Q^{id}_1(V,T)^N}{N!}$, is the ideal-gas contribution already considered
 in the previous section and $Z_N(V,T)$ is the configurational integral defined
as 
\begin{equation}\label{eq:partition}
 Z_N(V,T)=\int_V...\int_VdS_1...dS_N \exp\left[-\frac{\beta}{2}\displaystyle\sideset{}{'} \sum_{i, j=1}^N u(r_{ij})\right],
\end{equation} with $dS_i$ given by Eq.~(\ref{eq:dS}) and the prime on the double sum indicating that the term $i=j$
is excluded. As was discussed
in  some  detail in  Sec.~\ref{sec3},   a rigorous definition  of  the
thermodynamic quantities may be subtle in curved  space due either to the
finiteness of the system (spherical geometry) of  to the importance of
the boundary conditions  (hyperbolic  geometry). In addition   to this
fundamental issue, there is  a  practical difficulty when  considering
spaces which are not homogeneous and isotropic, for instance manifolds
with spatially varying   curvature. Thermodynamics may then still   be
defined but the intrinsic inhomogeneity or anisotropy of the embedding
space  makes practical calculations very   complex: for instance,  the  
harmonic analysis  (Fourier   analysis, convolution  theorem,  Laplace
eigenvalues, etc)  is only defined on symmetric spaces, namely manifolds
of constant  curvature. As far  as we  know, there are  virtually  no existing
statistical-mechanical studies   of this  kind.   Therefore, in what
follows,  and with   the goal  to   illustrate  the  influence  of the
substrate curvature    on  the  behavior  of the fluid,  we  only
consider homogeneous manifolds of constant curvature.

From the canonical partition function, one has access to the free-energy,
 the energy and the thermodynamic pressure. The latter is defined as
\begin{equation}\label{eq:pressure}
 \beta P=\left(\frac{\partial \ln(Q_N)}{\partial V}\right)_{T,N}.
\end{equation} 
As noted  in  Sec.~\ref{sec3}, the above  expression
corresponds to  the ''spreading pressure``  for a  spherical substrate.
In the case  of  a  hyperbolic geometry,   we  will focus on  the
''bulk  thermodynamic pressure`` which is   defined either through the
use    of  periodic boundary   conditions      or by restricting   the
configurational integrals to the deep   interior  of the system,   far
enough from the boundary (see however the discussion in Sec.~\ref{sec52}).
In the canonical ensemble, one can also define as usual the $n-$particle correlation (or distribution)
functions  through
\begin{equation}\label{eq:correl}
 g^{(n)}_N({\bf r}_1,{\bf r}_2,...,{\bf r}_n)=\frac{V^N}{Z_N}\int_V...\int_V
dS_{n+1}...dS_{N}\exp\left[-\frac{\beta}{2}\displaystyle\sideset{}{'} \sum_{i, j=1}^N u(r_{ij})\right].
\end{equation}
The above expression is valid in the spherical geometry for a large enough number $N$ of particles 
(otherwise, there are $1/N$ corrections: see Refs\cite{Kratky1980,Post1986}) and in the hyperbolic geometry for the
``bulk'' functions (see above). In both cases, the embedding space can be considered as homogeneous and 
isotropic so that $g^{(1)}_N({\bf r}_1)=1$, $g^{(2)}_N({\bf r}_1,{\bf r}_2)$ only depends on the geodesic distance
$r_{12}$, etc.

Having defined the thermodynamic quantities and the correlation functions for fluids constrained in curved spaces 
(of constant curvature), one can derive the various relations among them. These relations take simple
forms in the case where the interactions are pairwise additive (as considered here) and then only involve the pair 
correlation function.
\subsection{Thermodynamic pressure and equation of state}\label{sec42}
Whereas the expression for the excess internal energy in terms of the pair correlation function is a straightforward
extension of the standard formula in Euclidean space\cite{Hansen1986}, that for the thermodynamic pressure requires
more caution in its derivation. We treat the case of positive curvature (spherical geometry) and negative curvature
(hyperbolic geometry) separately. Indeed, in the former situation, a change of the total volume, as required from the
thermodynamic definition in Eq.~(\ref{eq:pressure}), implies a change of curvature while the total volume can be changed
at constant curvature in the latter case. For ease of exposition, we deal with two-dimensional manifolds, but the
reasoning is easily extended to higher dimensions.

Let us consider first a spherical substrate formed by the surface $S_2$ of a sphere of radius $R$. The spherical
metric can be conveniently expressed in polar coordinates with two angles $\theta$ (colatitude) and $\phi$ (longitude)
as 
\begin{equation}
 ds^2=R^2\sin(\theta) d\theta d\phi,
\end{equation}
or equivalently, by introducing the geodesic distance to the north pole $r=R\theta$ ($0\leq r\leq \pi R$),
 as $ds^2=\left[R\sin\left(\frac{r}{R}\right)\right]dr d\phi$, which makes the comparison with the Euclidean
and hyperbolic cases more direct. The configurational integral in Eq.~(\ref{eq:partition}) can be reexpressed as
\begin{equation}\label{partitiontheta}
 Z_N(A,T)=R^{2N}\int_0^{\pi}d\theta_1 \sin(\theta_1)\int_0^{2\pi}d\phi_1...\int_0^{\pi}d\theta_N \sin(\theta_N)\int_0^{2\pi}d\phi_N
\exp\left[-\frac{\beta}{2}\displaystyle\sideset{}{'}\sum_{i, j=1}^N  u(R\theta_{ij})\right],
\end{equation}
where $\theta_{ij}=\theta_{i}-\theta_{j}$ and the total area $A=4\pi R^2$. By using Eq.~(\ref{eq:partitionini}) and 
Eq.~(\ref{eq:pressure}) with the derivative with respect to the ``volume'' replaced by $\partial A= 
2\left(\frac{A}{R}\right)\partial R$, one finds
\begin{equation}
 \beta P =\frac{R}{2A Z_N}\left.\frac{\partial}{\partial R}Z_N\right|_{N,T},
\end{equation}
which, after inserting Eq.~(\ref{partitiontheta}) and the definition of the pair correlation  function in Eq.~(\ref{eq:correl}),
gives
\begin{align}
 \beta P&= \frac{N}{A}-\frac{\beta R}{4A}\displaystyle \sideset{}{'}\sum_{i, j=1}^N \frac{ R^4}{A^2}
\int_0^{\pi}d\theta_i \sin(\theta_i)\int_0^{2\pi}d\phi_i \int_0^{\pi} 
d\theta_j \sin(\theta_j)\nonumber\\&
\int_0^{2\pi}d\phi_j \,u'(R\theta_{ij})\,
\theta_{ij}\, g^{(2)}_N(R\theta_i,\phi_i,R\theta_j,\phi_j),
\end{align}
where $u'(r)=\frac{d}{dr}u(r)$. Finally,  by using the homogeneity and isotropy of space, introducing the particle
 density $\rho=N/A$, and considering $N\gg 1$, one arrives at the following expression for the equation of 
state\cite{Kratky1980,Post1986}
\begin{equation} \label{eq:pressureS2}
 \frac{\beta P}{\rho}=1-\frac{\beta\pi \rho R}{2}\int_0^{\pi R} dr \,r \sin\left(\frac{r}{R}\right)u'(r)g(r),
\end{equation}
where $g(r)\equiv g^{(2)}_N(r)$ is the radial distribution function depending on the geodesic distance $r$ between pairs of
particles. The expression can be generalized to small $N$ systems and to high dimensions as 
well\cite{Kratky1980,Kratky1982,Schreiner1983,Fanti1989}.

We  now move on to the hyperbolic plane $H_2$. The total area can be varied at constant (negative) curvature 
$K=-\kappa^2$, but to avoid boundary problems (see Sec.~\ref{sec3}) in computing the bulk thermodynamic pressure, we use
a variant of the Green-Bogoliubov method\cite{widomrowlinson82,Sausset2009}. The derivative in Eq.~(\ref{eq:pressure}) is performed via 
an affine transformation of the elementary area element,
\begin{equation}
 dS'=(1+\xi)dS,
\end{equation}
with $\xi$ an infinitesimal parameter. Due to the form of the hyperbolic metric (see Sec.~\ref{sec3} and Appendix A), this
transformation leaves the polar angle $\phi$ unchanged, whereas the radial coordinate $r$ becomes
\begin{equation}\label{eq:gbtrans}
 r'=r\xi\frac{(\cosh(\kappa r)-1)}{\kappa \sinh (\kappa r)}+O(\xi^2),
\end{equation}
which, contrary to the Euclidean case, amounts to a nonlinear transformation of the coordinate. We then consider the 
infinitesimal variation of the configurational integral which is generated by the above transformation.
Taking advantage of the homogeneity and isotropy of space (in the bulk), one finds
\begin{equation}
 \delta \ln(Z_N(A,T))=N\xi -\frac{\beta \rho^2}{2}A\int dS \,g(r) \,\delta u(r)
\end{equation}
where $dS=\kappa^{-1}\sinh(\kappa r)dr d\phi$ and $\delta u(r)$ is the infinitesimal change of the pair
potential which, by using Eq.~(\ref{eq:gbtrans}), can be written as
\begin{equation}
 \delta u(r)= \xi \frac{(\cosh(\kappa r)-1)}{\kappa \sinh (\kappa r)} u'(r)+O(\xi^2).
\end{equation}
The final result reads\cite{Sausset2009}
\begin{equation}\label{eq:pressureH2}
 \frac{\beta P}{\rho}=1-\frac{\beta\pi \rho}{2\kappa^2}\int_0^{\infty} dr (\cosh(\kappa r)-1) u'(r)g(r),
\end{equation}
where we have taken the thermodynamic limit by letting the range of integration go to infinity. The above
expression is the equation of state for the bulk thermodynamic pressure in $H_2$. In the specific case of Coulombic
systems, other types of pressures can be defined\cite{Fantoni2003}: this will be discussed 
in section\ref{sec52}. 

Note that when the curvature goes to zero, i.e. when $R\rightarrow \infty$ in Eq.~(\ref{eq:pressureS2}) and when 
$\kappa\rightarrow 0$ in Eq.~(\ref{eq:pressureH2}), both the spherical and hyperbolic equations of state reduce to the
Euclidean one, 
\begin{equation}
  \frac{\beta P}{\rho}=1-\frac{\beta\pi \rho}{4}\int_0^{\infty} dr r^2 u'(r)g(r).
\end{equation}
However, the results on $H_2$ and $S_2$ are not simply related by replacing $\kappa$ by $iR^{-1}$ (as it is true
for the metric, see Sec.~\ref{sec3} and Appendix A).

To illustrate the influence of curvature, one may consider the
virial expansion of the equation of state, which describes the low-density fluid: see Eq.~(\ref{eq:eospost}). The virial
coefficients can be derived by a direct expansion of the partition function or by inserting the density expansion
of the radial distribution function in the equation of state, Eqs.~(\ref{eq:pressureS2}) or (\ref{eq:pressureH2}).
The first correction to ideal-gas behavior is obtained by setting $g(r)=e^{-\beta u(r)}$, which leads after integrating
by parts to 
\begin{equation}\label{eq:B2S2}
 \left.B_2\right|_{S_2}=-\frac{\pi R}{2}\int_0^{\pi R} dr \left(\sin\left(\frac{r}{R}\right)+
\frac{r}{R}\cos\left(\frac{r}{R}\right)\right)
\left(e^{-\beta u(r)}-1\right)
\end{equation}
and 
\begin{equation}\label{eq:B2H2}
 \left.B_2\right|_{H_2}=-\frac{\pi }{\kappa}\int_0^{\infty} dr \sinh\left(\kappa r\right)\left(e^{-\beta u(r)}-1\right).
\end{equation}
One checks that both expressions reduce to the Euclidean formula for the second virial coefficient when the 
curvature goes to zero. By introducing as in Sec.~\ref{sec3} the usual Mayer integral with the appropriate metric,
\begin{equation}
 B_{2,M}(R \mbox{\rm{ or }$\kappa^{-1}$})=-\frac{1}{2}\int dS f(r) 
\end{equation}
with $f(r)=e^{-\beta u(r)}-1$, it is easy to check that $B_2|_{H_2}$ is equal to the corresponding Mayer integral, whereas
$B_2|_{S_2}$ verifies Eq.~(\ref{eq:B2post}) 
(with, here, $N\gg 1$). The difference of behavior between $H_2$ and $S_2$ reflects the property that the spherical 
manifold is finite and that its total area can only be varied by changing the curvature, i.e. the radius of the 
sphere.

\subsection{Correlation functions and integral equations}\label{sec43}
In liquid-state theory, it is well established that density expansions are only of limited use for describing the
liquid phase and that a more fruitful approach consists in deriving approximate integral equations for the pair
correlation functions\cite{Hansen1986}. This is based on the Ornstein-Zernike equation that relates the radial distribution function $g(r)$,
 or more precisely the so-called total pair correlation function $h(r)=g(r)-1$, to the direct correlation function
$c(r)$:
\begin{equation}\label{eq:convo}
 h(r)=c(r)+\rho \int_\Sigma dS' h(r')\,c(t({\bf r},{\bf r}')),
\end{equation}
where $\Sigma$ is either $S_2$ and $H_2$ (again, the formalism is easily generalized to higher-dimensional 
non-Euclidean Riemannian manifolds of constant curvature) and where $t({\bf r},{\bf r}')$ is the modulus of the (geodesic)
displacement associated with an element of the spherical or hyperbolic translation group. Eq.~(\ref{eq:convo}) should be
considered as a ``bulk'' property in the case of $H_2$, with all boundary effects removed when taking the 
thermodynamic limit. In the case of $S_2$, where finiteness is a source of additional difficulties compared to the
Euclidean space, we consider large enough systems so that the number of particles $N$ can be taken as a 
continuous variable and the explicit $1/N$ corrections can be neglected (see Sec.~\ref{sec3}). The Ornstein-Zernike
equation can then be interpreted as expressing the Legendre transform between canonical and grand-canonical ensembles,
the direct correlation function $c(r)$ being the second functional derivative of the grand potential with respect
to local density fluctuations\cite{Hansen1986}. Again, this interpretation, as well as the equivalence between thermodynamic
 ensembles, is rigorous in the case of  bulk properties in hyperbolic geometry but requires some caution in
 spherical geometry when small systems are investigated\cite{hill1963,Post1986}. With this proviso in mind, one can derive
the compressibility relation,
\begin{equation}\label{eq:oz}
 \frac{\rho \chi_T}{\beta}=1+\rho \int_\Sigma dS \,h(r)=\frac{1}{1-\rho \int_\Sigma dS \,c(r)},
\end{equation}
where $\chi_T$ is the isothermal compressibility and the second relation is a consequence of the Ornstein-Zernike equation.

For homogeneous spaces of constant curvature, it is convenient to reexpress the Ornstein-Zernike equation by using
a generalization of the Fourier transform. Harmonic analysis on such non-Euclidean spaces has been developed and, in
a nutshell, the plane waves used in conventional Fourier transform for a Euclidean space  are replaced by the 
eigenfunctions of the appropriate Laplace-Beltrami operator in  curved space (see Appendix A). On the sphere $S_2$, one makes use
of the Fourier expansion in terms of  spherical harmonics expressed with the two angles $\theta$ (colatitude) and
$\phi$ (longitude). Any function $f(\theta,\phi)$ (recall that $r=R\theta)$ can be written as
\begin{equation}\label{eq:fourierS2}
 f(\theta,\phi)=\sum_{k= 0}^\infty\sum_{|l|\leq k} \hat{f}(k,l)Y_{kl}(\theta,\phi),
\end{equation}
where 
\begin{equation}
 \hat{f}(k,l)=\int_0^{\pi}d\theta \sin(\theta) \int_0^{2\pi}d\phi f(\theta,\phi)
Y_{kl}^*(\theta,\phi)  
\end{equation}
and a star denotes the complex conjugate. For a function which is independent of $\phi$ (namely an isotropic function),
Eq.~(\ref{eq:fourierS2}) reduces to a Legendre transform
\begin{equation}
 f(\theta)=\sum_{k=0}^\infty \hat{f}_k P_k(\cos(\theta)),
\end{equation}
where 
\begin{equation}\label{eq:ozS2}
 \hat{f}_k=\left(\frac{2k+1}{2}\right)\int_{-1}^1 dx P_k(x) f(\arccos(x))
\end{equation}
and $P_k(x)$ is the \emph{k}th Legendre polynomial. Convolutions are easily calculated so that the Ornstein-Zernike equation
takes the form\cite{Lishchuk2006}
\begin{equation}\label{eq:ozfourierH2}
 1+\rho \frac{4\pi R^2}{(2k+1)}\hat{h}_k=\frac{1}{1-\rho \frac{4\pi R^2}{(2k+1)}\hat{c}_k}\,\,\,.
\end{equation}
It is easily checked that $4\pi R^2\hat{h}_0=\int_{S_2}dS \,h(r)$, so that the $k=0$ term of Eq.~(\ref{eq:ozS2}) provides
the compressibility via Eq.~(\ref{eq:oz})\cite{Lishchuk2006}.

The counterpart of the Fourier transform on the hyperbolic plane $H_2$ is the Fourier-Helgason 
transform\cite{Terras:1985,Helgason2005}. For a generic function $f$ of the polar coordinates $(r,\phi)$, it is defined
 as\cite{Terras:1985} 
\begin{equation}\label{eq:fourierH2}
\tilde{f}(n,t)=\kappa^{-1}\int_0^\infty dr \int_0^{2\pi} d\phi \sinh(\kappa r) e^{in\phi}
P^n_{-1/2+it}(\cosh(\kappa r))f(r,\phi),
\end{equation}
whereas the  inverse Fourier-Helgason transform reads
\begin{equation}\label{eq:fourierinvS2}
f(r,\phi)=\frac{\kappa^2}{2\pi}\sum_{n=-\infty}^{+\infty}(-1)^n\int_0^\infty dt \, t\, \tanh(\pi t) e^{-in\phi}
P^{-n}_{-1/2+it}(\cosh(\kappa r))\tilde{f}(n,t).
\end{equation}
For an isotropic function $f(r)$, the Fourier-Helgason function reduces to a Mehler-Fock transform\cite{gonzalez97}; the dependence on
$n$ disappears and, after rewriting $t=\frac{k}{\kappa}$, one has
\begin{equation}\label{eq:fourierH2radial}
\tilde{f}(k)=2\pi\kappa^{-1}\int_0^\infty dr  \sinh(\kappa r) P_{-1/2+ik/\kappa}(\cosh(\kappa r))f(r),
\end{equation}
and
\begin{equation}\label{eq:fourierinvH2radial}
f(r)=\frac{1}{4\pi}\int_0^\infty dk\,  k \tanh\!\left(\frac{\pi k}{\kappa}\right) 
P_{-1/2+ik/\kappa}(\cosh(\kappa r))\tilde{f}(k).
\end{equation}
In the above expressions, $P^n_{-1/2+it}(x)$ is a Legendre function of the first kind (conical function) and 
$P_{-1/2+it}(x)=P^0_{-1/2+it}(x)$. The Fourier-Helgason transform satisfies the convolution theorem, so that 
the Ornstein-Zernike equation, Eq.~(\ref{eq:convo}), can be rewritten as
\begin{equation}\label{eq:ozfourierS2}
 1+\rho \tilde{h}_k=\frac{1}{1-\rho \tilde{c}_k}.
\end{equation}
Both Eqs.(\ref{eq:ozfourierS2}) and (\ref{eq:ozfourierH2}) converge to the usual Euclidean equation with 
$\frac{4\pi R^2}{(2k+1)}\hat{h}_k$ and $\tilde{h}(k)$  reducing to the standard Fourier transform when the curvature goes to 
zero.

One again encounters the peculiarities of curved spaces that have already been pointed out. We have mentioned above
the difficulty associated with finiteness in the spherical geometry. A quite different feature occurs in hyperbolic
space. One indeed finds that the $k=0$ component of the Fourier-Helgason/Mehler-Fock transform, 
\begin{equation}
 \tilde{h}(k=0)=2\pi\kappa^{-1}\int_0^\infty dr  \sinh(\kappa r) P_{-1/2}(\cosh(\kappa r))f(r),
\end{equation}
is different than the integral of $h(r)$ over the whole space, i.e.,
\begin{equation}\label{eq:helgasonnot0}
 \tilde{h}(k=0)\neq 2\pi\kappa^{-1}\int_0^\infty dr  \sinh(\kappa r) f(r).
\end{equation}
Actually, $ \tilde{h}(k=0)$ is always smaller than the integral. As a result, the $k=0$ component of 
Eq.~(\ref{eq:ozfourierS2}) does {\it not} give the compressibility of the fluid. We shall dwell more on this issue when 
considering the liquid-gas critical point.

The starting point of the integral equation approach to the structure and the thermodynamics of liquids is that
 the direct  correlation function $c(r)$ is simpler and shorter-ranged than $h(r)$. Is is therefore a better 
candidate for devising approximations. Common approximations that can also be used in homogeneous curved spaces are
the Percus-Yevick (PY) and hypernetted chain (HNC) closures, in which 
\begin{eqnarray}\label{eq:PY}
 c(r)=&(1+\gamma(r))[\exp(-\beta u(r))-1]& \mbox{\hspace{2cm}	\rm (PY)}\\
\label{eq:HNC}
 c(r)=&\exp(-\beta u(r)+\gamma(r))-(1+\gamma(r))& \mbox{\hspace{2cm}	\rm (HNC)}
\end{eqnarray}
where $\gamma(r)=h(r)-c(r)$. Once the solution of the Ornstein-Zernike equation is obtained, there are several routes
to compute the thermodynamics from the pair correlation functions: the expression of the excess internal energy, the
compressibility relation (Eq.~(\ref{eq:oz})) and the equation of state for thermodynamic pressure (Eqs.~\ref{eq:pressureS2})
 or (\ref{eq:pressureH2})). As usual, the approximate nature of the approach leads to thermodynamic inconstancy and
the thermodynamic pressure obtained by  thermodynamic integration of the compressibility relation
\begin{equation}\label{eq:eoscompress}
 \frac{\beta P}{\rho}=1-\frac{1}{\rho}\int_0^{\rho} d\rho' \rho' \int_\Sigma dS \,c(r,\rho'),
\end{equation}
where $c(r,\rho)$ is the direct correlation function at density $\rho$, differs form the ``virial'' expression in 
Eqs. (\ref{eq:pressureS2}) or (\ref{eq:pressureH2}). The numerical procedures for approximate integral equations have been given
 in Ref~.\cite{Lishchuk2006} for the sphere $S_2$ and in Ref~.\cite{Sausset2009} for the hyperbolic plane $H_2$.
Results will be discussed in Sec.~\ref{sec5}.

\subsection{Interaction potentials}\label{sec44}
We still restrict ourselves to pairwise additive interactions and spherically symmetric potentials. If needed,
these restrictions can be lifted, but it would lead here to unnecessary complications. We have already mentioned one 
subtlety concerning interactions in curved space: when the manifold in which the fluid is confined is embedded in
higher-dimensional Euclidean space, distance-dependent pair potentials can either depend on the geodesic distance 
(``curved line of force'') or the Euclidean distance, e.g. the chord between  two points on a sphere (``Euclidean
line of force''). To our knowledge, work on the latter type is however very 
limited\cite{Post1986,Post1988}. Note that the alternative does not arise
in hyperbolic geometry which is not embeddable in Euclidean space. Another subtlety in spherical geometry, and more generally, on compact manifold, comes
with the definition of interacting pairs. Two particles on a sphere (or a hypersphere) are joined by two segments of 
a geodesic (great circle) and can therefore interact ``twice'' with pair potentials associated with the two geodesic
distances. For hard spherical particles, this clearly corresponds to the physical situation. For longer-range potentials
 and for a more formal use of ``spherical boundary conditions'' as a way to approach the thermodynamic limit in
Euclidean space (see Sec.~\ref{sec3} and Ref.\cite{Kratky1980}) this ``double distance convention''\cite{Kratky1980,Schreiner1983} is
to be contrasted with the minimum image convention used for the ``toroidal condition'' corresponding to the standard
implementation of periodic boundary conditions in Euclidean space. Note again that this feature is absent in the case
of the hyperbolic plane $H_2$ (although the use of periodic boundary conditions will also bring in compact toroidal 
conditions).

Before moving on to a discussion of the nature of intermolecular forces in curved spaces, it is worth considering 
the restriction put on the form of the pair interaction potentials by the requirement that the various configurational
integrals are well defined. In spherical geometry, the geodesic distance between two particles is bounded by $\pi R$
so that the interaction potentials $u(r)$ are meaningless for $ \pi R$. (Even hard core systems make no sense if
the hard-core diameter $\sigma$ in the spherical manifold is of the order of the (hyper)sphere radius $R$.)
In hyperbolic geometry, a constraint arises from the exponential nature of the metric at long distances 
(see Sec.~\ref{sec3} and Appendix A). At large separations, the correlations die out and $g(r)\rightarrow 1$. A typical
configurational integral that appears in the calculation of the excess internal energy or of the pressure in $H_2$ is
then 
\begin{equation}\label{eq:integral}
 I=\int_{r_c}^\infty dr \sinh(\kappa r) u(r) \simeq \frac{1}{2}\int_{r_c}^\infty dr e^{\kappa r} u(r), 
\end{equation}
where $r_c$ is chosen much larger than the radius of curvature $\kappa^{-1}$. The integral in Eq.~(\ref{eq:integral}) 
converges if the pair interaction decreases faster than $\exp(-\kappa r)$ at long distance. One therefore concludes
that the thermodynamic limit only exists in $H_2$ (and higher-dimensional hyperbolic manifolds) if and only if the spherically
symmetric pair potentials vanish exponentially fast with distance above the radius of curvature, i.e. faster than 
$\exp(-\kappa r)$. For instance, this precludes using nontruncated power-law potentials such as the celebrated 
Lennard-Jones atomic model. 

After this instructive detour, let us come back to the basics of intermolecular forces\cite{Buckingham1970}. On
top of the short-range repulsive interactions whose origin lies in the overlap of the outer electron shells of the atoms,
longer-range interactions can be derived from electrostatics, leading to Coulombic, dipolar, etc... interactions, as 
well as from the multipole dispersion interactions between the instantaneous electric moments in one atom and those
induced in the other\cite{Buckingham1970}. Roughly speaking, all the longer-range interactions can be obtained from
the knowledge of the Coulomb potential. In infinite Euclidean and hyperbolic space (we consider the two-dimensional
case for simplicity), the Coulomb potential $v(r)$ at a geodesic distance $r$ of a unit point charge is obtained from
 the Poisson equation
\begin{equation}\label{eq:PoissonS2}
\Delta v(r) =-2\pi\delta^{(2)}(r,\phi), 
\end{equation}
where $\Delta$ is the Laplace-Beltrami operator in polar coordinates and $\delta^{(2)}(r,\phi)$ is the Dirac
distribution in the appropriate manifold. The boundary  condition is that the potential vanishes at infinity.
In $H_2$ (see Appendix A), the  solution of Eq.~(\ref{eq:PoissonS2}) satisfying the boundary conditions is\cite{Jancovici1998} 
\begin{equation}
v(r)=-\ln\left(\tanh\left(\frac{\kappa r}{2}\right)\right),
\end{equation}
which behaves as  $2 \exp(-\kappa r)$ as $\kappa r  \rightarrow \infty$.

In spherical geometry, the Poisson equation, has no acceptable solution as it leads to a singularity for particles on opposite poles. A way out is
to define the Coulomb potential for a ``pseudo-charge'' corresponding to a unit positive point charge and a uniform background of total 
charge $-1~$\cite{Caillol1991,Caillol1982,Caillol1992,Caillol1992Caillol1992,Caillol2000} (another definition consists in taking a $+1$ point charge and a  $-1$ point charge 
located at the antipodal position\cite{Caillol2000}). The resulting potential on a sphere $S_2$ of radius $R$\cite{Caillol1991} is 
\begin{equation}\label{eq:CoulombS2}
v(r)=-\ln\left(\sin\left(\frac{ r}{2R}\right)\right).
\end{equation}
Note that the hyperbolic and spherical expressions of the Coulomb potential converge to the two-dimensional Euclidean logarithmic potential, $-\ln(r/cst)$, when the curvature goes to zero. 
(The extension of Eq.~(\ref{eq:CoulombS2})  to the hypersphere $S_3$ is given in Ref.\cite{Caillol1991,Caillol1993}).)

The above expressions of the Coulomb potential in curved space indicate the direction that one should take to properly define interatomic forces in the ``curved line-of-force'' 
interpretation. In the hyperbolic geometry, for which this is crucial, one should envisage multipolar and dispersion interactions as generated by 
appropriate derivatives of the 
Coulomb potential. As a consequence, all such pair interactions decrease exponentially at long distance as $\exp(-\kappa r)$ or faster. 
Consider for instance  a dipole made of two opposite charges $\pm q$ located at  positions ${\bf r}_1$ and ${\bf r}_2 $. The distance separating the two charges is denoted 
by $d$.
By using the relation between geodesic distances (see Appendix A) and taking the limit when $d\rightarrow 0$, $q\rightarrow \infty$ with $p=qd$ finite,  one obtains
\begin{equation}
 v_d(r,\phi)=-\frac{p\cos(\phi)}{\sinh(\kappa r)},
\end{equation}
where $p$ is the dipole moment an d$\phi$ the angle between the dipole moment and the  direction of the geodesic joining the position of the dipole and the center $O$.
 A crude reasoning, based on a one-dimensional  Drude model, 
suggests that  dispersion forces give rise to a pair interaction potential that goes as 
\begin{equation}
 u(r)\sim -\frac{\cosh(\kappa r)^2}{\sinh(\kappa r)^4}.
\end{equation}
Therefore,  at short distance, the potential behaves as $1/r^{4}$, as expected in two-dimensional Euclidean space, 
whereas decays exponentially as $\exp(-2\kappa r)$ when the distance $r$ is larger than the radius of curvature $\kappa^{-1}$.
In Euclidean space, it is known that the  simple Drude model captures the leading dependence on $r$, more sophiscated and realistic treatments only changing the prefactor of the interaction. 

If one is interested in studying generic properties of simple liquid models in hyperbolic space, with no direct connection to realistic interatomic forces,
 one should
make sure, at least,  that the model pair potentials are truncated beyond some cut-off so that the long-distance behavior remains well defined. 

\subsection{Computer simulations}\label{sec45}
Computer simulations play a major role in liquid-state studies in Euclidean space.
Unsurprisingly, they have also been developed for investigating fluid and condensed phases in curved space. 
Monte Carlo simulations for short-range\cite{Kratky1982,Schreiner1983,PhysRevB.30.6592,PhysRevB.34.405,Post1986,Tobochnik1988,Fanti1989,Giarritta1992,Giarritta1993,Caillol1998}
and long-range Coulombic-like\cite{PhysRevB.33.499,Caillol1992Caillol1992,Caillol1992,Caillol1999} potentials in 
two and three-dimensional  geometry,  and more seldom Molecular
Dynamic studies in two-dimensional spherical\cite{PhysRevLett.82.4078,PhysRevB.58.9677,PhysRevB.62.17043} and hyperbolic manifolds\cite{modes:235701,modes:041125,sausset:155701,PhysRevLett.104.065701,PhysRevE.81.031504}, have been performed . Early on, numerical
packing protocols have also been considered in spherical\cite{Kraschrei1982,Wille1987} and hyperbolic\cite{PhysRevB.28.6377} geometries.
In this section, we do not intend to give details on the methods that have been implemented but rather to stress the differences with the Euclidean case and the technical 
difficulties encountered.

A simplification coming with spherical backgrounds is that there is no need for periodic boundary conditions as the whole space is closed and finite. As already mentioned, this is 
the rationale underlying the use of ``spherical boundary conditions`` in order to approach the thermodynamic limit in Euclidean space. The price to pay is that in Monte Carlo algorithms
one must ensure a proper sampling of the (positively) curved manifold, with e.g. no bias toward the poles or the equator\cite{Kratky1982}, and in Molecular Dynamic studies, one has to be careful 
with the equations of motion in curved space. The technical difficulties however are reasonable.

The case of the hyperbolic plane $H_2$ (higher-dimensional hyperbolic manifolds have not been studied by computer simulation), which, contrary to spherical manifolds, cannot
be embedded in higher-dimensional Euclidean space, is much trickier. First, parallel transport in $H_2$ is complex, and implementing an algorithm for  solving the equations of motion
either for hard-core particles (where collisions must be handled\cite{modes:235701,modes:041125}) or continuous interaction potentials (where forces must computed and
vector quantities added and transported\cite{sausset:155701,PhysRevLett.104.065701}) is by no means obvious. However, the main challenge lies in the necessary use of boundary conditions. As discussed in 
Sec.~\ref{sec3} and further developed in the preceding subsections~\ref{sec41}, \ref{sec42}, \ref{sec43}, one is indeed interested in the ``bulk'' behavior of liquids in the hyperbolic plane. In
a simulation that is anyhow constrained to finite systems, the only way to limit the boundary effects, which are expected to be strong in hyperbolic geometry (see Sec.~\ref{sec3}),
is to implement periodic boundary conditions\cite{Sausset:2007}. This amounts to choosing a primitive cell (which contains the physical system) such that it can be infinitely replicated to tile the whole
plane. To ensure smoothness and consistency, the edges of the cell must be paired in a specific way. In $H_2$, the smallest such primitive cell equipped with an edge-pairing (or 
``fundamental polygon'') is an octagon. As the area of the primitive cell is fixed for a given choice of boundary condition, one must change the latter, and increase the number of
edges of the fundamental polygon, to study finite-size effects at constant curvature and particle density\cite{Sausset:2007}. This of course does not occur in Euclidean space 
where, due to the absence of any metric-related length scale, the system size can be changed while keeping the same type of boundary condition. Hyperbolic boundary conditions have been
developed and discussed in Ref.\cite{Sausset:2007} and implemented in Molecular Dynamics in Refs.\cite{sausset:155701,PhysRevLett.104.065701,PhysRevE.81.031504}. More details are given in Appendix B.

\subsection{Self-motion and diffusion equation}\label{sec46}
Up to this point, we have focused on static properties. We now briefly consider aspects pertaining to the dynamics of fluids in curved manifolds, more specifically the self-diffusion
of particles. In Euclidean space, the self-motion of the particles in a liquid is described by space-time correlation functions. The simplest one is the so-called self-intermediate
scattering function $F_s(k,t)$ which is the Fourier transform of the self part of the density-density time 
correlation function (or self van Hove function $G_s(r,t))$\cite{Hansen1986}),
\begin{equation}
 F_s(k,t)=\frac{1}{N}\sum_{j=1}^N\langle e^{i{\bf k}({\bf r}_j(t)-{\bf r}_j(0))}\rangle 
= \frac{1}{N}\sum_{j=1}^N\langle\cos(kd_j(0,t))\rangle,
\end{equation}
where $d_j(0,t)$ is the distance traveled by atom $j$ during times $0$ and $t$ and the bracket denotes an equilibrium
canonical average\cite{Hansen1986}. When $k\rightarrow 0$ and $t \rightarrow \infty$, $F_s(k,t)$ is well described by a Gaussian, 
corresponding to a truncated cumulant expansion, $\exp\left[\frac{-k^2}{2d}\langle d(0,t)^2\rangle\right]$, and in the same
long wavelength and long time limit, one expects that the self van Hove function obeys a diffusion equation similar to 
Fick's macroscopic law, which leads to $ F_s(k,t)=\exp\left[-k^2 Dt\right]$ with $D$ the coefficient of diffusion\cite{Hansen1986}.
Identification of the two expressions gives the Einstein relation,
\begin{equation}
 2dD=\lim_{t\rightarrow \infty}\left\lbrace \frac{\langle d(0,t)^2\rangle}{t}\right\rbrace.
\end{equation}

The generalization of the self-intermediate scattering function in spherical and hyperbolic geometries makes use 
of the appropriate extensions of the Fourier transform (see Sec.~\ref{sec43}). For $S_2$, one finds
\begin{equation}
  F_s(k,t)=\frac{1}{N}\sum_{j=1}^N\langle P_{kR}\left(\cos\left(\frac{d_j(0,t)}{R}\right)\right)\rangle,
\end{equation}
where $kR \in \mathbb{N}$, and on $H_2$
\begin{equation}
  F_s(k,t)=\frac{1}{N}\sum_{j=1}^N\langle
 P_{-\frac{1}{2}+i \frac{k}{\kappa}}\left(\cosh\left(\kappa d_j(0,t)\right)\right)\rangle.
\end{equation}
We recall that $P_{kR}(x)$ is a Legendre polynomial and $P_{-\frac{1}{2}+i \frac{k}{\kappa}}(x)$ is a Legendre function
of the first kind, while $d_j(0,t)$ is the geodesic distance traveled by atom $j$ between $0$ and $t$ and calculated with the 
appropriate metric.

One may then wonder what is the small $k$ and long time limit of the intermediate scattering function in non-Euclidean
space. One expects that it converges to a result predicted by the associated diffusion equation,
\begin{equation}
 \frac{\partial G_s(r,t)}{\partial t}=D \,\Delta G_s(r,t),
\end{equation}
with $\Delta$ the Laplace-Beltrami operator acting in the curved manifold. (Actually, by isotropy one expects $G_s$ to
only depend on $r$ so that only the ``radial'' part of the operator is required.) We keep focusing on the 
two-dimensional manifolds $S_2$ and $H_2$ for simplicity. The diffusion problem has been solved in both cases with
the solution in $S_2$ given by\cite{Caillol2004} 
\begin{equation}\label{eq:diffusionS2}
 G_s^{(S_2)}(r,t)=\sum_{k=0}^\infty \left(\frac{2k+1}{k}\right)\sin\left(\frac{r}{R}\right)
P_k\left(\cos\left(\frac{r}{R}\right)\right)e^{-\frac{k(k+1)Dt}{R^2}},
\end{equation}
and that in $H_2$ by\cite{Monthus1996a}
\begin{equation}\label{eq:diffusionH2}
 G_s^{(H_2)}(r,t)=\frac{e^{-\frac{\kappa^2Dt}{4}}}{2\sqrt{2\pi}(Dt)^{\frac{3}{2}}}\sinh(\kappa r)\int_r^\infty 
dy \frac{y\,e^{-\frac{y^2}{4Dt}}}{\sqrt{\cosh(\kappa y)-\cosh(\kappa r)}}.
\end{equation}
One can check that when the curvature goes to zero, i.e. $R\rightarrow +\infty$ in Eq.~(\ref{eq:diffusionS2}) and 
$\kappa \rightarrow 0$ in Eq.~(\ref{eq:diffusionH2}), both expressions reduce to the standard two-dimensional Euclidean 
formula,
\begin{equation}\label{eq:diffusionE2}
 G_s^{(E_2)}(r,t)=\frac{e^{-\frac{-r^2}{4Dt}}}{4\pi Dt}.
\end{equation}
Generalizations to the three-dimensional manifolds $S_3$ and $H_3$ are  given in Ref.\cite{Nissfolk2003,Caillol2004}
and Ref.\cite{Monthus1996a}, respectively.

\begin{figure}
\resizebox{10cm}{!}{\includegraphics{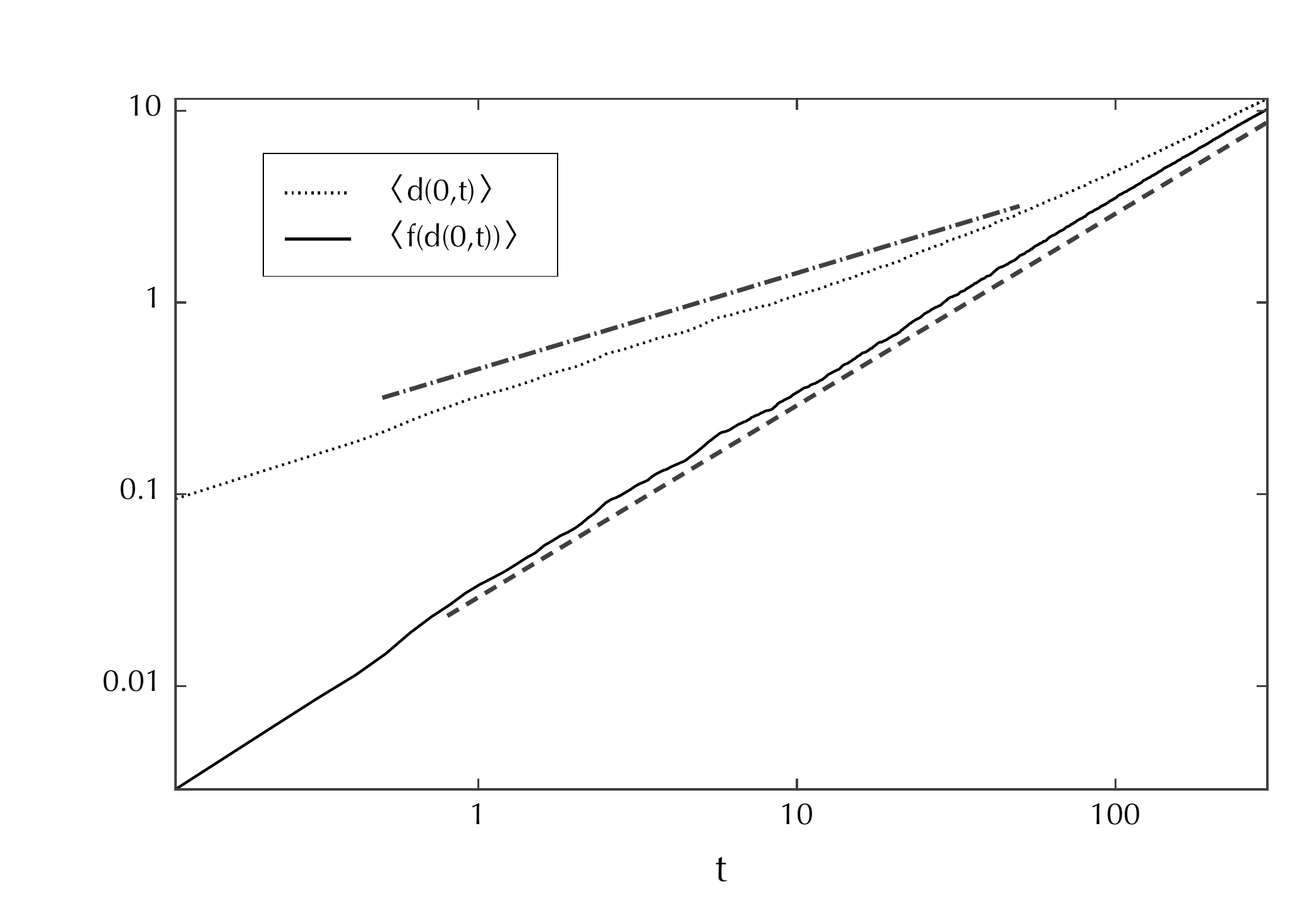}}
 \caption{Log–log plot of the mean absolute displacement $\langle d(0,t)\rangle$  in units of $\kappa^{-1}$ and 
of $\langle f(d(0,t))\rangle $ (see
Eq.~(\ref{eq:diffu})). The dashed line has a slope equal to $1$ and the dotted-dashed, a slope of $1/2$. To good approximation,
 $\langle f(d(0,t))\rangle$ has a  linear time dependence at all times, 
which corresponds to a diffusive regime in the
hyperbolic sense. The parameters are $\rho\sigma^2=0.91$, $T=2.17$, and $\kappa\sigma=0.2$.
 The system with octagonal periodic boundary conditions comprises $287$ atoms.}\label{fig:diffu}
\end{figure}

Diffusion on a sphere (or a hypersphere) is such that the particles can never escape from the closed space they are
embedded in and, as a result, the mean square displacement $\langle d(0,t)^2\rangle$ goes to a constant value when 
 $t\rightarrow +\infty$. Procedures to fit simulation data for extracting a self diffusion coefficient from the 
$\langle d(0,t)^2\rangle$ have been given in Refs.\cite{Nissfolk2003,Caillol2004}. Diffusion on Riemann manifolds 
(with a non constant curvature) has 
been recently investigated in the short-time limit\cite{castro2010}. For the hyperbolic plane,
the diffusion motion at long times is characterized by two regimes: when the mean traveled distance is large, but 
smaller than the radius of curvature $\kappa^{-1}$, one has an ordinary diffusion regime with 
$\langle d(0,t)^2\rangle\sim Dt$ and $\langle d(0,t)\rangle\sim \sqrt{Dt}$; on the other hand, when the mean traveled
distance is large compared to $\kappa^{-1}$, one finds a ``ballistic'' diffusion regime with 
$\langle d(0,t)^2\rangle\sim (Dt)^2$ and $\langle d(0,t)\rangle\sim Dt$\cite{Monthus1996a,Sausset2008b}. It has been
shown in Ref.\cite{Sausset2008b} that the self-diffusion coefficient can be conveniently extracted from simulation 
data of the function $\frac{1}{N}\sum_{j=1}^N\langle f(d_j(0,t))\rangle$ where 
\begin{equation}\label{eq:diffu}
 f(r)=\kappa^{-2}\ln\left(\frac{1+\cosh(\kappa r)}{2}\right).
\end{equation}
Indeed, when calculated with the solution of the hyperbolic diffusion equation, this function behaves as $Dt$ for 
{\it both} the ordinary and ballistic diffusion regimes. In Fig.~\ref{fig:diffu}, we show the Molecular Dynamics simulation result for 
a fluid with  truncated Lennard-Jones interactions on $H_2$ (see below)\cite{Sausset2008b} at a rather high temperature. It is 
clearly found that the atomic motion becomes diffusive at long enough time, and the self-diffusion coefficient can be
determined without having to worry about the crossover between ordinary and ballistic diffusion.

\section{Curvature effect on the thermodynamics and structure of simple fluids}\label{sec5}
\subsection{Short-range interaction potentials}\label{sec51}
In order to illustrate the influence of a nonzero curvature on the properties of a fluid embedded in a non-Euclidean 
space, we begin with the two dimensional hard-disk fluid on a sphere $S_2$ of radius $R$ and 
a hyperbolic plane $H_2$ of radius of curvature $\kappa^{-1}$. (We recall that the Gaussian curvature $K$ is constant
in both cases and equal to $1/R^2$ for $S_2$ and $-\kappa^{-2}$ for $H_2$.) The ``virial'' equation of state giving
the thermodynamic pressure, i.e. the spreading pressure for $S_2$ and the bulk pressure for $H_2$ (see preceding sections),
is obtained from Eqs. (\ref{eq:pressureS2}) and (\ref{eq:pressureH2}) after using the fact that for hard disks of 
diameter $\sigma$ in curved space, $\frac{d}{dr}\exp(-\beta u(r))=\delta(r-\sigma)$. For $S_2$\cite{Kratky1980,Post1986},
\begin{equation}\label{eq:eoshardS2}
 \left.\frac{\beta P}{\rho}\right|_{S_2}=1+\rho\frac{\pi R^2}{2}\left(\frac{\sigma}{R}\right)
\sin\left(\frac{\sigma}{R}\right)g(\sigma^+),
\end{equation}
and for $H_2$\cite{Sausset2009}
\begin{equation}\label{eq:eoshardH2}
 \left.\frac{\beta P}{\rho}\right|_{H_2}=1+\rho\frac{\pi}{2\kappa^2}
\left(\cosh\left(\kappa\sigma\right)-1\right)g(\sigma^+),
\end{equation}
where in both cases $g(\sigma^+)$ is the radial distribution function at contact.

In the low density limit, $g(\sigma^+)$ goes to $1$ and the second viral coefficient (see also Eqs.(\ref{eq:B2S2}) and
(\ref{eq:B2H2})) reads
\begin{equation}\label{eq:B2hardS2}
 \left. B_2\right|_{S_2}=\frac{\pi R^2}{2}\left(\frac{\sigma}{R}\right)\sin\left(\frac{\sigma}{R}\right),
\end{equation}
and 
\begin{equation}\label{eq:B2hardH2}
 \left. B_2\right|_{H_2}=\frac{\pi }{\kappa^2}\left(\cosh\left(\kappa\sigma\right)-1\right),
\end{equation}
where we have considered $N\gg 1$ for $S_2$ and Eq.~(\ref{eq:B2hardS2}) is only defined for $\sigma<\pi R$. For $H_2$, 
even if nothing a priori  prevents one from taking the particle diameter much larger than $\kappa^{-1}$, the discussion 
in Sec.~\ref{sec44} indicates that the physical interactions decrease  exponentially with  distance beyond $\kappa^{-1}$.
Particle diameters $\sigma \gg\kappa^{-1}$, which lead to an exponentially growing $\left. B_2\right|_{S_2}$, have therefore no physical
significance. Higher-order virial  coefficients have been calculated for $S_2$\cite{Kratky1980,Post1986} and for 
$H_2$\cite{modes:041125}. (The three-dimensional spherical case $S_3$ has been considered 
in Refs.\cite{Kratky1982,Tobochnik1988,Fanti1989}.)

As already noted, there is no simple  symmetry $\kappa \leftrightarrow iR^{-1}$ between the $H_2$ and $S_2$ pressure expressions,
due to the finite size of the sphere whose total area can only be changed by varying the curvature. For instance, when 
the curvature goes to zero in $S_2$ and $H_2$, the equation of state and virial coefficients approach the Euclidean
result, albeit in a non-symmetric way. As an illustration,
\begin{equation}
  \left. B_2\right|_{S_2}= \left. B_2\right|_{E_2}\left(1-\frac{1}{6}K\sigma^2 +O(K^2\sigma^4)\right),
\end{equation}
\begin{equation}
  \left. B_2\right|_{H_2}= \left. B_2\right|_{E_2}\left(1+\frac{1}{12}K\sigma^2 +O(K^2\sigma^4)\right),
\end{equation}
where $\left. B_2\right|_{E_2}=\pi \sigma^2/2$. The asymmetry between $S_2$ and $H_2$ is not compensated if one express the virial expression in terms of the
 packing fraction $\eta$ 
(or rather ``surface coverage`` in two dimensions) instead of the density $\rho$. Indeed, 
\begin{equation}
 \left. \eta\right|_{S_2}= 2\pi R^2\left(1-\cos\left(\frac{\sigma}{2R}\right)\right)\rho,
\end{equation}
\begin{equation}
 \left. \eta\right|_{H_2}= \frac{2\pi}{\kappa^2} \left(\cosh\left(\frac{ \kappa\sigma}{2}\right)-1\right)\rho,
\end{equation}
so that to first order,
\begin{equation}\label{eq:eosB2S2}
\left.\frac{\beta P}{\rho}\right|_{S_2}=1+\eta\left(\frac{\sigma}{2R}\right)
\frac{\sin\left(\frac{\sigma}{2R}\right) \cos \left(\frac{\sigma}{2R}\right)}{1-\cos \left(\frac{\sigma}{2R}\right)}+O(\eta^2)
\end{equation}
and 
\begin{equation}\label{eq:eosB2H2}
\left.\frac{\beta P}{\rho}\right|_{H_2}=1+\eta
\frac{\sinh\left(\frac{\kappa\sigma}{2}\right)^2 }{\cosh \left(\frac{\kappa\sigma}{2}\right)-1}+O(\eta^2).
\end{equation}

To go beyond the low-density expansion, several routes have been taken for exploring the hard-disk fluid in spherical and hyperbolic geometries. One  can build approximate equations
of state from scaled particle theory\cite{Post1988,Lishchuk2009}, free area 
theory\cite{modes:235701,modes:041125}, or by constructing rational approximants
from the first virial coefficients\cite{modes:041125}, with possible additional input from some close packing 
density\cite{Post1986,Tobochnik1988,haro:116101}. One can also compute the radial distribution function $g(r)$, either
from approximate self-consistent integral equations\cite{Lishchuk2006,Sausset2009} of from
 simulations\cite{Post1986,Tobochnik1988,Giarritta1992,modes:235701,modes:041125}. The thermodynamic pressure can be
obtained through the ``virial'', Eqs.(\ref{eq:eoshardS2}) or (\ref{eq:eoshardH2}), or the compressibility, Eq.~(\ref{eq:eoscompress}),
relations.

\begin{figure}[t]
\resizebox{10cm}{!}{\includegraphics{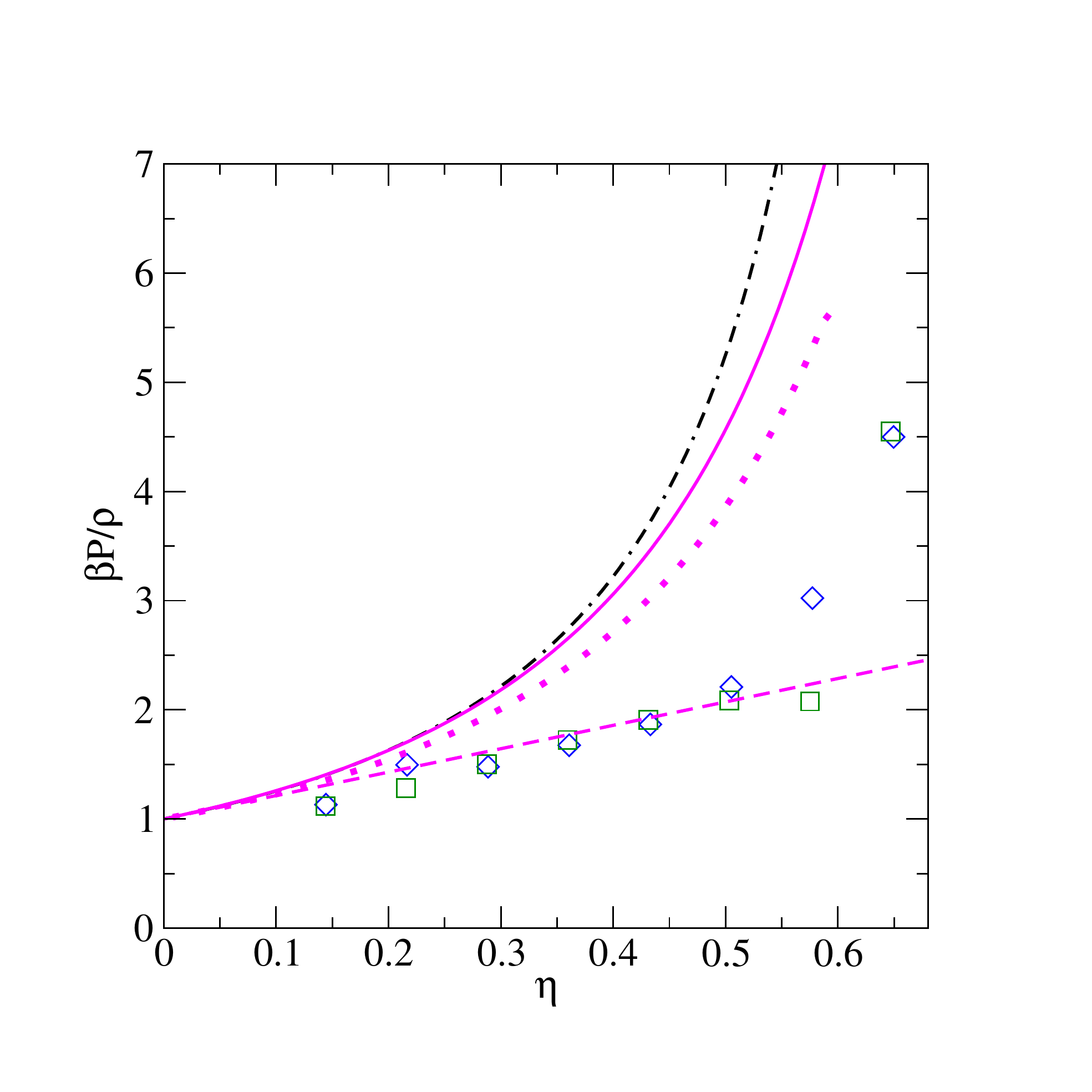}}
\caption{Equation of state of the hard-disk fluid in $H_2$ versus packing  fraction $\eta$ for a quite strong curvature $\kappa\sigma\simeq 1.06$. 
  Comparison between the PY results (compressibility route (dotted line) and virial route (full line)) for $\kappa \sigma=1.06$ and the simulation results of Ref.~\cite{modes:041125} 
for $\kappa \sigma=1.060$ (green squares) and $\kappa \sigma=1.062$ (blue diamonds). The dashed-dotted line is the prediction of Ref.~\cite{haro:116101} and 
the dashed line corresponds to the equation of state truncated at the second virial coefficient. The disagreement between predictions and simulation data is
significant alrealdy for $\eta \simeq 0.2$. }\label{fig:3}
 \end{figure}

As in Euclidean space, all approximate methods, including integral equations, are limited to moderate densities, i.e.
densities that are less than the ordering transition in the Euclidean plane. The exact location, the nature and the
order of the latter are known to be strongly system-size dependent (see Sec.~\ref{sec6}), so we typically mean here
densities $\rho\sigma^2\lesssim 0.8$ and packing fractions $\eta \lesssim 0.6$. In Fig.~\ref{fig:3}, we compare for the case of
the hyperbolic plane with a quite strong curvature $\kappa\sigma\simeq 1.06$ the equations of state obtained by 
simulation\cite{modes:235701,modes:041125} and predicted by several approximate methods (approximate equations of
state\cite{haro:116101,modes:235701,modes:041125} and results from the Percus-Yevick integral equation with both 
the virial and the compressibility routes\cite{Sausset2009}  As in the Euclidean case, the compressibility route gives a higher bulk thermodynamic pressure than the virial route.
An empirical recipe that has proven very accurate in Euclidean space\cite{Hansen1986} is to consider a linear 
combination
\begin{equation}
 P=\frac{2P_c+P_v}{3}
\end{equation}
where $P_c$ and $P_v$ are the ``compressibility'' and ``virial'' pressures. In Fig.~\ref{fig:3}, one can see that all 
predictions, except for the free-area method (which for this reason not reproduced here), approach at low
density the exact expression truncated at the second virial coefficient, Eq.~(\ref{eq:eosB2S2}),  whereas their validity rapidly deteriorates for even moderate coverage. 
One should however keep in mind
that the corresponding simulations\cite{modes:235701,modes:041125}
 have been carried out with a very small number of atoms, always less than $10$, and that the curvature is quite strong.

\begin{figure}[t]
\resizebox{10cm}{!}{\includegraphics{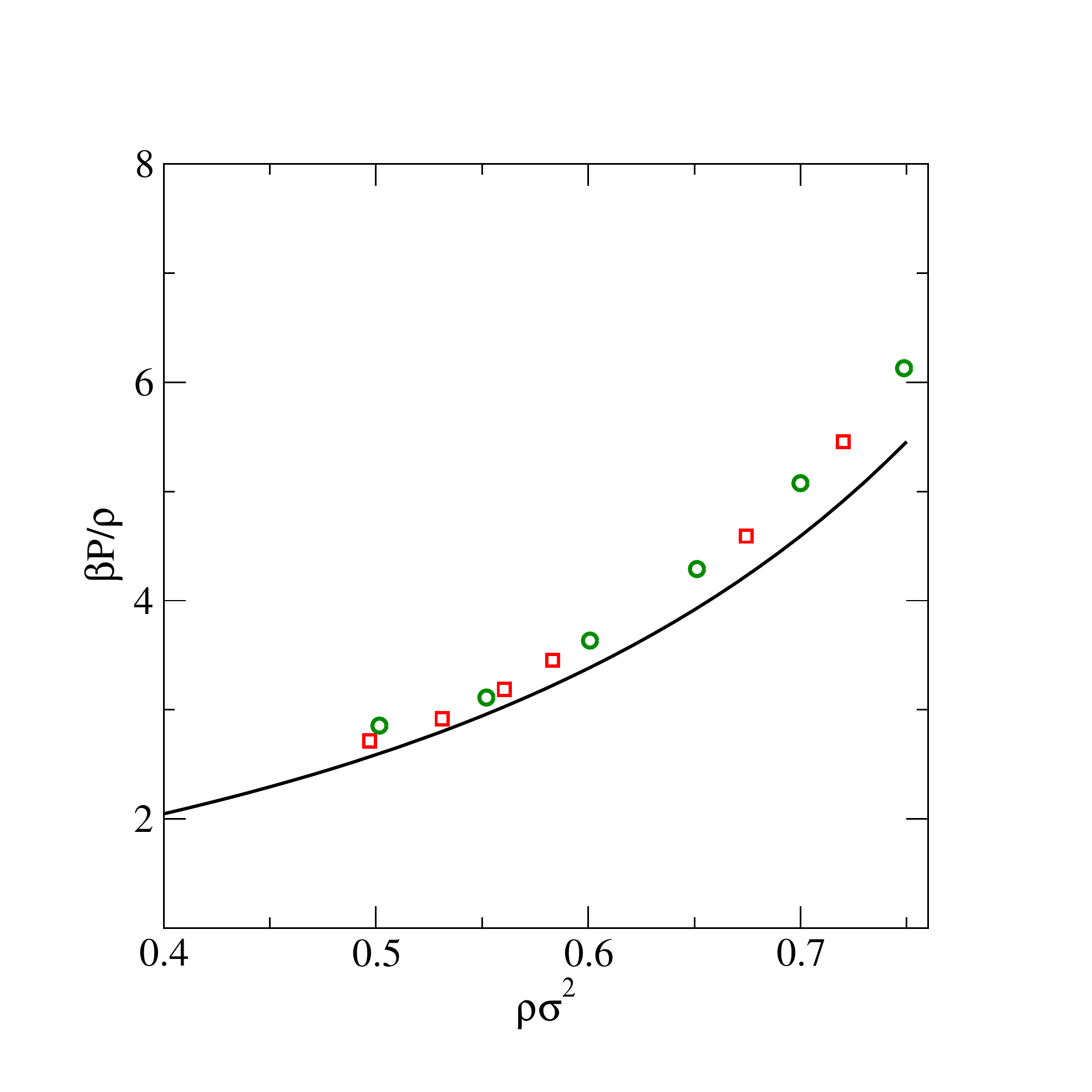}}
\caption{Equation of state of the hard-disk fluid in $S_2$ and in $H_2$ for the same magnitude of the curvature 
$\kappa\sigma=\frac{\sigma}{R}\simeq 0.45$. The symbols denote simulation results in $S_2$ (squares) and in $E_2$ (circles). 
The full curve corresponds to the PY approximation in $H_2$. The PY result in $S_2$ \cite{Lishchuk2006} is virtually 
indistinguishable from the simulation results (on this scale).}\label{fig:h2s2}
 \end{figure}

For the spherical substrate $S_2$, most of the existing data concern the approach to the Euclidean plane by 
following the equation of state $\beta P/\rho$ versus $\eta$ or $\rho \sigma^2$ at constant number $N$ of atoms (which
implies that the curvature varies with density, as $\rho\sigma^2=\left(\frac{N}{4\pi}\right)\left(\frac{\sigma}{R}\right)^2$).
We nonetheless display in Fig.~\ref{fig:h2s2} a comparison between the equations of state obtained from the the Percus-Yevick 
integral equation via the compressibility route for the hard-disk fluid in $S_2$\cite{Lishchuk2006} and $H_2$\cite{Sausset2009}
for the same ``radius of curvature'' $\sigma/R=\kappa\sigma=0.45$, as well as simulation results in $S_2$ and $E_2$. Despite possible 
limitations of the Percus-Yevick approximation, it appears that the results in $S_2$ almost coincide with those in the Euclidean $E_2$ whereas those in
$H_2$ are significantly different for $\rho\sigma^2\gtrsim  0.55.$

\begin{figure}[t]
\resizebox{10cm}{!}{\includegraphics{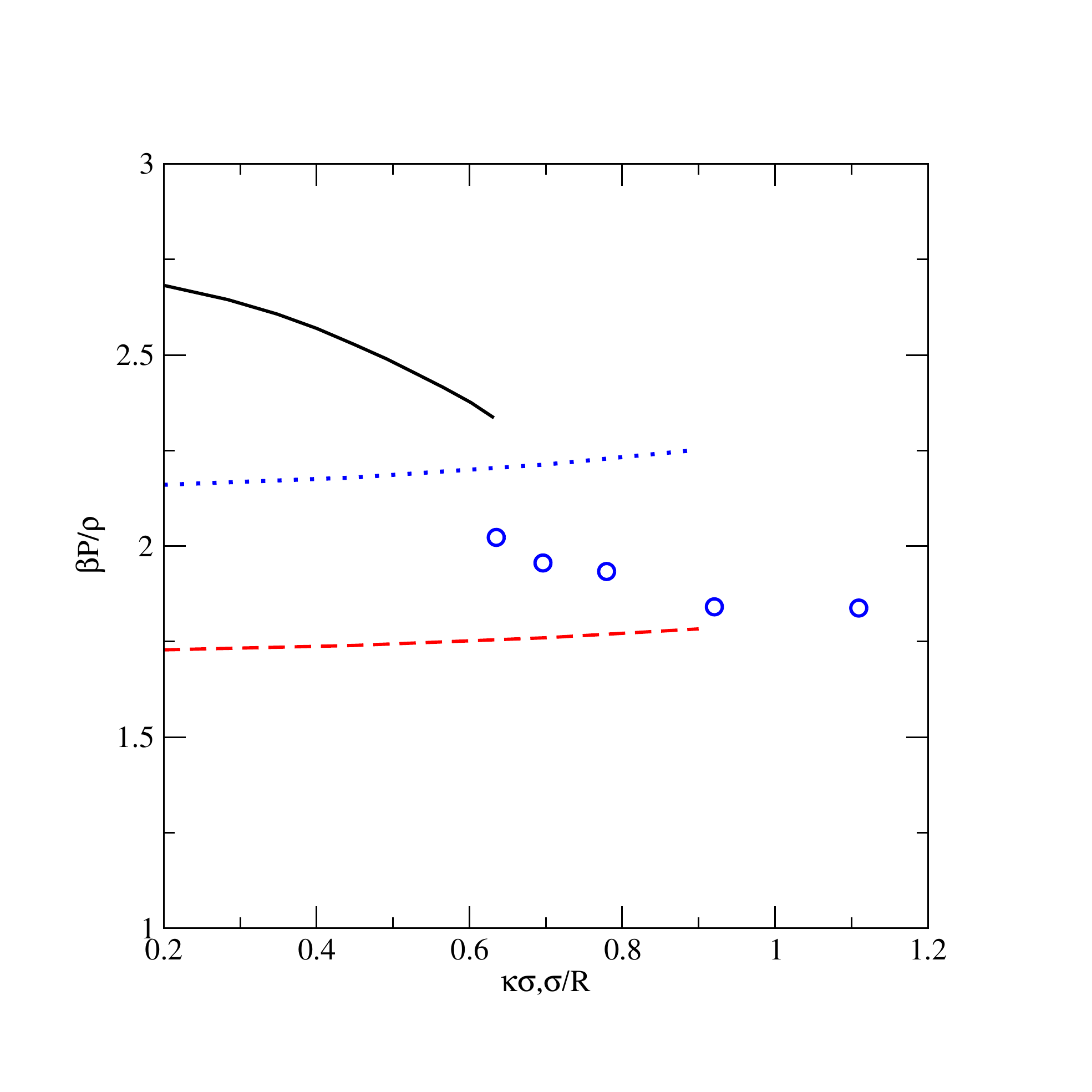}}
\caption{Compressibility factor $\beta P/\rho$ of the hard-disk fluid in $S_2$ and $H_2$ as a function of the curvature parameter, $\sigma/R$ or $\kappa\sigma$.
For $S_2$, are plotted the result from the PY approximation for $\rho\sigma^2=0.8$\cite{Lishchuk2006} (full line, where for convenience ($\beta P/\rho-5$) is plotted) 
and the simulation data for $\eta=0.4$ (which corresponds to $\rho\sigma^2$ varying between $0.385$ and $0.392$ \cite{Post1986} (symbols). For $H_2$, the two curves are 
obtained from the PY approximation with $\rho\sigma^2=0.2$ (dotted line) and $\rho\sigma^2=0.4$ (dashed line). }\label{fig:h2s2b}
 \end{figure}

In addition, to illustrate the influence of curvature, i.e. both its magnitude and its sign, on the thermodynamic properties
of a fluid, we have collected in Fig.~\ref{fig:h2s2b} simulation data and integral-equation results for the compressibility factor 
$\beta P/\rho$ of hard disks on $S_2$ and $H_2$, which we plot as a function of increasing curvature, 
$|K|^{-1/2}\sigma$,  for
different values of $\rho\sigma^2$. Whereas the pressure increases with curvature in $S_2$ 
for a given density $\rho\sigma^2$ (or packing fraction $\eta$), the opposite trend is observed in $H_2$.
This effect can already  be seen as very small densities when considering only the correction to the ideal-gas behavior due to the second virial coefficient (see above).

\begin{figure}[t]
\centering
\resizebox{10cm}{!}{\includegraphics{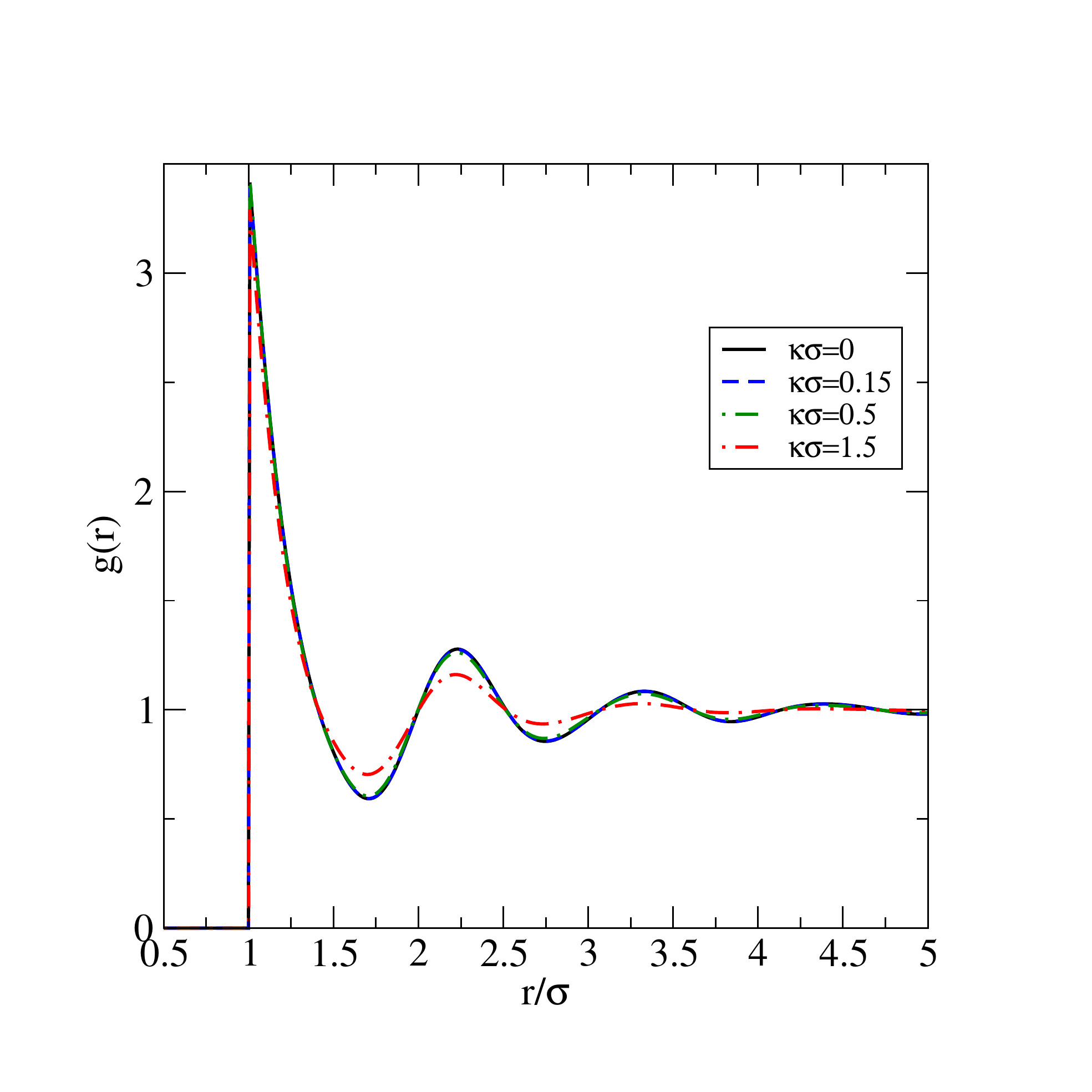}}
\caption{Radial distribution function $g(r)$ of the hard-disk fluid in $H_2$, as obtained from the PY equation,
 at a packing fraction $\eta =0.55$
for     various       values   of     the      curvature:      $\kappa
\sigma=0,0.15,0.5,1.5$. Only  for  $\kappa=1.5$,   namely a    radius of
curvature  smaller than the  particle diameter, does the structure display
significant deviations from the Euclidean  case. }\label{fig:STV1} \end{figure}

The main advantage of computing the radial distribution function $g(r)$ is that it also provides some information
on local structural order in the fluid. We focus here on the fluid or liquid regime at densities and temperatures
(for continuous interaction potentials) such that the system in flat space has not undergone its ordering transition 
(and if attractive interactions are present, away from the gas-liquid critical point). Higher densities and lower
temperatures, which correspond to interesting new physics, will be considered in Secs\ref{sec6} and \ref{sec7}, and 
critical behavior in Sec.~\ref{sec53}. In the fluid/liquid regime at ``moderate'' densities and temperatures (the
qualifier ``moderate'' having the above discussed meaning), it is found that curvature has only a weak influence, at least 
when the curvature is not too large, i.e. when $\sqrt{|K|}\sigma$ is significantly less than $1$. This can already
be inferred from the thermodynamic data (see above). We illustrate the effect of a negative curvature on the $g(r)$ 
of the hard-disk fluid, as obtained from the Percus-Yevick equation, in Fig.~\ref{fig:STV1}\cite{Sausset2009}. For a packing fraction
$\eta=0.55$, it is found that the curves are essentially superimposable on the Euclidean one for a range of 
curvature parameter  $\kappa \sigma$ between $0$ and $0.5$. Only when going to higher curvatures, 
e.g. $\kappa\sigma=1.5$, can one distinguish the influence of the curvature in the structure. (Note that on the sphere,
strong curvatures with $\sigma\gtrsim R$ correspond to small systems for which finite-size and discreteness effects
are very important\cite{Post1986}.) 

The same conclusion about the curvature effect
has been reached for a truncated Lennard-Jones model in $H_2$, with no detectable  changes from flat-space results
up to $\kappa\sigma \simeq 0.5$\cite{Sausset2009}. The  model is defined by a pair potential
\begin{equation}
 u(r)=4\epsilon\left[\left(\frac{\sigma}{r}\right)^{12}-\left(\frac{\sigma}{r}\right)^{6}\right]+u_c
\end{equation}
that is truncated for $r\geq r_c$ with $r_c=2.5\sigma$; $u_c$ is the shift obtained from the relation $u(r_c) =0$.
In this study, a large range of temperature, density and curvature
parameter has been investigated through the solution of the Percus-Yevick and HNC integral equations. As far as we know,
this is also the only work in which  the validity of such integral equations in curved space has been tested. As shown in Fig.~ 7,
of Ref.\cite{Sausset2009}, the predictions compare well with Molecular Dynamics simulation data with a slight 
advantage to the Percus-Yevick approximation.

\subsection{Coulombic systems}\label{sec52}
We have briefly reviewed in Sec.~\ref{sec44} the way to define the Coulomb potential in curved space 
(following a ``curved line of force'' interpretation) via the solution of the relevant Poisson equation. We have 
also stressed the difference between hyperbolic geometry where space can be infinite and spherical geometry in 
which the finiteness of space requires the introduction of pseudo-charges in order to properly define the Coulomb 
potential. The main advantage of spherical geometry being the absence of boundary, ``spherical'' and ``hyperspherical''
boundary conditions have been widely used by Caillol, Levesque and 
coworkers\cite{Caillol1982,PhysRevB.33.499,Caillol1992Caillol1992,Caillol1992,Caillol1999,Caillol2000,Caillol1991,PhysRevLett.43.979}
to approach the thermodynamic limit of Coulombic systems in Euclidean space (as the radius of the sphere or the 
hypersphere goes to infinity). Multipolar interactions deriving from the Coulomb potential have also been 
considered \cite{Caillol1992Caillol1992}. Here, as in most  of the article, we focus on two-dimensional manifolds of constant positive ($S_2$)
or negative ($H_2$) curvature and we consider systems of charges interacting through the Coulomb potential,
\begin{equation}\label{eq:potcoulombS2}
 \left.u(r)\right|_{S_2}=-q_1  q_2 \ln\left(\sin\left(\frac{r}{2R}\right)\right),
\end{equation}
\begin{equation}\label{eq:potcoulombH2}
 \left.u(r)\right|_{S_2}=-q_1  q_2 \ln\left(\tanh\left(\frac{\kappa r}{2}\right)\right),
\end{equation}
for two charges $q_1, q_2$ (or pseudo-charges in $S_2$) separated by a geodesic distance $r$ (with $r\leq \pi R$ in
$S_2$).

The two simplest Coulombic systems are the one-component plasma (OCP) and the two-component plasma (TCP). 
The former is a monodisperse system of point charges of equal sign and magnitude $q$ embedded in a charged uniform
background that maintains global electroneutrality. Conventionally, one uses the coupling constant 
$\Gamma=\beta q^2$ as control parameter. At small and intermediate coupling, $\Gamma=O(1)$ or less, the system is in a 
fluid phase whereas at zero temperature ($\Gamma\rightarrow +\infty$), its ground state in the Euclidean plane is
a hexagonal crystal forming a triangular lattice. The questions which remain debated\cite{PhysRevLett.82.4078} 	are (i)
whether there is a freezing transition at some large but finite coupling and (ii) if indeed freezing occurs, whether the 
transition is a first-order one or proceeds via two continuous transitions separated by a hexatic phase as predicted
by the KTNHY theory\cite{Nelson1979,Young1979,Nelson:2002}. As will be more extensively discussed in Secs.\ref{sec6} and   \ref{sec7}, placing
the system in curved geometry ``frustrates'' hexagonal order by forcing in an irreductible number of topological defects.
We just mention here that it has then been argued that approaching the thermodynamic limit of the OCP in the Euclidean
plane via spherical boundary conditions avoids the artifacts generated by periodic boundary conditions on the ordering
behavior\cite{PhysRevLett.82.4078}.

The TCP consists of a binary mixture of oppositely charged point particles (with charges $\pm q$) in equal concentration. No neutralizating background is needed.
 In the Euclidean plane,
the system is stable in a conducting phase up to a coupling $\Gamma=2$\cite{Hansen1985,Cornu1987}. 
Above this value, a collapse of pairs of opposite 
charges occurs. This can be regularized by introducing an additional hard-core repulsion 
(the system is then also known as the restricted primitive model). The system then undergoes a 
continuous Kosterlitz-Thouless transition at $\Gamma=4$ to a dielectric phase in which all charges are bounded 
into dipoles in the limit of an infinitesimal particle density. 
For larger densities, the critical temperature of the Kosterlitz-Thouless transition decreases and terminates  in a first-order transition slightly above the  
gas-liquid critical point. The location of the 
latter remains quite difficult to obtain precisely, as both the temperature  and the density are very small\cite{Orkoulas1996}.

In the following, we restrict ourselves to the OCP and TCP in their fluid, conducting phase at small and intermediate coupling in spherical ($S_2$) and hyperbolic
geometry ($H_2$). Several properties found in the Euclidean plane carry over to $S_2$ and $H_2$: (i) the small coupling/high temperature regime is (asymptotically) 
described by the linearized Debye-Hückel approximation, (ii) exact solutions for the 
thermodynamics and the structure are obtained for the special value of the coupling constant $\Gamma=2$, and (iii) in the conducting phase, 
exact sum rules generalizing the Stillinger-Lovett relations are satisfied.

We begin with the OCP. Assuming a perfect compensation between the point charges and the background amounts to replacing $g(r)$ by the pair correlation function of the point 
charges, $h(r)=g(r)-1$, in the equation of state for the thermodynamic pressure, Eqs.(\ref{eq:pressureS2}) and (\ref{eq:pressureH2}).
 After inserting Eqs.(\ref{eq:potcoulombS2}) and 
(\ref{eq:potcoulombH2}), one finds 
\begin{equation}\label{eq:eosOCPS2}
 \left.\frac{\beta P}{\rho}\right|_{S_2}=1+\rho\left(\frac{\pi\Gamma}{4}\right)
\int_0^{\pi R}dr \,r \left(1+\cos\left(\frac{r}{R}\right) \right) h(r),
\end{equation}

\begin{equation}\label{eq:eosOCPH2}
 \left.\frac{\beta P}{\rho}\right|_{H_2}=1+\rho\left(\frac{\pi\Gamma}{\kappa}\right)
\int_0^{\infty}dr  \left(\frac{\cosh(\kappa r)-1}{\sinh(\kappa r)} \right) h(r),
\end{equation}
which both converge to the Euclidean formula
\begin{equation}
  \left.\frac{\beta P}{\rho}\right|_{E_2}=1+\rho\left(\frac{\pi\Gamma}{2}\right)\int_0^{\infty}dr \,r \, h(r)
\end{equation}
when the curvature goes to zero. It should be stressed that defining a pressure for the OCP is far from trivial
 as both the way one treats the uniform background and the choice
of embedding conditions may matter. This has been carefully discussed in the case of the hyperbolic plane $H_2$ for which no less than $5$ different pressures have been 
considered\cite{Fantoni2003}. The above defined ``bulk thermodynamic pressure'' reduces to the standard thermodynamic pressure 
(then also equal to the ``Maxwell pressure'') with
vanishing curvature.

It has been shown for  $S_2$\cite{Choquard1987} and $H_2$\cite{Jancovici1998} that the OCP in the conducting phase satisfies 
generalized Stillinger-Lovett sum rules\cite{RevModPhys.60.1075},
\begin{equation}
 \rho\int_{S_2 \, \mbox{\rm \small or }H_2}dS \,h(r)=-1,
\end{equation}
which expresses the strictly enforced electroneutrality, and 
\begin{equation}
 \pi R^2 \rho^2\Gamma \int_{S_2}dS\,h(r)\left(1-\cos\left(\frac{r}{R}\right)\right)=-1
\end{equation}

\begin{equation}
 \frac{4\pi \rho^2\Gamma}{\kappa^2} \int_{H_2}dS\, h(r)\ln\left(\cosh\left(\frac{\kappa r}{2}\right)\right)=-1,
\end{equation}
which both express the screening property, with $\sqrt{2\pi\rho\Gamma}$ the inverse of the Debye  screening length. Note that the sum rules in the case of $H_2$ should be 
interpreted as valid in the bulk (see Secs.~\ref{sec3} and \ref{sec4} and Ref.\cite{jancovici2004}).

In the high-temperature or small-coupling limit, the linearized Debye-Hückel approximation becomes asymptotically exact and analytical expressions for the pair correlation 
function can be derived. The direct correlation function $c(r)$ becomes equal to $-\beta u(r)$, so that by using the Ornstein-Zernike relation, Eq.~\ref{eq:ozfourierS2},
one obtains  in $S_2$
\begin{equation}\label{eq:debyeS2}
 h(\theta)=-\Gamma \left(\frac{1}{2(\chi^2+1)}+\sum_{k=1}^\infty  \left(\frac{2k+1}{2}\right)\frac{1}{\chi^2+k(k+1)}P_k(\cos(\theta))\right).
\end{equation}
(No explicit expression of $h(\theta)$ has been obtained, contrary to the case of
the hypersphere\cite{Caillol1991}.)

In $H_2$, one finds\cite{Jancovici1998,jancovici2004}
\begin{equation}\label{eq:debyeH2}
 h(r)=-\Gamma Q_\nu( \cosh\left(\kappa r\right))
\end{equation}
where
\begin{equation}\label{eq:nuS2}
\nu= -\frac{1}{2}+\sqrt{\frac{1}{4}+\frac{2\pi \rho \Gamma}{\kappa^2}}.
\end{equation}

As anticipated, one can show that the above Debye-Hückel expressions  satisfy the generalized
 Stillinger-Lovett sum rules\cite{Jancovici1998}. Eqs.(\ref{eq:debyeS2}) and (\ref{eq:debyeH2}) can then be inserted in 
Eqs. (\ref{eq:eosOCPS2}) and (\ref{eq:eosOCPH2}) to derive the equation of state\cite{Sausset2009}. For illustration, 
we consider the $\Gamma \rightarrow 0$ limit in the hyperbolic plane. At fixed curvature, Eq.~(\ref{eq:nuS2}) shows that 
$\nu\rightarrow 0$ and, by using the definition $Q_0(x)=\frac{1}{2}\ln\left(\frac{1+x}{1-x}\right)$ and the property
that $\int_1^\infty dx \frac{Q_0(x)}{1+x}=\frac{\pi^2}{3}$, one finally obtains that\cite{Sausset2009}
\begin{equation}\label{eq:eosH2debye}
  \left.\frac{\beta P}{\rho}\right|_{H_2}\simeq 1-\left(\frac{\rho\pi^3\Gamma^2}{12\kappa^2}\right)
\end{equation}
as $\Gamma\rightarrow 0$. On the other hand, at fixed (small) $\Gamma$ and $\kappa\rightarrow 0$, one finds
that the Debye-Hückel expression in $H_2$ leads to 
\begin{equation}
 \left.\frac{\beta P}{\rho}\right|_{H_2}=1-\frac{\Gamma}{4},
\end{equation}
which is the exact result for the thermodynamic pressure.  It is also worth noting that in $H_2$, the exponential decay
of the Coulomb potential restores an analytical virial expansion in the Debye-Hückel limit for a finite radius of curvature
 $\kappa^{-1}$ whereas in Euclidean spaces, the pressure  is  independent of  density in $E_2$ and has a nonanalytical behavior 
in $E_3$\cite{Baus1980}. 

As mentioned above, for the specific value of the coupling constant $\Gamma=2$, exact analytical expressions 
of the thermodynamic quantities and the pair correlation function $h(r)$ can be obtained in 
Euclidean\cite{PhysRevLett.46.386}, spherical\cite{Caillol1981,Choquard1987}, and hyperbolic 
geometries\cite{Hastings1998,Jancovici1998} (as well as on a 
''Flamm's paraboloid``\cite{Fantoni2008}). For spherical and hyperbolic geometries, the pair correlation function is  given by
\begin{equation}\label{eq:hderS2}
 \left.h(r)\right|_{S_2}=-\left(\frac{1+\cos \left(\frac{r}{R}\right)  }{2}\right)^{4\pi \rho R^2},
\end{equation}
and
\begin{equation}\label{eq:hderH2}
 \left.h(r)\right|_{H_2}=-\left(\cosh\left(\frac{\kappa r}{2}\right)\right)^{-2+\frac{8\pi \rho}{\kappa^2}}.
\end{equation}
From the above expressions, one can calculate the thermodynamic pressure with the following results:
\begin{equation}\label{eq:eosg2S2}
 \left.\frac{\beta P}{\rho}\right|_{S_2}= \frac{1}{2}
\end{equation}
and 
\begin{equation}\label{eq:eosg2H2}
 \left.\frac{\beta P}{\rho}\right|_{H_2}= \frac{2\pi \rho +\kappa^2}{4\pi \rho +\kappa^2}.
\end{equation}
Note that the low-density limit of Eq.~(\ref{eq:eosg2H2}) coincides, as it should, with Eq.~(\ref{eq:eosH2debye}) when $\Gamma=2$. When
the curvature goes to zero, Eq.~(\ref{eq:eosg2H2}) reduces to the Euclidean result, $\frac{\beta P}{\rho}=\frac{1}{2}$, whereas
 the spherical expression always coincides with it, regardless of the (positive) curvature. An interesting 
observation concerning the hyperbolic plane is that if one takes the limit of an infinite (negative) curvature, $
\kappa \rightarrow \infty$, the pressure goes  to the ideal-gas limit. This result is expected to be quite general
(see Eq.~(\ref{eq:eosg2S2}) and Sec.~\ref{sec52}): the influence of the interactions appears to vanish in the limit of large
negative curvature, if however this limit makes any sense.

Having discussed the OCP in two-dimensional curved manifolds (more details can be found in
 Refs.\cite{Caillol1981,Caillol1982,Choquard1987} for $S_2$ and 
Refs.\cite{Hastings1998,Jancovici1998,Fantoni2003,jancovici2004,Sausset2009} for $H_2$), we now  move on to the case of 
the TCP model. In this system, there is no background and one has to consider the correlation functions between
pairs of equal charges, $h_{++}(r)=h_{--}(r)$, and of opposite charges, $h_{+-}(r)=h_{-+}(r)$. Electroneutrality implies
\begin{equation}
 \rho \int dS \,(h_{++}(r)-h_{+-}(r))=-2,
\end{equation}
where $\rho=2\rho_+=2\rho_-$ is the total particle density. Higher-order Stillinger-Lovett types of sum rules are 
also satisfied by $h_{++}(r)$ and $h_{+-}(r)$ and the thermodynamic pressure from Eqs.(\ref{eq:pressureS2}) and 
(\ref{eq:pressureH2}) is now given by equations similar to Eqs.~(\ref{eq:eosOCPS2}) and (\ref{eq:eosOCPH2}) with $\rho$ 
replaced by $\rho/2$ and $h(r)$ by ($h_{++}(r)-h_{+-}(r)$).

In the small coupling limit, the Debye-Hückel approximation gives the first nontrivial term in $\Gamma$. As the
linearized Poisson equation is similar to that of the OCP, the pair correlation functions are given by 
$h_{++}(r)=-h_{+-}(r)$ with $h_{++}(r)+h_{+-}(r)$ equal to the solution in Eqs.(\ref{eq:hderS2}) and (\ref{eq:hderH2}).
Consequently, the equation of state as the same of the OCP given above. Exact analytical results are also
obtained for the special coupling value of $\Gamma=2$. However, the calculations become quite involved and tedious. Details
concerning the hyperbolic case can be found Refs.\cite{Hastings1998,Jancovici1998,jancovici2004}.

\subsection{Liquid-gas critical behavior}\label{sec53}

Critical behavior implies fluctuations on all spatial scales and a diverging correlation length. Strictly speaking,
critical phenomena cannot exist in finite systems and therefore not for systems confined to spherical substrates. 
(As stressed before, spherical geometry can however be used as a trick to approach the thermodynamic limit in 
Euclidean space and to study finite-size scaling by decreasing the curvature and concomitantly increasing the system 
size\cite{Caillol1998}.) On the other hand, {\it bona fide} critical points can be present in systems embedded in 
hyperbolic geometry as space can then be infinite. We consider in this section the gas-liquid critical behavior of atomic
fluids in the hyperbolic plane $H_2$. 

Drastic changes in critical behavior are expected in hyperbolic geometry. This is known from field theoretic studies: 
a negative curvature acts as an ``infrared regulator'', suppressing fluctuations on wavelengths larger than the
radius of curvature\cite{Callan1990}. As a consequence, critical behavior in statistical systems on a hyperbolic
manifold is expected to be mean-field-like, with classical values of the critical exponents 
($\eta=0$, $\gamma=1$, etc)\cite{Rietman1992,Angl`esd'Auriac2001,Doyon2004}. This mean-field character can be 
understood by going back to the analysis of section \ref{sec43} concerning the correlation functions and the 
compressibility. The latter is given by an integral over space, Eq.~(\ref{eq:oz}), which in the case of $H_2$ is 
expressed as
\begin{equation}\label{eq:compressibilityH2}
 \frac{\rho \chi_T}{\beta}=1+\rho \frac{2\pi}{\kappa }\int_0^\infty dr \sinh(\kappa r) h(r).
\end{equation}
For long distances, $\sinh(\kappa r)\sim (1/2)\exp(\kappa r)$, and, as already pointed out in the section concerning
pair interactions, the integral over the whole space $H_2$ is bounded if and only if $h(r)$ decreases at large
$r$ faster than $\exp(-\kappa r)$. Assume for instance that $h(r)$ decreases exponentially as $\exp(-r/\xi)$
with $\xi$ the correlation length. One then finds that Eq.~(\ref{eq:compressibilityH2}) becomes
\begin{equation}\label{eq:compressH2criti}
 \frac{\rho \chi_T}{\beta}=\frac{(\kappa \xi)^2}{1-(\kappa \xi)^2}+ finite,
\end{equation}
which diverges for $\kappa \xi\rightarrow 1$ (and is infinite for $\kappa \xi >1$). The compressibility
can therefore diverge with a finite correlation length $\xi=\kappa^{-1}$ (this is the meaning of the ``infrared 
regulator'' discussed above). The critical point being then characterized by a finite correlation length, one
expects $\xi$ to be a regular function of the control parameters $T$ and $\rho$ even in the vicinity of $(T_c, \rho_c)$,
\begin{equation}
 \xi(T,\rho)=\kappa^{-1}-A(T-T_c)-B(\rho-\rho_c)^2+...
\end{equation}
with $A,B>0$. As a result, along the critical isochore, 
\begin{eqnarray}
 \chi_T\sim |T-T_c|^{-1},& \mbox{\hspace*{2cm}\rm $T\rightarrow T_c^+$}
\end{eqnarray}
and along the critical isotherm
\begin{eqnarray}
 \chi_T\sim (\rho-\rho_c)^{-2},& \mbox{\hspace*{2cm}\rm $\rho\rightarrow \rho_c$}
\end{eqnarray}
which gives the classical values of critical exponents, $\gamma=1$, $\delta=3$, and $\eta=0$. Note also 
that, at the critical point,  the Fourier-Helgason transform of the pair correlation function $\tilde{h}(k)$ 
(see Eq.~(\ref{eq:fourierH2radial})), is {\it finite} for $k=0$, which is a dramatic illustration of 
Eq.~(\ref{eq:helgasonnot0}), and has a regular expansion in $\left(k/\kappa\right)^2$ when
 $\left(k/\kappa\right)\rightarrow 0$.

We stress that the above reasoning concerns the ``bulk'' behavior in $H_2$, that which is obtained by taking the
thermodynamic limit either with a succession of periodic boundary conditions or by removing the boundary regions
(see Sec.~\ref{sec3} and \ref{sec4}). The exponentially growing character of the hyperbolic metric otherwise induces
possible ``boundary transitions''\cite{Angl`esd'Auriac2001} which we do not consider here.

\begin{figure}[t]
\centering
 \resizebox{8cm}{!}{\includegraphics{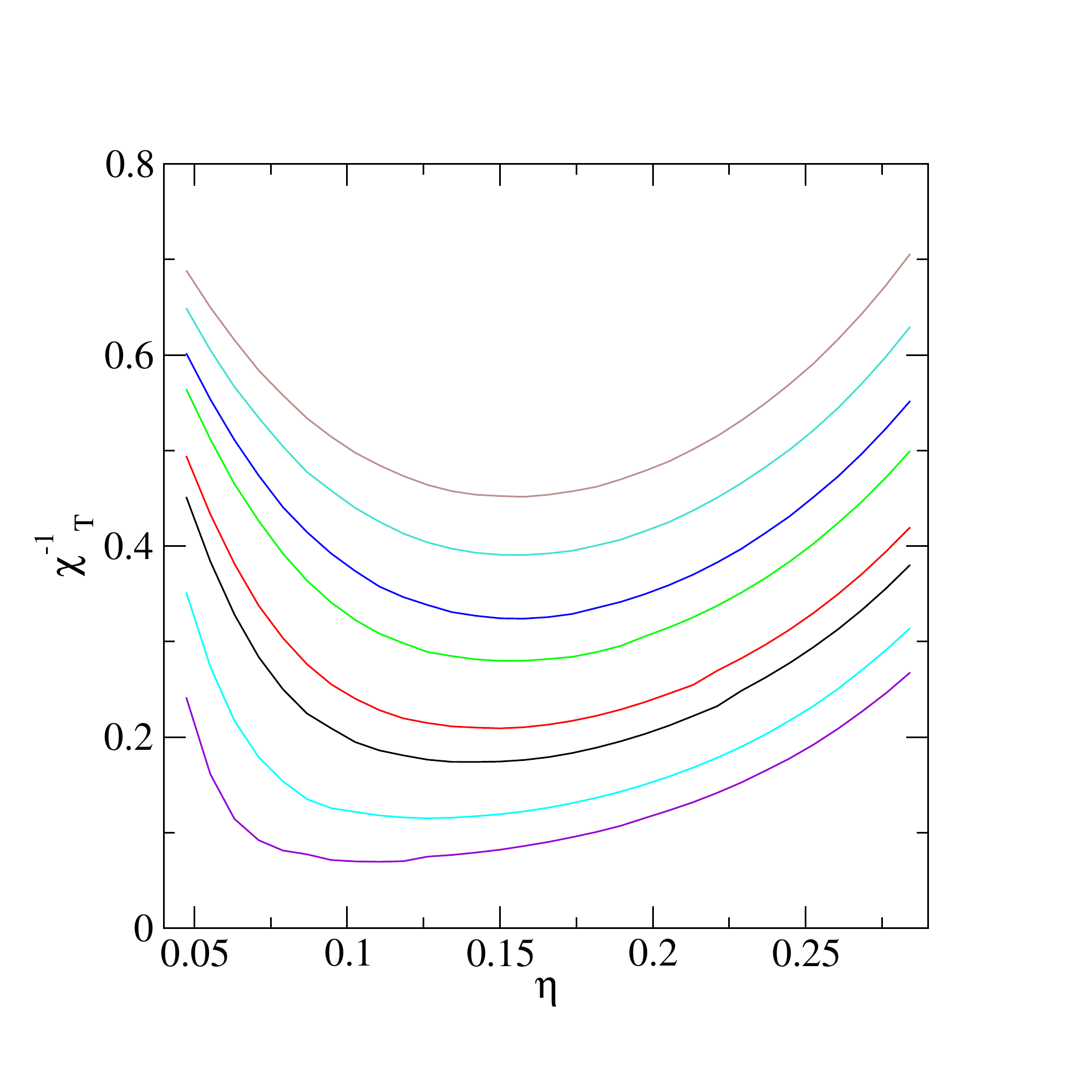}}
\caption{Inverse of the isothermal compressibility versus packing fraction $\eta$ for
the  truncated  Lennard-Jones fluid in $H_2$ within the PY  approximation near but
above the critical temperature: from top to  bottom , $T=0.640, 0.610,
0.580, 0.560, 0.530, 0.515, 0.490, 0.472$ (in the usual reduced units). 
The curvature parameter is $\kappa\sigma=0.5$. }\label{fig:compressibility} \end{figure}

\begin{figure}[t]
\centering
 \resizebox{7cm}{!}{\includegraphics{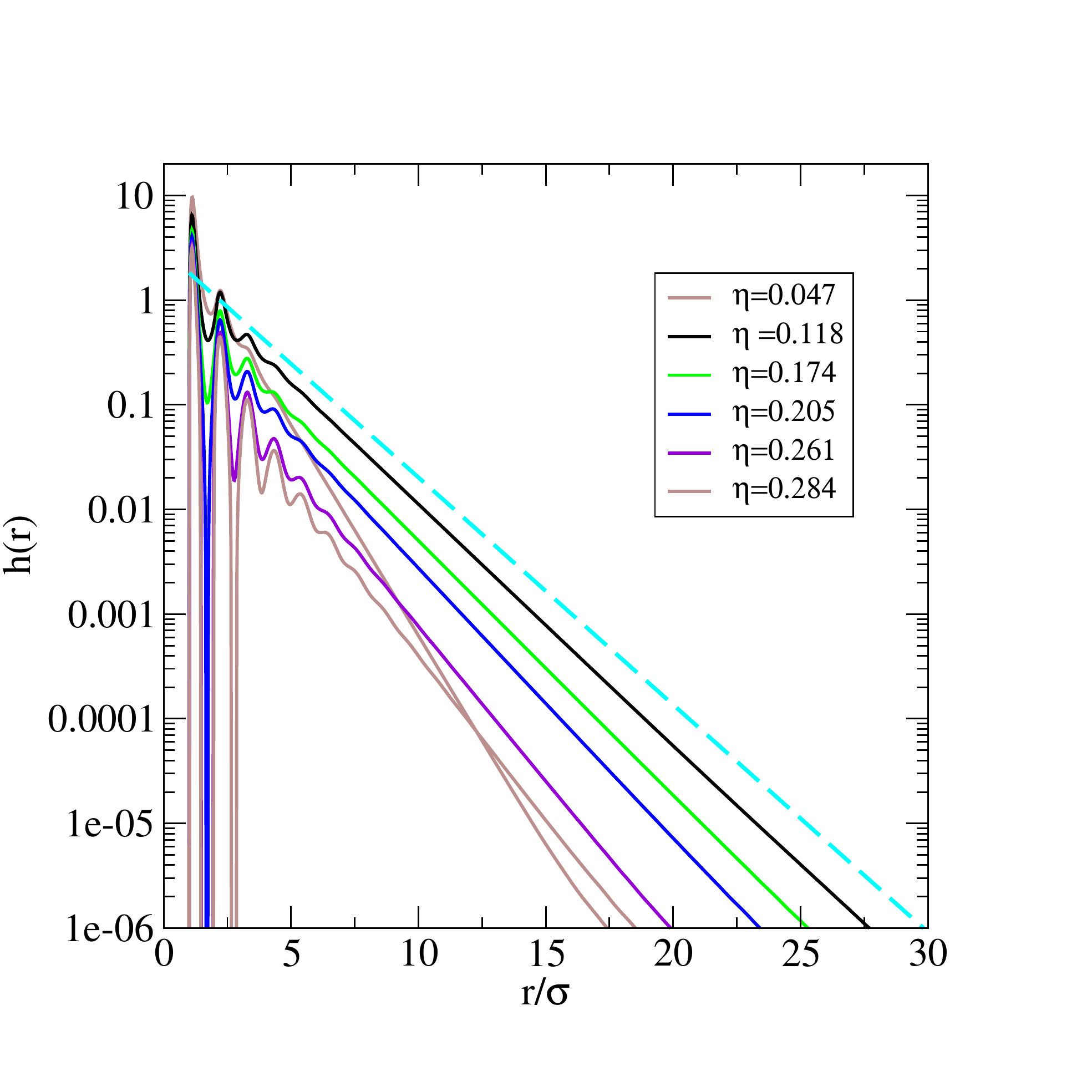}}
\resizebox{7cm}{!}{\includegraphics{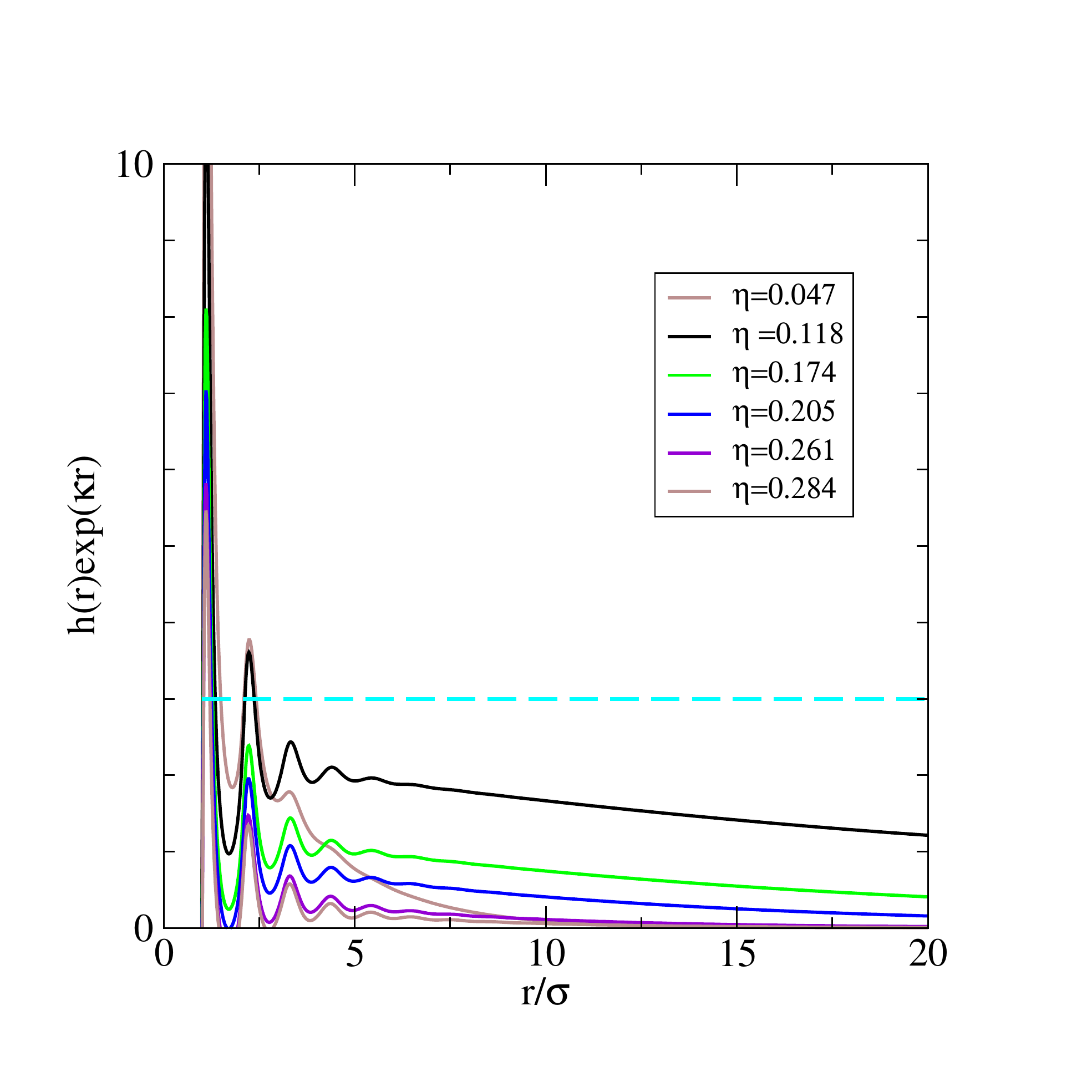}}
\caption{(a) Log-linear plot of the pair correlation function $h(r)=g(r)-1$ of the truncated Lennard-Jones fluid in $H_2$, as obtained 
in the PY approximation for $T=0.472$.
 The  behavior as  a function  of  $\eta$ is   nonmonotonic and  the
 slowest decay is for $\eta=0.118$, which  corresponds  to the maximum of
 the compressibility. The dotted line  is $\exp(-\kappa  r)$. (b) Same   data
 multiplied by $\exp(\kappa   r)$: a  convergence  towards a  constant
 plateau  at  large $r$  is  clearly  visible  as  one  approaches the
 critical point. The curvature parameter is $\kappa\sigma=0.5$.}\label{fig:gdercri} \end{figure}

The gas-liquid critical point of the truncated Lennard-Jones fluid in $H_2$ has been studied through the integral
equation approach\cite{Sausset2009} that has been described in Sec.~\ref{sec43}. The results are illustrated in Figs.\ref{fig:compressibility} and \ref{fig:gdercri}
for the Percus-Yevick closure and a curvature parameter $\kappa\sigma=0.5$. Fig.~\ref{fig:compressibility} displays the inverse compressibility
along different isotherms as one approaches the critical one. It is well known that approximate integral equations
such as the Percus-Yevick one fail to capture nonclassical critical behavior when it is present, as in the Euclidean
plane. The relevant conclusions that can be drawn from the Percus-Yevick equation study is therefore not about
critical exponents, but rather concerning the above discussed scenario of an exponentially decreasing pair correlation
function with a correlation length approaching the radius of curvature $\kappa^{-1}$ from below. A clear
 confirmation of the scenario can be seen in Fig.~\ref{fig:gdercri} where both $h(r)$ and $\exp(\kappa r) h(r)$ are plotted in 
the vicinity of the critical point. Note that the gas-liquid critical point for $\kappa \sigma=0.5$ is found at a
lower temperature than that in the Euclidean plane.

An interesting question concerning the critical point in the presence of negative curvature is whether the location
of the point in the $(T,\rho)$ diagram goes continuously to the value in the Euclidean plane or goes to a lower, 
possibly zero-temperature point as the curvature is reduced to zero. The former possibility is natural if one
thinks of the curvature as introducing a mere finite-size cut-off. However, the latter has been conjectured by Angles
d'Auriac {\it et al}\cite{Angl`esd'Auriac2001} in their study of the Ising model on hyperbolic lattices; these authors have also
suggested the existence of a crossover line, emanating from the critical point in Euclidean space and extending to 
small nonzero curvature below which the pair correlation function has an algebraic decay for $r<\kappa^{-1}$ and an exponential decay at long distance.
Approximate integral equations presumably predict a continuous behavior, and no crossover is seen around the location
of the flat space critical point, but this may come from their intrinsic limitation. We have undertaken preliminary
Molecular Dynamics simulations of the truncated Lennard-Jones model in $H_2$ to try to locate its gas-liquid critical
point and possible crossover behavior. 
\begin{figure}[t]
\centering
 \resizebox{10cm}{!}{\includegraphics{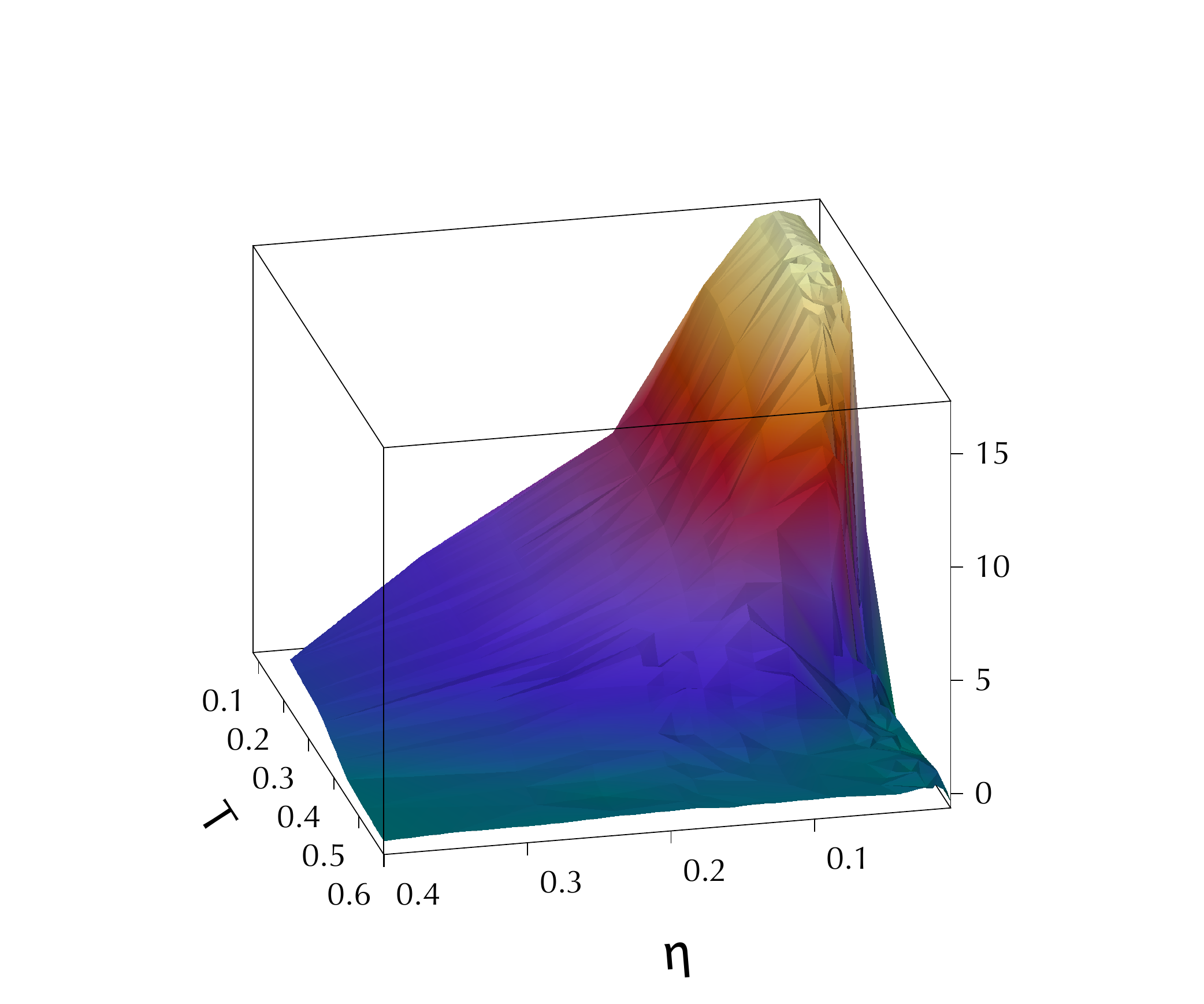}}
\caption{MD simulation of the critical behavior of the truncated Lennard-Jones model in $H_2$ for 
$\kappa\sigma=0.2$. Three-dimensional plot of the integral of $h(r)$ over the system size as a function of $T$ 
and $\eta$. The critical point of the model in $E_2$ obtained by Monte Carlo simulation\cite{Singh1990} is around 
$T_c=0.472$ and $\eta=0.33$. }\label{fig:hdercri} \end{figure}
Fig.~\ref{fig:hdercri} displays a three-dimensional plot of the integral over space of $h(r)$ in the $(T,\phi)$ 
plane for a curvature parameter 
$\kappa\sigma=0.2$. Periodic boundary conditions with a fundamental polygon of $14$ edges (see Appendix B and 
Ref.\cite{Sausset:2007}) are used and the system size is unfortunately rather small (from $8$ to $320$ atoms).
One observes that the integral indeed starts to rise very steeply as one reaches the vicinity of the Euclidean-space
critical point and saturates at lower $T$ and $\rho$ due to  system-size limitations. However, without studying 
finite-size effects by changing the fundamental polygon of the periodic boundary conditions, no clear-cut conclusion
can be reached. Further work is needed to elucidate this question.

\section{Freezing, jamming and the glass transition}\label{sec6}
\subsection{Geometric frustration}\label{sec61}

Starting with the work of Frank\cite{Frank:1952}, Bernal\cite{bernal1959} and others in the fifties, a whole line of research has developed, 
trying to understand liquids, glasses and amorphous packings, i.e. systems with no apparent structural long-range order, from a geometric point of view. 
An explanation for the avoidance of crystal formation when  cooling a liquid, with the resulting glass formation, as well as for the structure of glasses and amorphous materials
has been put forward in terms of ``geometric frustration''\cite{Sadoc:1999}. The latter describes a competition between a short-range tendency for the extension of a locally preferred
order and global constraints that preclude tiling of the whole space by a periodic repetition of the local structure. A prototypical and well documented example is that of 
icosahedral order in  three-dimensional systems in which particles interact through spherically symmetric pair potentials: despite being locally more favorable, icosahedral order 
built  from tetrahedral units cannot freely propagate in space to give rise to long-range crystalline order. Icosahedral order therefore must come with {\it topological defects}. A step 
forward in the geometric description has been provided by the ``curved space approach'' developed by Kl{\'e}man, Sadoc, Mosseri on the one hand\cite{Kl'eman1979,Sadoc:1999} and by Nelson,
 Sethna,  and coworkers on
the other\cite{Sethna1985,Nelson:2002}. In the late seventies, Kleman and Sadoc\cite{Kl'eman1979} realized that perfect icosahedra could tile three-dimensional space if the metric of the latter were modified
to introduce a constant positive curvature. On the hypersphere $S_3$ with a radius $R$ equal to the golden number $\left(\frac{1+\sqrt{5}}{2}\right)$ times the particle size,
$120$ particles form a perfect icosahedral tiling known as the $\{3,3,5\}$ polytope\cite{Sadoc:1999}. Such an unfrustrated crystallization is then expected to take place more
easily, i.e. at higher temperature, than crystallization in Euclidean space which, due to frustration of icosahedral order, must involve a reorganization of the local order and 
a different, hexagonal close-packed or face-centered cubic, long-range order\cite{PhysRevB.30.6592,PhysRevB.34.405}. The template or ``ideal order'' can then be used to describe real physical
 systems in Euclidean space, mostly metallic glasses, provided that one can described how it evolves when ``flattening'' space: roughly speaking, one observes in Euclidean space the 
remains of ideal icosahedral order that is broken up by the necessary appearance of topological defects, 
essentially disclination lines, which in some sense carry the curvature needed
for compensating the flattening of the template. The same approach can be taken for a variety of local orders that are frustrated in Euclidean space but can tile space in curved
spherical or hyperbolic geometries\cite{Sadoc:1999}. A theory of the glass transition has also been developed along these lines\cite{Kivelson:1995,Tarjus:2005}.

Quite generally, curvature can be used as a tool to either frustrate or generate long-range order. In the above discussed example of icosahedral order, curvature allows one to build a 
perfect tetrahedral/icosahedral tiling in the form of a polytope in $S_3$. Generically, such polytopes can be found for special values of the radius of curvature expressed in 
units of the polytope edge: there are a finite number of them in spherical geometries and an infinite number in hyperbolic geometries (one could also consider geometries with 
spatially varying curvature)

Consider now two-dimensional space and fluids of particles interacting via spherically symmetric pair potentials. Such systems of disks on the Euclidean plane are not subject
to geometric frustration: the locally preferred structure is a regular hexagon, with one atom at the center and $6$ neighbors at the
 vertices, and this structure can be
periodically repeated in space to form a triangular lattice. The system crystallizes extremely easily, either through a sequence of two continuous transitions
 with an intermediate
hexatic phase as in the KTNHY\cite{Nelson1979,Young1979,Nelson:2002} or through a weak first-order transition, and glass formation never occurs. 
Curving space then frustrates hexagonal order and 
forces in topological defects which are point-like in two dimensions. (The topic of hexagonal order and associated defects will be considered 
in more detail in Sec.~\ref{sec7}.) Again, for 
specific values of the radius of curvature compared to the particle size, the spherical and hyperbolic manifolds 
($S_2$ and $H_2$, respectively) allow crystalline-like tilings which are conventionally represented by the Schläfli 
notation $\{p,q\}$ with $q$ the number of edges of the elementary polygonal tile and $p$ the number of polygons
meeting at each vertex. Such $\{p,q\}$ tilings satisfy $(p-2)(q-2)=4$  in the 
Euclidean space $E_2$ ,  $(p-2)(q-2)<4$ on $S_2$ and $(p-2)(q-2)>4$ on $H_2$. This leaves the two dual 
triangular/hexagonal tilings $\{3,6\}$ and $\{6,3\}$ and the square tiling $\{4,4\}$ in $E_2$,  the five tilings $\{3,3\}$, $\{4,3\}$, $\{3,4\}$,
$\{5,3\}$ and $\{3,5\}$, corresponding to the platonic solids, in $S_2$ and an infinity of tilings in $H_2$\cite{Coxeter:1969}.

The flexibility offered by the multiple tilings in $H_2$ has for instance been used by Modes and 
Kamien\cite{modes:041125,modes:235701} to study ``isostatic'' packings of hard disks in $H_2$. Isostaticity means
that the number of constraints coming from force and torque balance equations is exactly equal to the number of
degrees of freedom in the system\cite{Maxwell1864}. It is a global requirement that is related to marginal stability
in solids\cite{Alexander1998,PhysRevE.72.051306}; it has recently received renewed attention in the context of ``jamming phenomena`` present in 
equilibrium and driven disordered assemblies of particles\cite{Liu1998}.   Isostaticity is a topological property that does not depend on
the curvature of space. For a two-dimensional system of $N$ hard disks, there are  $2N$ degrees of freedom and 
$\overline{z}N/2$ constraints coming form the number of contacts, where $ \overline{z}$ is the average number of 
contacts per disk, so that isostaticity corresponds to $\overline{z}=4$. Therefore, all allowed tilings $\{4,q\}$ in
$H_2$ are isostatic, which broadens the scope of systems that can be used to study the generic features associated
with isostaticity\cite{modes:041125,modes:235701}.

As stressed several times in this article, only hyperbolic geometry allows one to study macroscopic systems in the 
thermodynamic limit {\it at constant nonzero curvature}. Spherical geometry implies a finite system which can be 
investigated {\it per se}, as in studies of colloidal systems in spherical substrates, or used as trick to converge 
to the thermodynamic limit in the Euclidean plane by decreasing the curvature. The latter approach, corresponding to the
already mentioned spherical boundary conditions, has been taken to study dense disordered packings of hard 
particles\cite{Kraschrei1982,Tobochnik1988} as well as the nature of crystallization for particles interacting with
power-law and logarithmic pair potentials\cite{PhysRevB.58.9677,PhysRevLett.82.4078}. In both cases, the rationale is 
that spherical boundary conditions do not favor crystalline hexagonal arrangements as much as periodic boundary
conditions directly implemented in $E_2$, which then provides a less biased finite-size approach to phenomena taking place in flat space in the
thermodynamic limit. On the other hand, one may be interested in studying the influence of geometric frustration on
the structure and the dynamics of a system and therefore work with a constant nonzero curvature. Consider the 
hyperbolic plane $H_2$ which, as already stated, is of infinite spatial extent. For disks of diameter $\sigma$
embedded in $H_2$, the local order of the liquid changes as one increases the radius of curvature $\kappa^{-1}$. 
From hexagonal at zero and small curvature parameter $\kappa \sigma$, it becomes heptagonal at a larger $\kappa\sigma$,
then octogonal, etc\cite{PhysRevB.28.6377}, and for commensurate values of the curvature parameter, the locally
preferred structure can freely propagate to tile space; in such cases, corresponding to\cite{PhysRevB.28.6377}
\begin{equation}
 \kappa_n\sigma=2\cosh^{-1}\left[\frac{1}{2\sin\left(\frac{\pi}{n}\right)}\right]
\end{equation}
with $n=6,7,8,...$, there is no frustration. The effect of frustration can be investigated off these commensurate
curvatures. For instance, frustrated hexagonal order can be studied for $\kappa\sigma$ larger than zero 
($\kappa_6\sigma=0$) but significantly less than $\kappa_7\sigma\simeq 1.09055$\cite{PhysRevB.28.6377}.

Before moving on to a more extensive discussion of frustration and glass formation in a liquid model in the hyperbolic
plane, we briefly describe the various tools that have been used in practice to assess the structure of dense phases
in curved space. All of them are actually extensions of methods and observables developed for the Euclidean case.
For spherical particles, both ''positional`` (translational) and ''bond-orientational`` orders are of interest. 
''Positional`` refers to the distribution functions introduced in Sec.~\ref{sec4} that involve correlations between 
the particle centers. The most easily measured or computed is the radial  distribution function $g(r)$, with 
$r$ being the geodesic distance between two particle centers. ''Bond-orientational`` refers to the distribution functions
associated with the (artificial) ''bonds`` joining two nearest-neighbor particles. There are different ways of 
defining such bonds, a commonly used one being to assign nearest neighbors through a Voronoi-Dirichlet
construction (see below). In Euclidean space, one associates a unit vector to each bond and orientational order
refers to the average and the correlations of local order parameters defined in two dimensions as\cite{Nelson:2002}

\begin{equation}\label{eq:orderparameter}
 \Psi_n({\bf r}_{j})=\frac{1}{N_b}\sum_{k=1}^{N_b}\exp(in\theta_{jk}),
\end{equation}
where the sum is over the $N_b$ nearest neighbors of the particle located at ${\bf r}_j$ and $\theta_{jk}$ is the angle characterizing the ''bond`` between atoms $j$ and $k$
($n=6$ for instance is characteristic of $6-$fold hexagonal or hexatic order).  Similar quantities can be introduced in higher dimensions\cite{PhysRevB.28.784}.

The
difficulty that one faces when dealing with non-Euclidean space is that there is no global existence of vector fields: vectors are defined locally
 (in the  tangent Euclidean
manifold) and to be combined or compared to vectors at another point in space, they must be ''parallel transported`` along the geodesic joining the points (see Appendix A and Refs.\cite{goetz1970,Sadoc:1999}). 
As a result,
 the bond-orientational correlation functions involving an extension to non-Euclidean geometry of the local order parameters in Eq.~\ref{eq:orderparameter} are path-dependent in the 
presence of a nonzero curvature. A natural definition of the pair correlation functions is however to consider the geodesic between the two points under consideration, which leads to
\begin{equation}
 g_n(r)=\frac{1}{N}\sum_{i,j=1}^n\langle\tilde{\Psi}_n(i|j)\Psi_n^*(j)\rangle_{\Gamma_{ij}} \delta^{(2)}(r_{ij}-r),
\end{equation}
where $\tilde{\Psi}_n(i|j)$ is the order parameter when parallel transported from point ${\bf r}_i$ to point ${\bf r}_j$ along the geodesic $\Gamma_{ij}$, $r_{ij}$ is the geodesic
distance between the two points and $\delta^{(2)}$ is the delta function that is appropriate for the non-Euclidean metric (for $S_2$, see Ref.\cite{Giarritta1992} and for $H_2$
see Refs.\cite{PhysRevB.28.6377,PhysRevLett.104.065701}). Finally, the Voronoi tesselation or its dual, the Delaunay construction (or Dirichlet triangulation in two
dimensions), is a way to uniquely define nearest neighbors, hence bonds, coordination number, etc, which is most useful in dense liquid and amorphous phases. In particular,
it allows one to characterize at a microscopic level topological defects (especially disclinations which are orientational defects in the form of lines in three dimensions
 and points in two
dimensions) occurring in a putative order. These constructions can be extended to spherical\cite{Sadoc:1999,PhysRevB.30.6592,Giarritta1992} and hyperbolic
 geometry\cite{PhysRevB.28.6377,Leibon2000,PhysRevE.81.031504}.

\subsection{Glassforming liquid on the hyperbolic plane}\label{sec62}

\begin{figure}
        \begin{center}
                 \subfloat[]{\label{fig:fico}\includegraphics[draft=false,height=6cm]{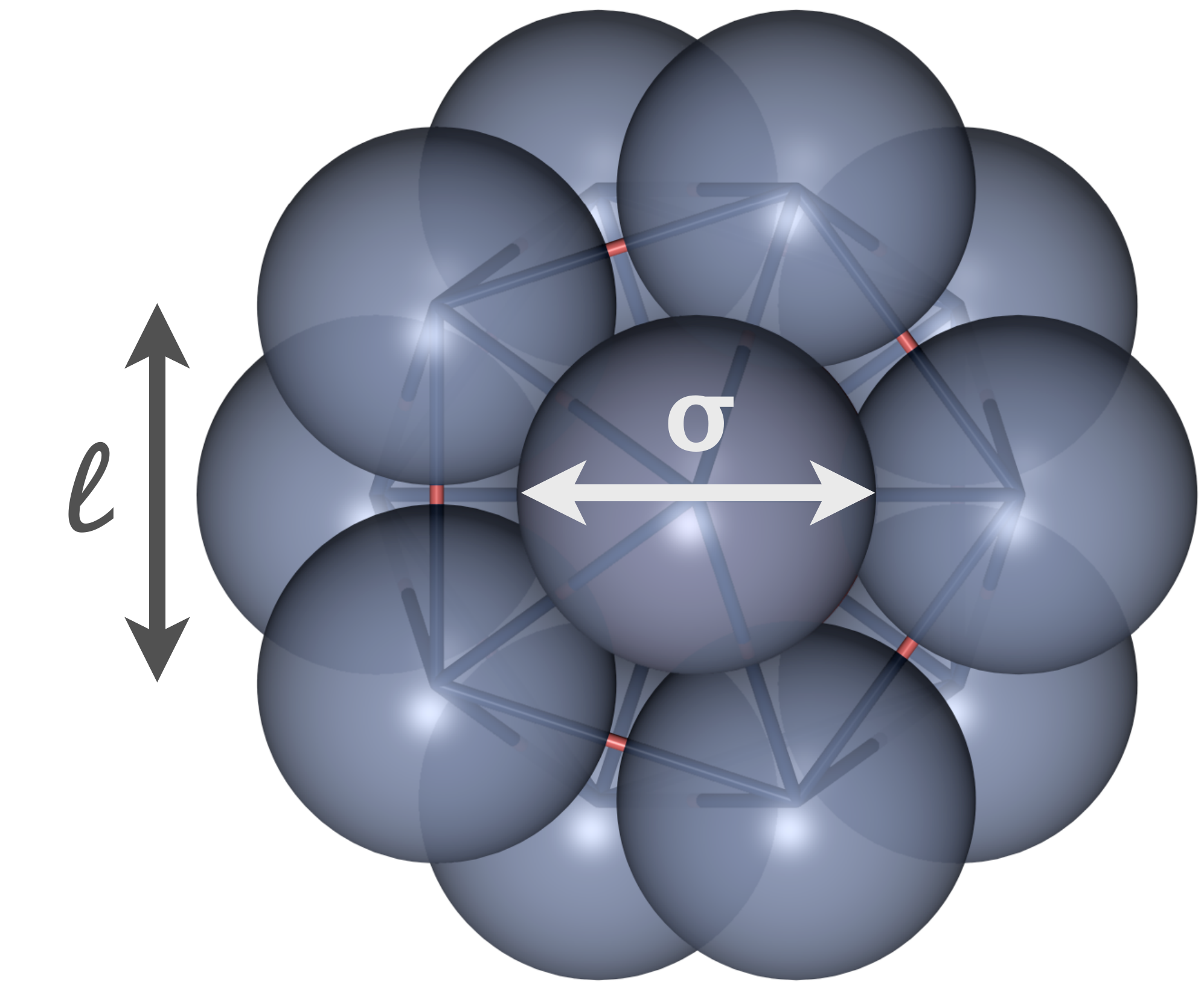}}
                \hspace{0.6cm}
                  \subfloat[]{\label{fig:fhex}\includegraphics[draft=false,height=6cm]{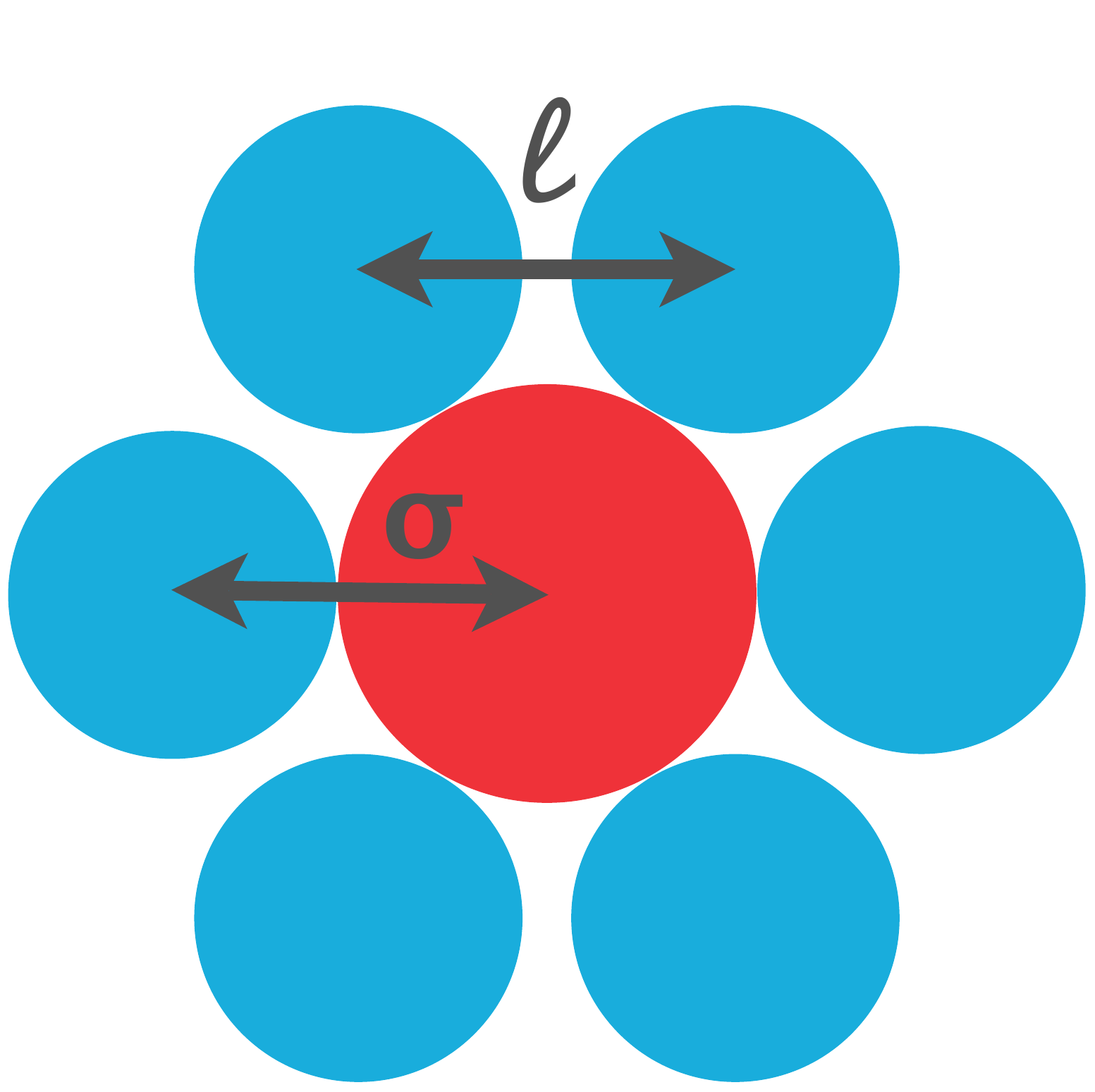}}
        \end{center}
\caption{ Locally preferred structure in atomic liquids: \subref{fig:fico} Frustrated icosahedral order in three-dimensional Euclidean space $E_3$: the distance $l$ between two neighboring outer spheres is slightly larger than the particle diameter  $\sigma$ which is also the distance between the central sphere and the $12$ outer ones. \subref{fig:fhex} Frustrated hexagonal order in the hyperbolic plane $H^2$: the nonzero curvature induces that, here too,
 $l$ is slightly larger than $\sigma$.}\label{fig:localorder}
\end{figure}

We briefly review here the extensive computer simulation study that we have recently carried out on the structure and the dynamics of the truncated Lennard-Jones liquid 
(see above)
embedded in $H_2$ with curvature parameter $\kappa \sigma$ small enough  that the local order in the liquid is hexagonal/hexatic and is therefore frustrated at long distance. 
The motivation behind this study was to assess the validity of the frustration-based theory of the glass transition\cite{Kivelson:1995,Tarjus:2005}. In the latter, frustration is hypothesized to be 
ubiquitous in liquids. The salient features of the phenomenology of glassforming liquids, above all the spectacular ''super-Arrhenius`` increase of the viscosity and the relaxation
time as one lowers the temperature\cite{Tarjus:2005}, are then attributed to the frustration-limited extension of the locally preferred liquid order; this extension is driven
by the proximity to an ''avoided ordering transition`` that would take place in the absence of frustration\cite{Kivelson:1995,Tarjus:2005}. Frustrated icosahedral order in three dimensions with $S_3$ as the 
unfrustrated space\cite{Sadoc:1999,Nelson:2002} is one example that we have already discussed. A simpler model however is provided by frustrated hexagonal order in two-dimensional 
negatively curved space\cite{Nelson:2002,PhysRevB.28.6377,Nelson1983}, as illustrated in Fig.~\ref{fig:localorder}. In this case, ''ideal ordering`` takes place in flat space and this ordering
transition observed at a temperature $T^*$ is avoided as soon as one introduces a nonzero curvature, with the curvature parameter $\kappa\sigma$ playing the role of the 
frustration strength.

\begin{figure}
        \begin{center}
                \includegraphics[draft=false,width=10cm]{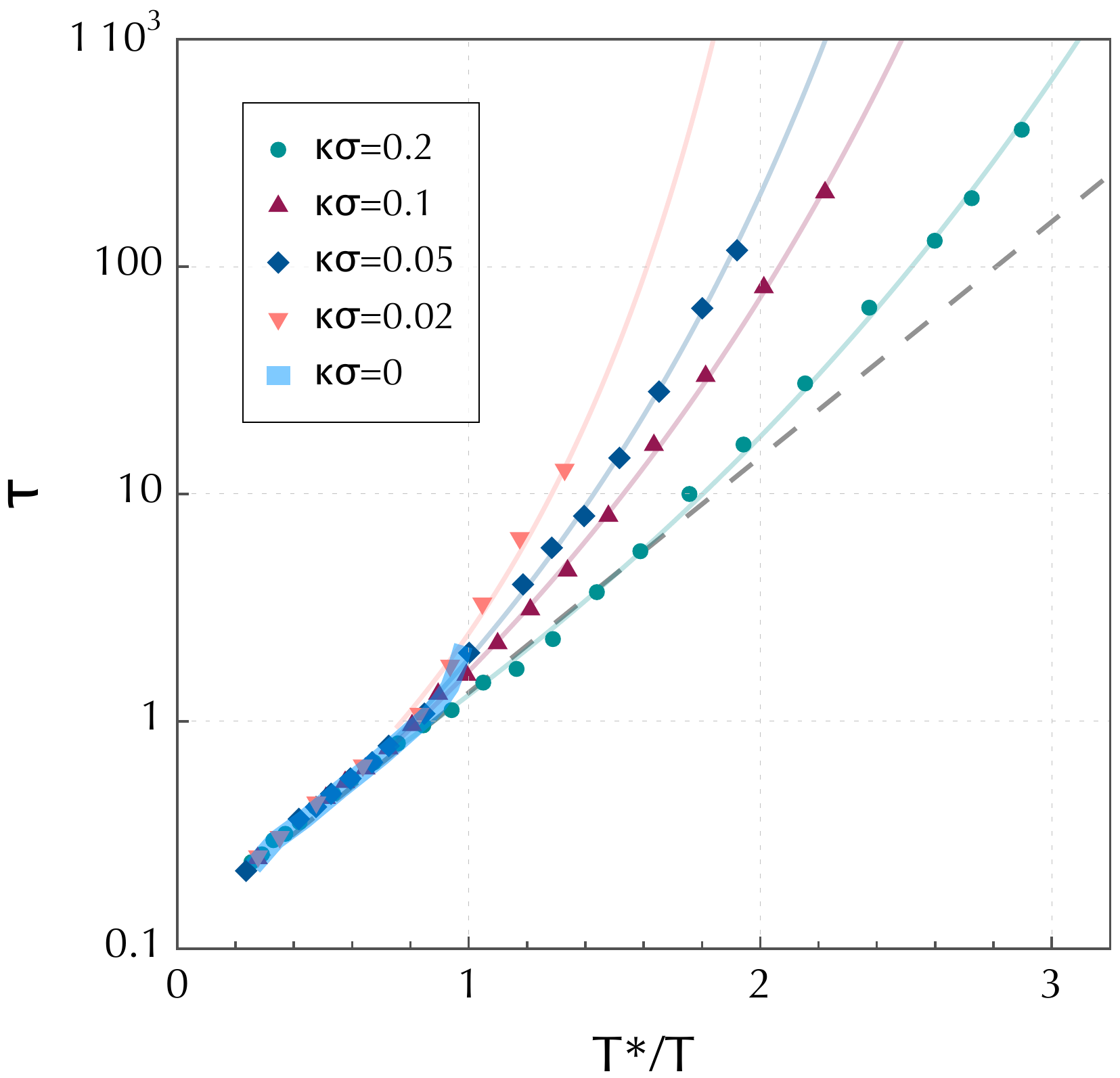}
        \end{center}
        \caption{Slowing down of relaxation in the truncated Lennard-Jones model in $H_2$ (MD simulation\cite{sausset:155701}): logarithm of the translational
 relaxation time  $\tau$ versus $T^*/T$ for $\rho \simeq 0.85$ and for various curvature parameters 
$\kappa \sigma$  ($T^*$ is the ordering temperature in $E_2$). The dotted line is the Arrhenius $T$-dependence roughly observed at temperatures 
above $T^*$.  When $\kappa\sigma >0$, the system remains liquid below $T^*$ until it forms a glass.
The deviation from Arrhenius  behavior increases when curvature (hence frustration) decreases.
 }\label{fig:Arrhenius}
\end{figure}

Glass formation which is preempted by ordering at $T^*$ in the Euclidean plane becomes possible in the hyperbolic plane and the liquid phase can be kept in equilibrium at
 temperatures
 below $T^*$. This is shown in Fig.~\ref{fig:Arrhenius} where we plot the logarithm of the relaxation time versus $1/T$ for several curvatures, as obtained from Molecular 
Dynamics simulation. (Details on the Molecular 
Dynamics simulation technique in hyperbolic geometry and on the generalization of the time-dependent correlation functions are given
 in Refs.\cite{sausset:155701,PhysRevE.81.031504,PhysRevLett.104.065701}.)
One observes that curvature plays virtually no role for temperatures above $T^*$ and that super-Arrhenius behavior, i.e. a deviation from simple Arrhenius $T-$dependence, becomes
significant around $T^*$ and is more pronounced as one decreases the curvature parameter $\kappa\sigma$. These observations support the theoretical predictions that the 
avoided
transition controls the slowing down of the relaxation and that the ''fragility`` of a glassformer, which quantifies how much it deviates from simple Arrhenius behavior,
 decreases as one 
increases the frustration, i.e here the curvature. Note that this trend should also apply to liquids of colloidal particles on a spherical 
substrate: at least for
small enough curvature such that the local order remains hexagonal, decreasing the curvature should lead to a stronger slowing down of the dynamics.

\begin{figure}
        \begin{center}
   \subfloat[]{\label{fig:gderliquid}\includegraphics[draft=false,width=12cm]{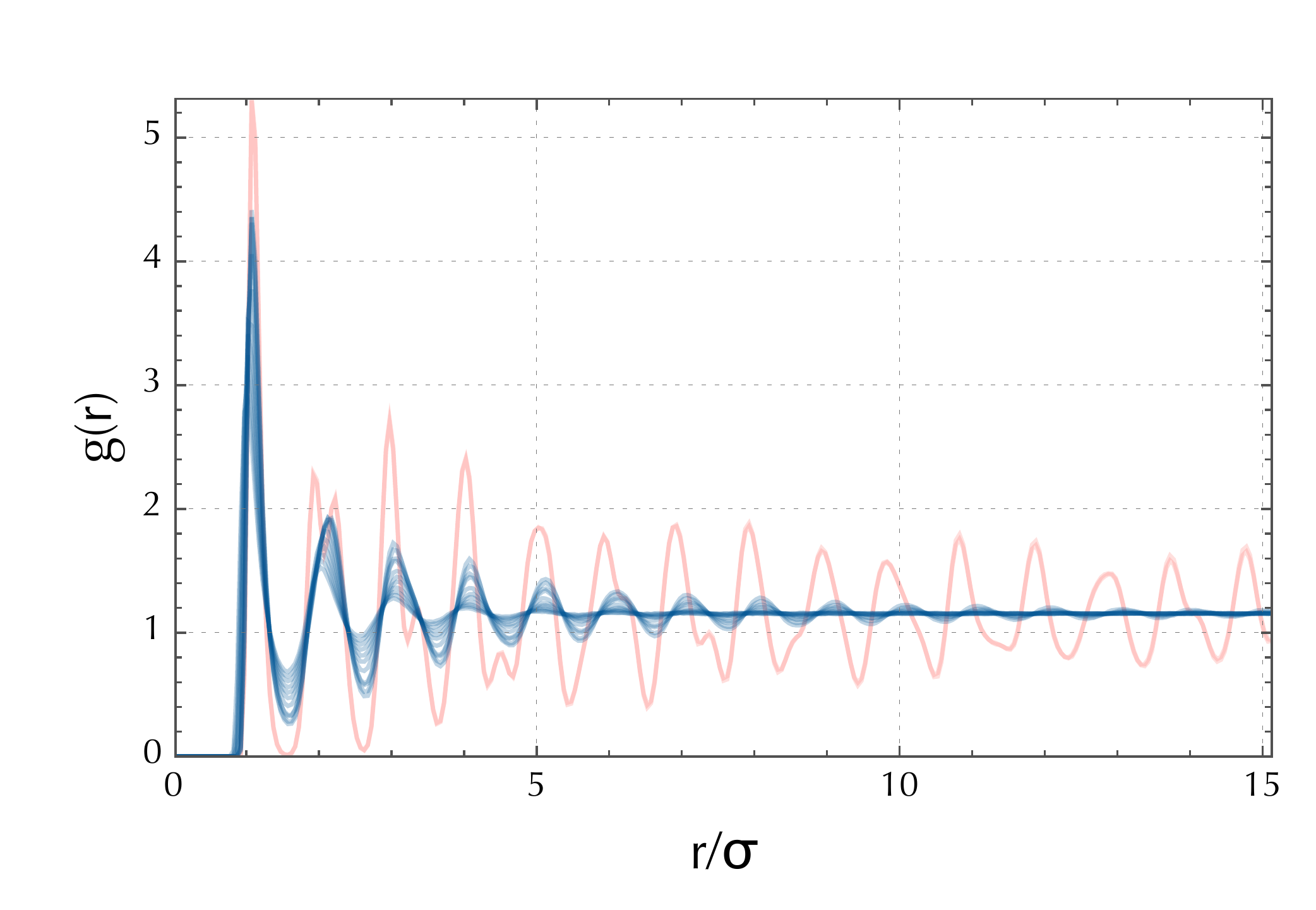}}\\
 \subfloat[]{\label{fig:gderglass}\includegraphics[draft=false,width=12cm]{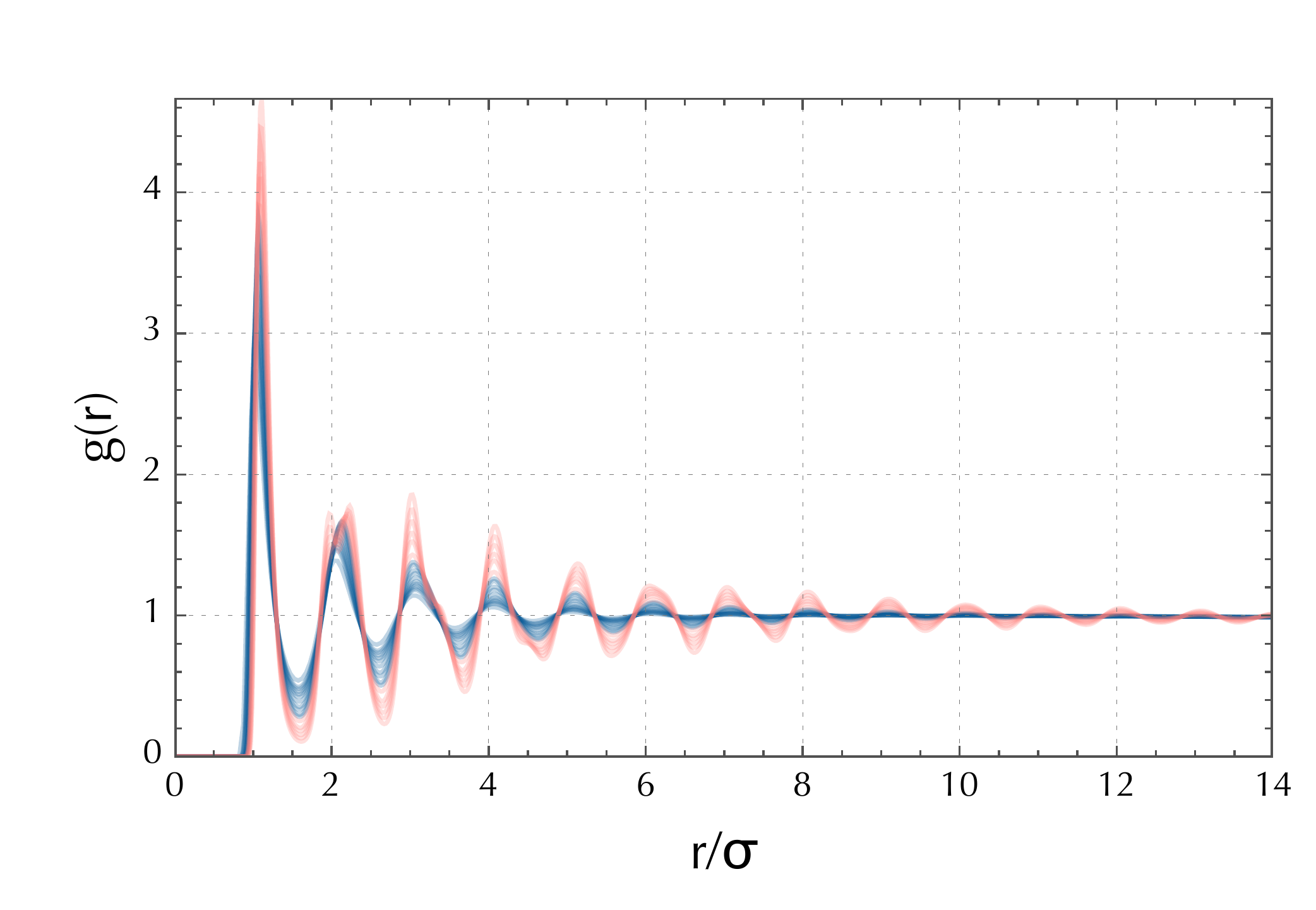}}                            
        \end{center}
        \caption{Radial  distribution function  of the truncated Lennard-Jones model in $H_2$. 
The (blue)  curves correspond to temperatures  above the flat-space ordering 
 transition   ($T > T^*$). \subref{fig:gderliquid}   Euclidean plane ($\kappa\sigma=0$): when  $T < T^*$ (red curves) $g(r)$  displays many peaks corresponding 
to the existence of a 
quasi-long range positional order.  \subref{fig:gderglass} Hyperbolic plane with  $\kappa \sigma = 0.1$: when 
   $T < T^*$ (red curves), there is no signature of quasi-long range order and the system remains in a liquid phase.}
\label{fig:g_r}
\end{figure}

The effect of the curvature can also be seen on the structure of the liquid. This is illustrated in Fig.~\ref{fig:g_r} where the radial distribution function $g(r)$ of the 
Lennard-Jones model in the Euclidean plane is compared to that in $H_2$ with $\kappa \sigma=0.1$. The (quasi) long-range order present below $T^*$ in flat (unfrustrated) space is 
no longer observed in curved space, thereby confirming that the one-component system remains a disordered liquid in the latter.

On general grounds, one expects that curvature-generated frustration induces three different regimes in a liquid as temperature decreases. In a first regime, above $T^*$, the
structure and the dynamics only involve local properties; as, locally, curved space looks  flat,  there is no significant influence
of curvature (at least for small curvatures that accommodate the same local order). A second regime near and below $T^*$ is controlled by the proximity of the
 avoided transition:  
the locally preferred structure extends in space and the associated correlation length grows with decreasing temperature (as can be checked by studying the bond-orientational 
order correlation function $G_6(r)=g_6(r)/g(r)$, see Ref.\cite{PhysRevLett.104.065701}). A final regime is reached when the correlation length associated with growing (bond-orientational)
 order saturates due to 
frustration: the spatial extent of the frustrated order cannot grow beyond the radius of curvature and the final regime is dominated by the presence of an irreducible density
of topological defects. More will be said below about this regime. This three-regime scenario has been observed in computer simulations of liquids in both 
negative\cite{PhysRevLett.104.065701,PhysRevE.81.031504} and positive\cite{Giarritta1992} curvature manifolds.

\section{Ground-state properties, order and defects}\label{sec7}
So far, we have mostly dealt with liquid and fluid phases in curved space. We briefly discuss now low-temperature (or high-density) phases and ground-state properties. In the
preceding section, we have stressed that a nonzero curvature can either induce long-range 
order\footnote{We recall the reader that the terms 'ordered phase' and 'ordering transition' for finite systems, as found in spherical geometries, should be taken with a grain 
of salt. It is possible that, viewed in configurational space, the topography of the energy 'landscape' changes at low energies\cite{Gross1997}, but it remains true that the energy 
barriers involved in equilibrating the system are finite, so that the system cannot freeze in a restricted set of configurations at nonzero temperature. Of course, just like in glass
formation, the equilibration time may be extremely large, which in practice leads to freezing phenomena. In hyperbolic geometry, for which the system may be of infinite extent, 
thermodynamic phase transitions are on the other hand well defined (see Sec.~\ref{sec4} and \ref{sec5}).}, allowing tiling of a homogeneous curved space by regular polytopes for 
specific values of the curvature, or, on the contrary, frustrate the long-range order present in Euclidean space. We focus on the latter situation and we moreover restrict the 
discussion to the two-dimensional case with spherical particles, in which the hexagonal/triangular order that forms the ground-state in flat space is frustrated by the introduction 
of curvature. The topic of order, curvature and defects in two dimensions has recently been nicely reviewed in a quite exhaustive  article by  Bowick and Giomi\cite{Bowick2009}.
Therefore, we only intend to give a brief survey of this aspect, for the sake of completeness of this article.

A first insight into frustrated hexagonal order in non-Euclidean geometries is provided by topological considerations. Consider the Delaunay triangulation of a dense assembly of particles 
on a two-dimensional manifold of genus $g$ (see Sec.~\ref{sec6}) and, if open, with $h$ boundaries. Its Euler characteristic is then  $\chi =2(1-g)$ and the Euler-Poincar\'e theorem
states that any ''tesselation`` (i.e. tiling) of the manifold satisfies
\begin{equation}
 V-E+F=\chi,
\end{equation}
where $V$, $E$ and $F$ are the numbers of vertices, edges and faces in the tesselation. In the case of the Delaunay tesselation by triangles, this leads to 
\begin{equation}\label{eq:chizave}
 \frac{N}{6}(6-\overline{z})=\chi,
\end{equation}
where $N\equiv V$ is the number of particle centers (i.e., of vertices) and $\overline{z}$ is the  average coordination number of the particles. If one defines the topological charge
$q_i$ of a particle with coordination number $z_i$ as $q_i=6-z_i$, one can rewrite Eq.~\ref{eq:chizave} as
\begin{equation}\label{eq:qchi}
 \sum_{i=1}^N q_i=6\chi=12(1-g)-6h.
\end{equation}
For the Euclidean plane, $g=1$ and $h=0$, so that $\chi=0$ and $\overline{z}=6$: perfect hexagonal tiling is possible. For the sphere, $S_2$, $g=2$ and $h=0$, so that the average 
coordination number must be strictly less than $6$. The minimal way to satisfy Eq.~\ref{eq:qchi} is then to have $12$ particles with configuration number number $5$ in an otherwise
$6-$fold coordinated configuration. Such particles appear as (point) topological defects in hexagonal order. They represent positively charged ($q=+1$) disclinations. For the
hyperbolic plane $H_2$, one may first take a detour via periodic boundary conditions (see Appendix B). When imposing a periodic boundary condition, the primitive cell 
containing the system corresponds to  a (compact)
quotient space of genus $g\geq 2$, hence with $\chi=-2(g-1)$. In consequence, the total topological charge in the primitive cell has to be equal to
$-12(g-1)$: there must be an excess of negatively charged disclinations in this case (e.g. of disclinations with $q=-1$ corresponding to $7$-fold coordinated particles). 
By using geometric input in 
the form of the Gauss-Bonnet theorem that relates curvature $K=-\kappa^2$ and characteristic $\chi$,
\begin{equation}\label{eq:gaussbonnet}
 \int_\Sigma dS\, K=2\pi \chi,
\end{equation}
where we recall that $dS=\sqrt{|g(x)|}d^2x $ (see Eq.~(\ref{eq:dS})) and $\Sigma$ is the primitive cell (or fundamental
polygon) of the periodic boundary condition, one finds that the area of the primitive cell is given by
 $A=-2\pi\chi\kappa^{-2}$. As a result, Eq.~(\ref{eq:qchi})  becomes
 \begin{equation}
\frac{1}{A}\sum_{i=1}^N q_i =-\left(\frac{3}{\pi}\right)\kappa^2, 
\end{equation}
which means that the density of topological charge in $H_2$ is only controlled by the curvature, irrespective of 
the choice of periodic boundary condition, thereby providing a well defined thermodynamic limit for a ''bulk`` 
property (see also the discussion in Secs.~\ref{sec3} and \ref{sec4}).

In curved two-dimensional space, hexagonal order must  come with an irreductible number (or density) of topological
defects even in the ground state. Actually, constructing the ground state of a system of particles on a two-dimensional
 manifold is highly nontrivial. The question goes back to Thomson who considered the ground state of repulsive
 charged  particles on a sphere\cite{thomson04}. (Another famous example is the explanation by Caspar and Klug\cite{caspar1962} of the
icosahedral symmetry of spherical virus capsids.) Brute-force
numerical approaches are plagued by the presence of multiple low-energy minima, which are induced by geometric frustration,
and finding the global minimum for a large number of particles is extremely 
difficult.To get around this problem,
coarse-grained approaches have been devised, most prominently a continuum elastic theory developed by Nelson, Bowick,
Travesset and their coworkers\cite{Nelson1987,PhysRevB.62.8738,Bowick2009,PhysRevE.72.036110,PhysRevE.75.021404,Vitelli2006a,PhysRevLett.89.185502}. In the latter,
one directly deals with the defect degrees of freedom, and all the microscopic information about particle interactions
is embedded in effective elastic constants and defect core energies\cite{Bowick2009}. (Note that an alternative route
to the continuum elastic theory has recently been proposed on the basis of a coarse-grained density functional theory:
see Ref.\cite{PhysRevE.81.025701}.)

In the Bowick-Nelson-Travesset approach\cite{PhysRevB.62.8738}, one focuses on the density of disclinations, which 
are the elementary defects from which other defects such as dislocations can be built, and treats the $6$-fold coordinated
particles through continuum elastic theory. At low temperature, the elastic free energy of an arbitrary disclination
density
\begin{equation}\label{eq:disclindensi}
 s({\bf x})=\frac{\pi}{3\sqrt{|g({\bf x})|}} \sum_{i=1}q_i \,\delta^{(2)}({\bf x}-{\bf x}_i),
\end{equation}
where $N_d$ is the total number of disclinations, is given by\cite{Bowick2009,PhysRevB.62.8738}
\begin{equation}\label{eq:disclin}
 F[s]=\frac{Y}{2}\int_\Sigma\int_\Sigma d^2x\sqrt{|g({\bf x})|}d^2y\sqrt{|g({\bf y})|}\,
(s({\bf x})-K({\bf x}))
G_2({\bf x},{\bf y})
(s({\bf y})-K({\bf y}))+\sum_{i=1}^{N_d} E_{core,i},
\end{equation}
where $Y$ is the Young modulus of the hexagonal crystal in flat space, $E_{core,i}$ is a disclination-core free energy
renormalized by thermal fluctuations\cite{PhysRevB.62.8738},  $K({\bf x})$ is the Gaussian curvature at point
 ${\bf x}$, and $G_2({\bf x},{\bf y})$  is the Green function of the bi-Laplacian (or biharmonic operator)
on the manifold $\Sigma$, i.e. satisfying
\begin{equation}
 \Delta^2G_2({\bf x})=\delta^{(2)}({\bf x}),
\end{equation}
with $\Delta$ the Laplace-Beltrami operator. The first term of the free-energy functional is minimized by having 
$s({\bf x})=K({\bf x})$ at each point, which means that the distribution density exactly cancels the effect of the 
Gaussian curvature.
If one considers homogeneous manifolds of constant Gaussian curvature, $K({\bf x})\equiv K$, such as  $S_2$ and $H_2$, the former
condition cannot be everywhere satisfied as the disclinations are discrete objects. Topology however requires a generalization
of the electroneutrality constraint, namely
\begin{equation}
 \int_\Sigma d^2x \sqrt{|g({\bf x})|} \,(s({\bf x})-K)=0,
\end{equation}
which by inserting Eq.~(\ref{eq:disclindensi}) and the Gauss-Bonnet theorem, Eq.~(\ref{eq:gaussbonnet}), is equivalent
to Eq.~(\ref{eq:disclindensi}). Therefore, on average, the disclination charges screen the Gaussian curvature.

\begin{figure}
\begin{center}
                \subfloat[]{
                \includegraphics[draft=false,width=7cm]{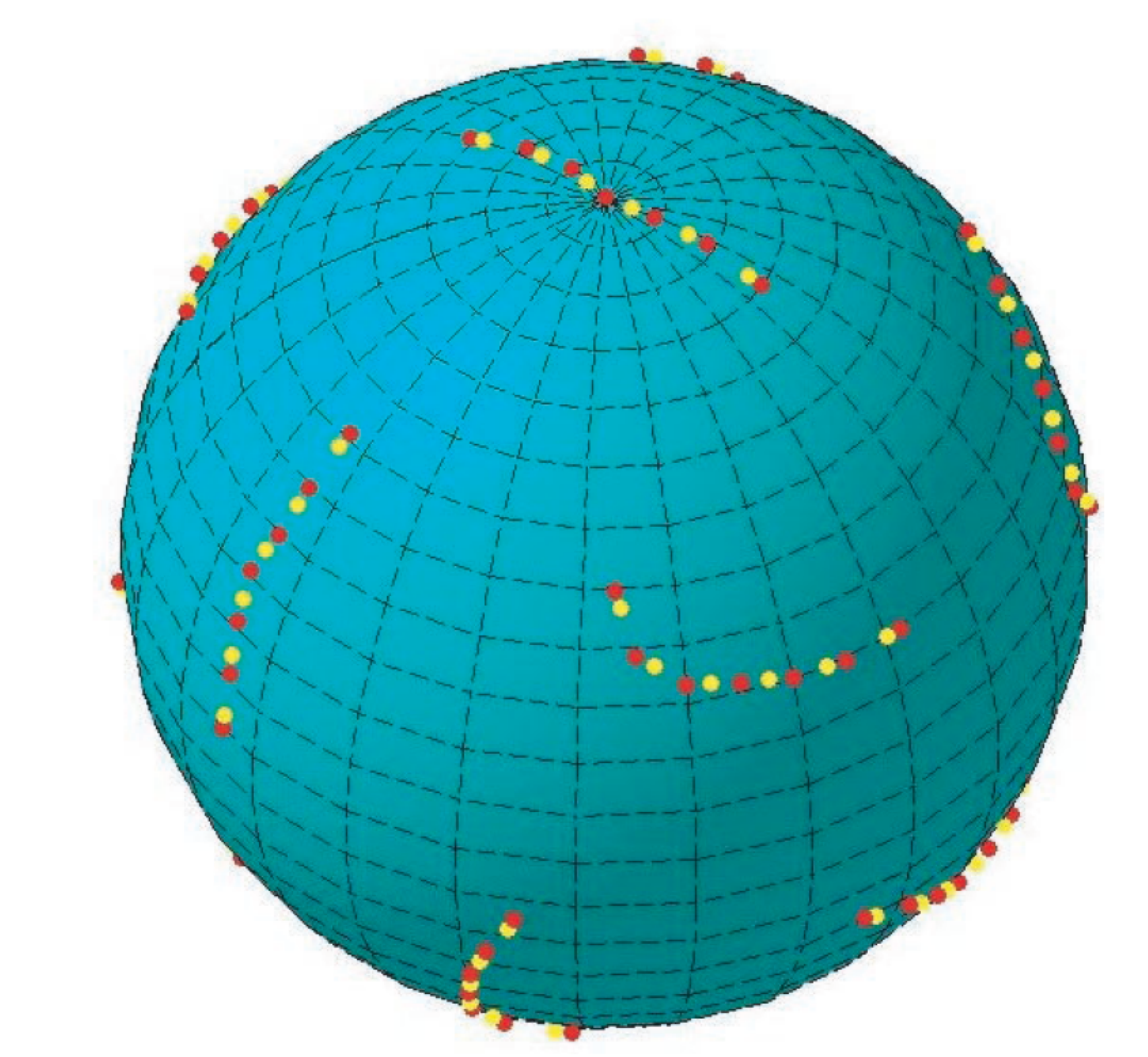}}
                \hspace{1cm}
                \subfloat[]{
                \includegraphics[draft=false,width=7cm]{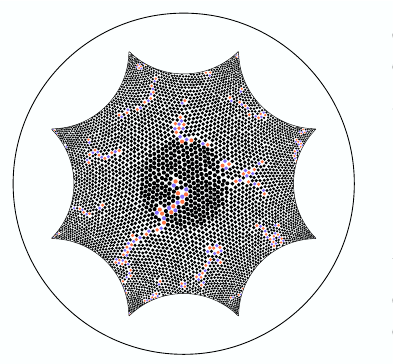}}
        \end{center}
 \caption{Illustration of grain boundary scars in a hexagonal background  in two-dimensional curved space. (a) Model grain boundaries  in $S_2$ obtained by 
minimization of Eq.~(\ref{eq:disclin}) \cite{Bausch2003}. (b) Low-temperature atomic configurations of the truncated Lennard-Jones liquid in the Poincar\'e disk representation of $H^2$\cite{PhysRevLett.104.065701}
 (see Appendix A for the definition of Poincar\'e disk).}\label{fig:scars}
\end{figure}

As the core energy of the dislocations increases with the magnitude of their topological charge, the second term of
the free-energy functional is minimized by having the smallest irreducible number of disclinations with elementary
topological charges, e.g. $12$ disclinations of  charge $q=+1$ on $S_2$ and a density of disclinations of charges
$q=-1$ equal to $\frac{3}{\pi}\kappa^2$ on $H_2$.

For small enough curvature (which for $S_2$ means a large enough number of particles if the particle size $\sigma$ is 
kept fixed), adding extra disclinations in the system on top of the irreducible number may lower the (free) energy
by screening more efficiently the curvature\cite{Bowick2009,PhysRevB.62.8738,PhysRevB.55.3816}. The total charge
of these extra disclinations is zero (so that Eq.~(\ref{eq:qchi}) remains satisfied) and the most economical way for them to organize
is by forming dislocations that consist of ''dipoles`` made by a positive and a negative disclination. The surprising
outcome of the theoretical studies on order on the sphere $S_2$ is that these dislocations form strings that radiate
from each of the irreductible disclinations and terminate in the hexagonal crystalline background at a 
{\it finite} distance\cite{Bowick2009,PhysRevB.62.8738,PhysRevB.55.3816}. These structures which are found found on 
the hyperbolic plane\cite{PhysRevE.81.031504}, but are forbidden  in Euclidean space, have been dubbed ''grain boundary
scars``\cite{PhysRevB.62.8738,Bausch2003}. We illustrate these structures on 
$S_2$\cite{Bausch2003} and $H_2$\cite{PhysRevE.81.031504} in Fig.~\ref{fig:scars}. The theory predicts the number of 
dislocations per irreductible disclination as well as the cut-off distance $r_c$ before which the strings of dislocations 
terminate. For instance, the latter is given in $S_2$ by\cite{PhysRevB.62.8738}

\begin{equation}
 r_c=R\arccos\left(\frac{5}{6}\right)\simeq 0.59 R,
\end{equation}
and in $H_2$ by\cite{PhysRevE.81.031504}
\begin{equation}
 r_c=\kappa^{-1}\mbox{\rm arccosh}\left(\frac{7}{6}\right)\simeq 0.57 R,
\end{equation}
regardless of the microscopic details of the system of particles. These predictions concerning grain boundary scars
in $S_2$ and $H_2$ are well supported by experimental\cite{Bausch2003,Einert2005} and computer simulation 
results\cite{Bowick2009,PhysRevE.81.025701,PhysRevE.81.031504}. The elastic theory based on defects has been generalized to long
range interactions between particles on a sphere, to other two-dimensional manifolds, with varying curvature and possibly open, and 
to other type of order: this is discussed in detail in the review of Bowick and Giomi\cite{Bowick2009}, where all
relevant references can  also be  found. In addition, the dynamics of the defects, essentially the diffusion of the dislocations,
has been studied theoretically and experimentally on the sphere\cite{Bowick2009,Lipowsky:2005,PhysRevE.75.021404} as well as 
theoretically and by computer simulation in the hyperbolic plane\cite{PhysRevE.81.031504}.

\section{Conclusion}\label{sec8}
In this article, we have reviewed the work on the structure and the dynamics of fluids, liquids, and more generally dense
phases, in curved space. We have stressed that the motivation for such studies is  twofold. On the one hand, 
curving space provides an additional control parameter, curvature, to shed light on the behavior of systems of 
interest in the flat, Euclidean space. Examples are provided by the use of spherical and hyperspherical boundary
conditions as well as by investigations of the glass transition in the context of geometric frustration. On the other
hand, there are many physical situations in physical chemistry, soft condensed matter and material science in which
a two-dimensional curved surface is coated by a layer of particles that can move and equilibrate on the curved
substrate. We have focused on cases where the geometry of the substrate is frozen, and most theoretical developments
have been presented for the case of homogeneous two-dimensional manifolds of constant Gaussian curvature, the sphere
$S_2$ and the hyperbolic plane $H_2$.

Non-Euclidean geometries bring in subtleties in the statistical mechanics of systems of particles when compared to 
the standard Euclidean case. We have discussed the main ones, concerning the thermodynamic limit and the effect of
the boundaries, the definition of the pressure and the relations involving correlation functions, the constraints on the
form of the interaction potentials, the properties of particle diffusion, or the nature and the organization of the
defects in dense and quasi-ordered phases. Additional results can be summarized as follows: curvature has  a negligible or weak effect
on the fluid behavior at high temperature and/or low density when  the local structure or dynamics are probed. On the
other hand, a nonzero curvature has strong consequences in dense phases where it can either induce or frustrate 
ordering. In the latter case, it slows down the relaxation and allows glass formation, as well as imposes in an 
irreducible number of topological defects in ground-state and low-temperature configurations. Curvature has also
a distinct influence when long-range interactions, as in Coulombic systems, or long-range correlations, as in the
vicinity of a critical point, are expected: there, the radius of curvature acts as a cutoff that imposes exponential 
decay at long distances.

Needless to say that progress should be made to describe liquids and fluids on substrates with varying curvature 
or even with  fluctuating curvature as encountered in membranes. Above all, one can hope that in the near future more experiments
will be performed on curved substrates to provide systematic information on the phase behavior and the dynamics of
particle systems for a significant range of temperature or density and for several curvatures or geometries.

\appendix
\section{A recap on Riemannian manifolds}
To describe liquids on curved surfaces, let us introduce some elements of
differential geometry\cite{Terras:1985,goetz1970,Sadoc:1999}.
A  $d-$dimensional Riemannian manifold   is defined by a set of coordinates ${\bf x}=(x_1,x_2,...,x_d)$ and a metric tensor
 $g_{ij}({\bf x})$. The
length element $ds$ is defined by 
\begin{equation}
 ds^2=\sum_{ij}^d g_{ij}({\bf x})dx_i dx_j,
\end{equation}
and  the  ''volume'' element is given by
\begin{equation}
 dS=\sqrt{|g({\bf x})|}\prod_{i=1}^d dx_i,
\end{equation}
where $|g({\bf x})|$ denotes the absolute value of the determinant of the metric tensor.

 Diagonalizing the  curvature tensor
provides the   principal directions of  the manifold (eigenvectors) at point ${\bf x}$
and the   eigenvalues are  the  radii of curvature, $R_i, i=1,...,d$.   The mean
curvature is defined as the algebraic average of the  curvatures,
\begin{equation}
 k({\bf x})=\frac{1}{d}\sum_{i=1}^d\frac{1}{R_i({\bf x})}.
\end{equation}
In the case where the manifold is two-dimensional, one defines
the Gaussian curvature as the product of the two curvatures,
\begin{equation}
 K({\bf x})=\frac{1}{R_1({\bf x})R_2({\bf x})}.
\end{equation}

If $X$ is a vector field (defined at each point ${\bf x}$ in the tangent Euclidean manifold), the divergence is given by
\begin{equation}
 div(X)=\frac{1}{\sqrt{|g({\bf x})|}}\sum_{i=1}^d \frac{\partial (\sqrt{|g({\bf x})|} X^i)}{\partial x_i}
\end{equation}
and the gradient of the scalar function $f$
\begin{equation}
 (grad(f))^i=\sum_{j=1}^d  g^{ij}\frac{\partial f}{\partial x_j}
\end{equation}
where $g^{ij}({\bf x})$ is the inverse tensor of $g_{ij}({\bf x})$, i.e. $\sum_{j=1}^d g_{ij}g^{jk}=\delta_i^k$ with  $\delta_i^k$  the Kronecker
symbol.

The Laplace-Beltrami operator $\Delta$ acts on a function $f({\bf x})$  as $\Delta f=div\, grad (f)$, i.e.,
\begin{equation}
 \Delta f =\frac{1}{\sqrt{|g({\bf x})|}}\sum_{i=1}^d   \frac{\partial }{\partial x_i}\left(\sqrt{|g({\bf x})|}g^{ij}({\bf x})  \frac{\partial f}{\partial x_j} \right)
\end{equation}

We mostly consider in this article homogeneous manifolds of constant nonzero curvature. In two dimensions, there are two associated geometries, 
the spherical and the hyperbolic ones.
For the sphere $S_2$, the metric tensor is diagonal in the angular coordinates  $\theta$ (colatitude) and $\phi$ (longitude) and one has
\begin{equation}
 g_{\theta\theta}=R^2,\,\,\; g_{\phi\phi}=R^2\sin(\theta)^2.
\end{equation}

The squared length element $ds^2$ is then equal to
\begin{equation}
 ds^2=R^2(d\theta^2+\sin(\theta)^2 d\phi^2),
\end{equation}
the area element is given by
\begin{equation}
 dS=R^2\sin(\theta) d\phi d\theta,
\end{equation}
and the two radii of curvature are equal and constant with $R_1=R_2=R$ and the Gaussian curvature $K=R^{-2}$. Finally, the Laplace-Beltrami operator is 
\begin{equation}
 \Delta  =\frac{1}{\sin(\theta)} \frac{\partial}{\partial \theta}\left( \sin(\theta) \frac{\partial}{\partial \theta}\right)+
\left(\frac{1}{\sin(\theta)}\right)^2 \frac{\partial^2}{\partial \phi^2}.
\end{equation}

For the hyperbolic plane $H_2$ (also called ``pseudosphere`` or ``Bolyai-Lobachevski plane``)\cite{Coxeter:1969,Hilbert:1952}, one can use the polar
coordinates $r$ and $\phi$.
The metric tensor is then diagonal with 
$g_{rr}=1$ and $g_{\phi\phi}=\kappa^{-2}\sinh(\kappa r)^2$,
which gives a squared length element 
\begin{equation}
 ds^2=dr^2+\left(\frac{\sinh(\kappa r)}{\kappa}\right) d\phi^2
\end{equation}
and an  area element 
\begin{equation}
 dS=\frac{\sinh(\kappa r)}{\kappa}dr d\phi.
\end{equation}
The two radii of curvature are of opposite signs, $R_1=-R_2=\kappa^{-1}$, so that $K=-\kappa^{2}$. In addition, the Laplace-Beltrami operator is given by
\begin{equation}
 \Delta  =\frac{1}{\sinh(\kappa r)}\left(\frac{\partial}{\partial r} \sinh(\kappa r) \frac{\partial}{\partial r}\right)+
\left(\frac{\kappa}{\sinh(\kappa r)}\right)^2   \frac{\partial^2}{\partial \phi^2}\,.
\end{equation}
The hyperbolic plane cannot be embedded in three-dimensional Euclidean space (contrary to $S_2$) and ''models``, i.e. projections must be used
for its visualization. A convenient one is the Poincar\'e disk model which projects the whole hyperbolic plane $H_2$ onto a unit disk. The 
projection is conformal (angles are conserved) but not isometric (distances are deformed and shrinks as one reaches the disk boundary). This 
representation is used in Fig.~\ref{fig:scars}b. If $x$ and $y$ are the Cartesian coordinates of a point on the unit disk, the relation to 
the above polar coordinate is as follows:
\begin{align}
 \sqrt{x^2+y^2}&=\tanh\left(\frac{\kappa r}{2}\right),\\
\frac{y}{x}&=\tan(\phi),
\end{align}
and the squared length element $ds^2$ is given by 
\begin{equation}
\mathrm{d} s^2=\kappa^{-2} \frac{4\left(\mathrm{d} x^2 +\mathrm{d} y^2\right)}{\left(1-\left(x^2 + y^2\right)\right)^2 }.
\end{equation}
From the above formulae, one can compute for instance the geodesic distance  $r_{12}$ between two points ${\bf r}_1$ and ${\bf r}_2$.
In $S_2$, one finds 
\begin{equation}\label{eq:trigoS2}
 \cos(\theta_{12})=\cos(\theta_{1})\cos(\theta_{2})-\sin(\theta_{1})\sin(\theta_{2})cos(\phi_1-\phi_2)
\end{equation}
where ($\theta_{1},\phi_{1}$) and ($\theta_{2},\phi_{2}$) are the coordinates of the two points and $R\theta_{12}$ the geodesic distance between these 
points. On the other
hand in $H_2$, one has
\begin{equation}\label{eq:trigoH2}
 \cosh(\kappa r_{12})=\cosh(\kappa r_1)\,\cosh(\kappa r_2)-\sinh(\kappa r_1)\,\sinh(\kappa r_2)\,\cos(\phi_1-\phi_2),
\end{equation}
where ($r_{1},\phi_{1}$) and ($r_2,\phi_{2}$) are the (polar) coordinates of the two points.
Note the symmetry between Eqs.~(\ref{eq:trigoS2}) and (\ref{eq:trigoH2}) in the 
exchange of $R\leftrightarrow i\kappa^{-1}$ . The same symmetry takes place in relating spherical and hyperbolic
trigonometries. So for instance, from the standard result on a sphere, one finds the following trigonometric
relations for the a general hyperbolic triangle with sides $a$, $b$, and $c$ and opposite angle $\alpha$, 
$\beta$ and $\gamma$\cite{Coxeter:1969}:
\begin{equation}
 \frac{\sinh(\kappa a)}{\sin(\alpha)}=\frac{\sinh(\kappa b)}{\sin(\beta)}=\frac{\sinh(\kappa c)}{\sin(\gamma)}
\end{equation}
\begin{equation}\label{eq:trigo2H2}
 \cosh(\kappa c)=\cosh(\kappa a)\,\cosh(\kappa b)-\sinh(\kappa a)\,\sinh(\kappa b)\,\cos(\gamma),
\end{equation}
\begin{equation}
 \cosh(\kappa c)=\frac{\mathrm{cos}(\alpha)\,\mathrm{cos}(\beta)+\mathrm{cos}(\gamma)}{\mathrm{sin}(\alpha)\,\mathrm{sin}(\beta)}\, .
\end{equation}
Note that the first two relations have Euclidean counterparts (obtained by letting  $\kappa\rightarrow 0$),
but not the third one which is specific to nonzero curvatures.

\section{Periodic boundary conditions on the hyperbolic plane}
Generically, implementing periodic boundary conditions consists in choosing a primitive cell containing the physical system such that it can be infinitely 
replicated to tile the whole space.
So, prior to building periodic boundary conditions, one needs to know the allowed tilings of the space under consideration.
Here, for simplicity, we will limit ourself to regular cells and so to regular tilings.

On the hyperbolic plane $H_2$, an infinity of regular tilings $\{p,q\}$ are allowed if $p$ (the number of edges of the primitive cell) and $q$ (the number of cells meeting at each vertex of the tiling) verify the following condition
\begin{equation}
	\label{eq:tiling}
	(p - 2) (q - 2) > 4.
\end{equation}
This, therefore, opens the possibility to have an infinite number of  possible periodic boundary conditions.

To ensure smoothness and consistency, the edges of the primitive cell of any periodic boundary condition have to be paired in a specific way: leaving the cell 
through one edge implies to come back by another edge, a process which should be equivalent to exploring the whole tiling of the plane. Constraints thus arise on 
how cell edges are paired 
together in addition to those on the shape of the cell.
We give in the following a rapid overview of how to classify and construct periodic boundary conditions in $H^2$ by describing cell shapes and edge pairings.

\begin{figure}
	\begin{center}
		\includegraphics[width=6cm]{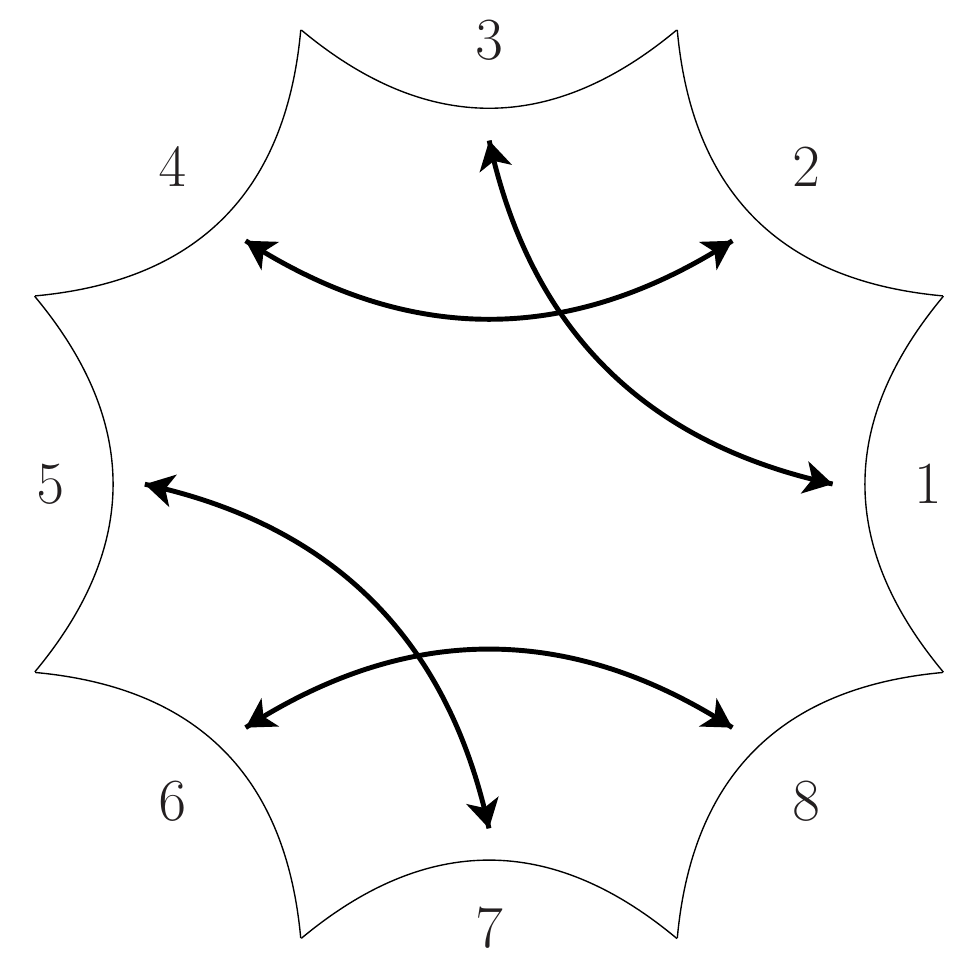}
	\end{center}
	\caption{Simplest fundamental polygon on the hyperbolic plane. It is associated to the $\{8,8\}$ tiling. The arrows indicate the way edges are paired.}\label{fig:poly}
\end{figure}

\begin{figure}
	\begin{center}
		\subfloat[]{\includegraphics[width=6cm]{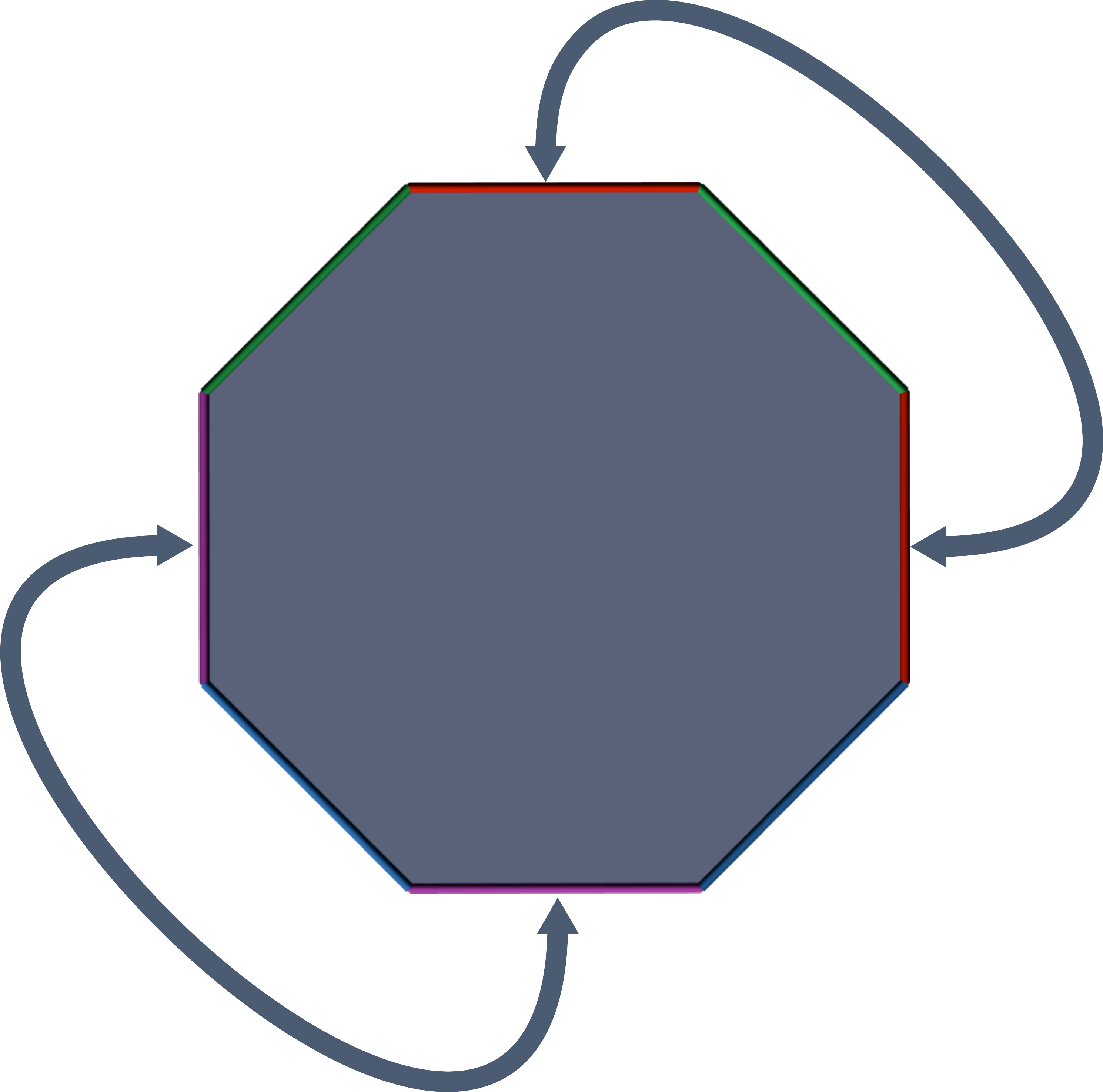}}
		\hspace{2cm}
		\subfloat[]{\includegraphics[width=6cm]{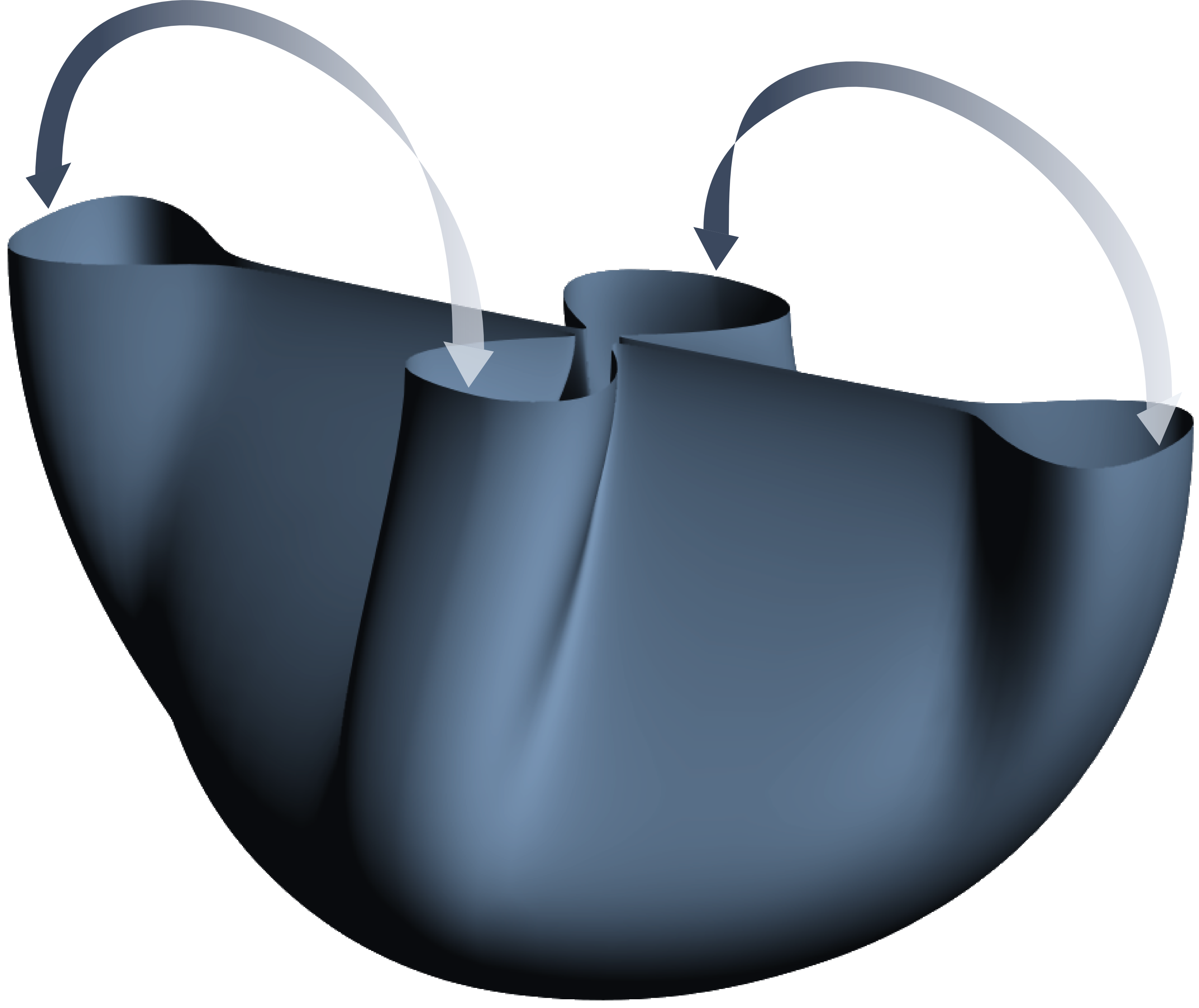}}
	\end{center}
	\caption{Schematic representation of the ``compactification" of the fundamental polygon shown in Fig.~\ref{fig:poly}. The paired edges are glued together: 1 with 3 and 5 with 7 in (a); 2 with 4 and 6 with 8 in (b). The final compact manifold is a two-hole torus
 represented in Fig.~\ref{fig:tor}.}\label{fig:rep}
\end{figure}

\begin{figure}
	\begin{center}
		\includegraphics[width=9cm]{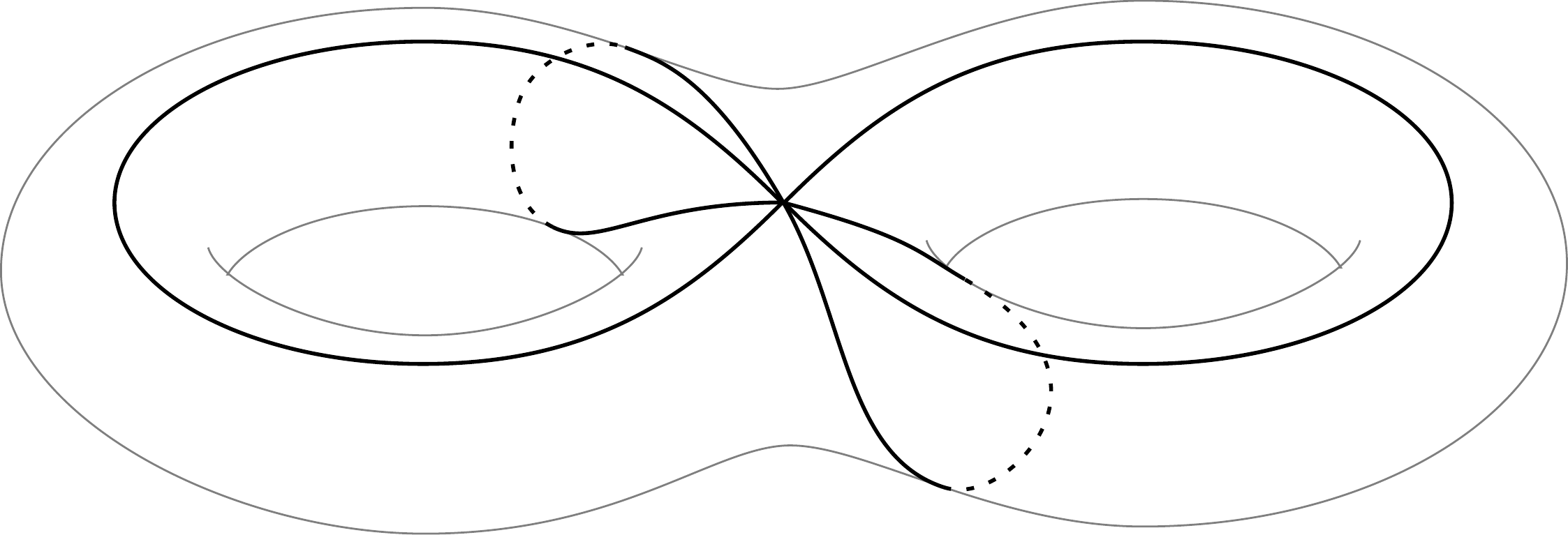}
	\end{center}
	\caption{Compact manifold and graph obtained by gluing the edges of the fundamental polygons of
 Fig.~\ref{fig:poly}.}\label{fig:tor}
\end{figure}

First, a \emph{fundamental polygon} (primitive cell with properly paired edges) encodes all the needed information to build periodic boundary conditions 
and to replicate the system in the entire space. The simplest (and smallest) fundamental polygon in $H^2$ is an octagon, corresponding to an $\{8,8\}$ tiling, 
with the edge pairing  shown in
 Fig.~\ref{fig:poly}.
 By gluing the paired edges together (see Fig.~\ref{fig:rep} for of visualization of intermediate states) it is possible to represent the fundamental polygon as a compact 
manifold, also corresponding to the ''quotient space'' (see Fig.~\ref{fig:tor}). The above octagonal periodic boundary condition  leads to a $2$-hole torus, whose genus 
(number of holes, here $g=2$ ) fixes the area of the fundamental polygon through the Gauss-Bonnet theorem (see Eq.~(\ref{eq:gaussbonnet}): 
$A=4\pi\kappa^{-2}(g-1)$). 
As the genus is an integer, compact manifolds embedded in the hyperbolic plane can only lead to a discrete set of areas. In more technical words, an homothety cannot 
be applied to hyperbolic manifolds without changing the curvature. Here, the octagonal fundamental polygon cannot be scaled at constant curvature to allow one to study 
a bigger system for instance. Therefore, to change the area of the fundamental polygon in $H^2$, one as to vary its genus, which in turn implies
to change its symmetry, more particularly its number of edges.

To classify all the possibilities (for regular polygons), one can use the properties of the graph formed by glued edges of the fundamental polygon and 
embedded in the associated $g$-hole torus
 (see Fig.~\ref{fig:tor} for an  example on the $2$-hole torus). The constraints on such graphs to obtain relevant periodic boundary conditions are detailed 
in~\cite{Sausset:2007} which allows one to classify and build all possible regular periodic boundary conditions 
in the hyperbolic plane~\footnote{A tool to build such periodic boundary conditions can be found at the following address: http://physics.technion.ac.il/$\sim$sausset/CLP.html}.
 The classification exhibits ``families'' that comprise graphs with a given number of vertices and a given pairing pattern but different values of the genus 
$g$ and that  share similarities when varying $g$. 
In this framework, the octagonal fundamental polygon shown in Fig.~\ref{fig:poly} 
can be seen as the direct generalization of the square
 periodic boundary condition encountered in the Euclidean plane (both are in the same ``family'').

%

\end{document}